\documentclass{cernyrep}
\usepackage[T1]{fontenc}
\usepackage[bookmarks, colorlinks=true, linktoc=page, pdftex, linkcolor=black, citecolor=black, urlcolor=blue]{hyperref}

\usepackage{fancyhdr}
\fancyhfoffset{4 mm}
\fancypagestyle{ARTTITLE}{%
\fancyhf{} 
\lhead{\hfill CERN Accelerator School Proceedings ---
{\it RF for Accelerators} ---  Berlin,  Germany, 2023\hfill}
\lfoot{\hspace{3mm} Available online at \url{https://cas.web.cern.ch/previous-schools}}
\rfoot{\thepage\hspace*{3mm}}
 
}

\usepackage{varwidth}
\usepackage{xcolor}
\usepackage{tikz}
\usepackage{amsmath}
\usepackage{amssymb}
\usepackage{epic,eepic,pspicture,mathptmx,times,graphpap}
\usepackage{tikz}				
\usetikzlibrary{decorations}
\usetikzlibrary{decorations.pathmorphing}
\usetikzlibrary{decorations.pathreplacing}
\usetikzlibrary{arrows,backgrounds,automata}
\usetikzlibrary{trees}
\usetikzlibrary{patterns}
\newcommand{\eqn}[3]{\parbox{9cm}{#1}\hspace{1cm}\parbox{5cm}{\raggedright\bf #2}\hfill\parbox[b]{8mm}{\begin{equation} \label{#3} \end{equation}}}

\begin{document}
\newcommand{\meqn}[3]{\noindent\parbox{9cm}{\begin{align*} #1 \end{align*}}\hspace{1cm}\parbox{5cm}{\raggedright\bf #2}\hfill\parbox[b]{8mm}{\begin{equation} \label{#3} \end{equation}}}
\newcommand{\dd}{\mathrm{d}}
\newcommand{\twovector}[2]{\begin{pmatrix} \displaystyle #1 \vspace{0.2cm} \\ \displaystyle #2 \end{pmatrix}} 
\newcommand{\threevector}[3]{\begin{pmatrix} \displaystyle #1 \vspace{0.2cm}\\ \displaystyle #2 \vspace{0.2cm}\\ \displaystyle #3 \end{pmatrix}} %
\title{RF Basics I \& II}
\author{F. Gerigk}
\institute{CERN, Geneva, Switzerland}

\begin{abstract}
This lecture starts with a brief historical introduction and an explanation how to get from Maxwell's Equations to a simple cavity. 
After simplifying and adapting the equations for their application to Radio Frequency problems, the~most important formulae and characteristics for cavities and wave-guides are derived. The most common figures of merit are explained and some of the~different cavity types are introduced. The alternative description of cavities as a~lumped circuit model is then introduced, which is often used to characterise the cavity-coupler-generator interplay. 
\end{abstract}

\maketitle
\thispagestyle{ARTTITLE}
\section{Historical introduction}
The first RF linear accelerator (linac) was proposed and tested by Rolf Wider\"{o}e in 1927/1928. He connected a single drift tube between two grounded electrodes to an RF source, which delivered 25\,keV at a frequency of 1\,MHz. In his experiment he accelerated potassium ions to an energy of 50\,keV and thus demonstrated the principle of RF acceleration. This meant that for the first time the maximum obtainable energy of an accelerator was no longer limited by the electrostatic breakdown voltage of DC machines, which were in operation at the time. With this principle it was now possible to multiply the available voltage of an RF source $V_{RF}$ by the number of gaps $N$ and thus to increase the maximum energy gain $\Delta E$ of a particle with charge $q$ to\\
	\eqn{\[\Delta E=qN_{gap}V_{RF}\mbox{.}\]}{energy gain}{eq:eneg}\\

The principle was soon extended to 30 drift tubes and a 10\,MHz RF source by Sloan and Lawrence in 1931 to accelerate mercury ions to $1.26$\,MeV. 

\begin{figure}[h!]
	\begin{center}
	\begin{tikzpicture}[>=stealth, x=5cm/5,red!70!black,thick]
	\draw[->,black,dashed, line width=1pt](-1.9,0) -- (-0.9,0) (-0.5,0) -- (0.3,0) (0.9,0) -- (1.7,0) (2.5,0) -- (3.3,0) (4.3,0) -- (5,0)node[right]{beam};
	\draw[black](5,-3)--(7,-3)--(7,-1)--(5,-1)--(5,-3);
	\draw[black](6,-2) circle (0.5cm); 
	\draw[black](6,-1.3) node {RF source};
	\draw[red,very thick,samples=100,domain=5.7:6.3] plot (\x, {0.2*sin(600*\x )-2});
	\draw[black] (-2.3,-0.4) -- (-2.3,0.4) -- (-1.7,0.4) -- (-1.7,0.2) arc (90:270:0.2) -- (-1.7,-0.4) -- (-2.3,-0.4);
	\draw[black] (-2.5,0.6) node {ion source}; 
	\draw[black] (5,-2.5) -- (-2,-2.5) (5,-1.5) -- (-1,-1.5) (-2,-0.4) -- (-2,-4) (-2.2,-4) -- (-1.8,-4);
	\fill [black] (-2,-2.5) circle (2pt);
	\foreach \x in {0.3,3.5}
		{\fill [black] (\x,-2.5) circle (2pt);
		\draw[black] (\x,-2.5) -- (\x,-0.8);}
	\foreach \x in {-1,1.8}
		{\fill [black] (\x,-1.5) circle (2pt);
		\draw[black] (\x,-1.5) -- (\x,-0.8);}
	\foreach \x in {-1.2,0,1.4,3}
		\draw[black] (\x,0) ellipse (0.3cm and 0.8cm);
	\foreach \x in {-0.8,0.6,2.2,4}
		\draw[black] (\x,-0.8) arc (-90:90:0.3cm and 0.8cm);
	\draw[black] (-1.2,0.8) -- (-0.8,0.8) node [above,midway] {1} (-1.2,-0.8) -- (-0.8,-0.8);
	\draw[black] (0,0.8) -- (0.6,0.8) node [above,midway] {2} (0,-0.8) -- (0.6,-0.8);
	\draw[black] (1.4,0.8) -- (2.2,0.8) node [above,midway] {$\ldots$} (1.4,-0.8) -- (2.2,-0.8);
	\draw[black] (3,0.8) -- (4,0.8) node [above,midway] {n} (3,-0.8) -- (4,-0.8);
	\draw[<->,black] (2.6,1.4) -- (4.2,1.4) node [above, midway] {$l_n$};
	\foreach \x in {-0.5,2.5}
		\draw[->,red, line width=6pt] (\x,0) -- (\x+0.8,0);
	\foreach \x in {-1.8,0.8}
		\draw[<-,red, line width=6pt] (\x,0) -- (\x+0.8,0);
	\end{tikzpicture}
	\caption{Principle of Wider\"{o}e's RF linear accelerator: RF electric fields between drift tubes accelerate the particles}
	\label{fig:wideroe}
	\end{center}
	\end{figure}
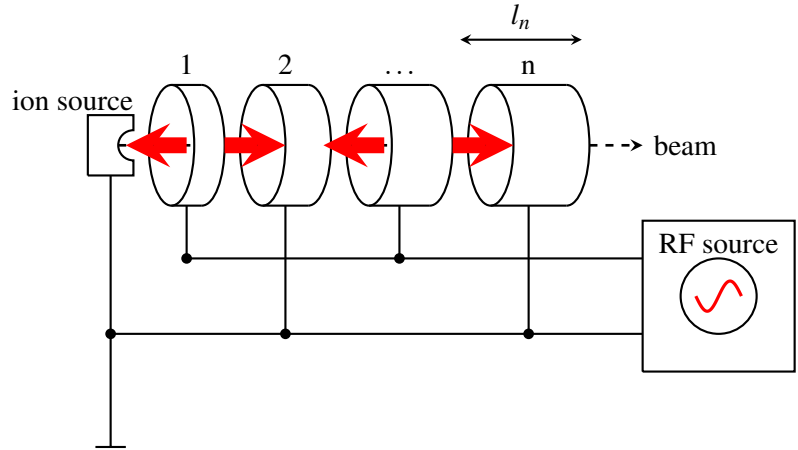

The basic idea of the Wider\"{o}e's linac is shown in Fig.~\ref{fig:wideroe}. Subsequent drift tubes are connected to opposite polarities of an RF source. This means that at a `frozen' point in time every second gap has an~RF phase suitable for acceleration. The idea is that while the particles move from one gap to the next the polarity of the RF source changes, so that particles always see an accelerating field in the gaps. In order for this to work, the RF oscillations have to be in synchronism with the passage of the accelerated particles. For a fixed RF frequency $f$ this leads to the following condition for the distance $l_{\,n}$ between the gaps of a Wider\"{o}e's linac:\\
	\eqn{\[l_n=\frac{v}{2 f}\mbox{,}\]}{synchronism condition}{eq:sync}\\
	
\noindent which means that the distance between the gap centres $l_{\,n}$ has to become larger with increasing particle velocity $v$. 

\begin{figure}[h!]
	\begin{center}
	\includegraphics[width=8cm]{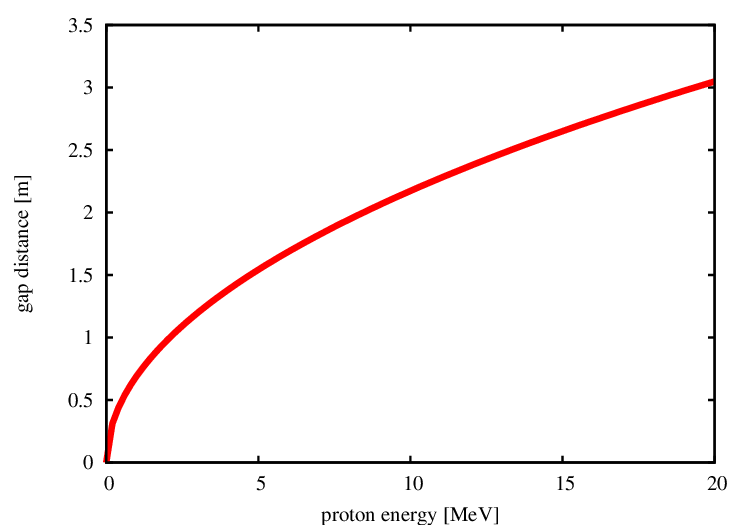}
	\caption{Gap distance for proton acceleration with a 10\,MHz RF source}
	\label{fig:dtprotons}
	\end{center}
	\end{figure}

Even though the Wider\"{o}e linac revolutionized the acceleration of particles, its maximum energy was still severely limited for two reasons: i) For higher velocities the length of the drift tubes has to increase, which means that there was a natural `practical' limit for these machines. Especially for light ions or protons the drift tubes would simply become too long (see Fig.~\ref{fig:dtprotons}). One can solve this problem by going to higher frequencies, but ii) since the accelerating structure is not enclosed by an electric boundary, the operation at higher frequencies ($> 10$\,MHz) meant that the drift tubes were basically becoming antennas. With increasing frequency they radiate more and more of the RF energy instead of using it for acceleration, thus leading to a very poor efficiency of the accelerator. 

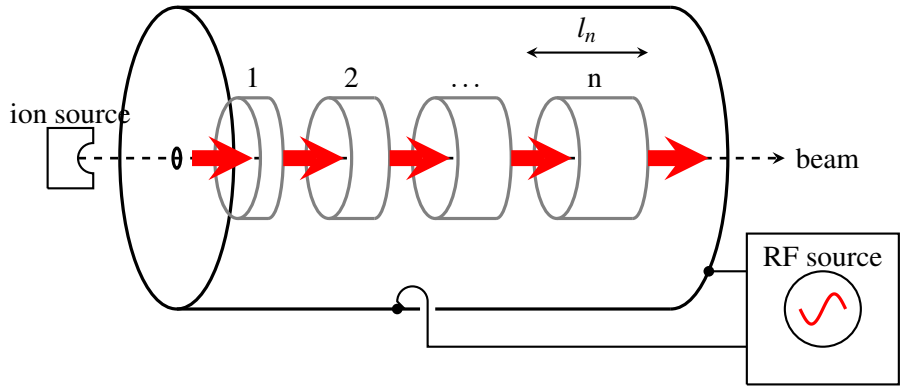
\begin{figure}[h!]
	\begin{center}
	\begin{tikzpicture}[>=stealth, x=5cm/5,red!70!black,thick]
	\draw[->,black,dashed, line width=1pt](-3.3,0) -- (-0.9,0) (-0.5,0) -- (0.3,0) (0.9,0) -- (1.7,0) (2.5,0) -- (3.3,0) (4.3,0) -- (6,0)node[right]{beam};
	\draw[black](5.5,-3)--(7.5,-3)--(7.5,-1)--(5.5,-1)--(5.5,-3);
	\draw[black](6.5,-2) circle (0.5cm); 
	\draw[black](6.5,-1.3) node {RF source};
	\draw[red,very thick,samples=100,domain=6.2:6.8] plot (\x, {0.2*sin(600*(\x-0.5) )-2});
	\draw[black] (-3.7,-0.4) -- (-3.7,0.4) -- (-3.1,0.4) -- (-3.1,0.2) arc (90:270:0.2) -- (-3.1,-0.4) -- (-3.7,-0.4);
	\draw[black] (-3.4,0.6) node {ion source}; 
	\draw[black] (5.5,-2.5) -- (1.3,-2.5) -- (1.3,-1.9) arc (0:180:0.2) -- (1.0,-2) (5.5,-1.5) -- (5,-1.5);
	\fill [black] (0.9,-2) circle (2pt);
	\fill [black] (5,-1.5) circle (2pt);
	\draw[black, very thick] (-2,2) -- (4.5,2) (4.5,-2) -- (1.4,-2) (1.2,-2) -- (-2,-2); 
	\draw[black, very thick] (-2,0) ellipse (0.75cm and 2cm);
	\draw[black, very thick] (4.5,-2) arc (-90:90:0.75cm and 2cm);
	\draw[black, very thick] (-2,0) ellipse (0.05cm and 0.133cm);
	\foreach \x in {-1.2,0,1.4,3}
		\draw[gray, very thick] (\x,0) ellipse (0.3cm and 0.8cm);
	\foreach \x in {-0.8,0.6,2.2,4}
		\draw[gray, very thick] (\x,-0.8) arc (-90:90:0.2cm and 0.8cm);
	\draw[gray, very thick] (-1.2,0.8) -- (-0.8,0.8) node [black, above,midway] {1} (-1.2,-0.8) -- (-0.8,-0.8);
	\draw[gray, very thick] (0,0.8) -- (0.6,0.8) node [black, above,midway] {2} (0,-0.8) -- (0.6,-0.8);
	\draw[gray, very thick] (1.4,0.8) -- (2.2,0.8) node [black, above,midway] {$\ldots$} (1.4,-0.8) -- (2.2,-0.8);
	\draw[gray, very thick] (3,0.8) -- (4,0.8) node [black, above,midway] {n} (3,-0.8) -- (4,-0.8);
	\draw[<->,black] (2.6,1.4) -- (4.2,1.4) node [black, above, midway] {$l_n$};
	\foreach \x in {-1.8,-0.6,0.8,2.4,4.2}
		\draw[->,red, line width=6pt] (\x,0) -- (\x+0.8,0);
	\end{tikzpicture}
	\caption{Principle and field profile of an Alvarez linac}
	\label{fig:Alvarez}
	\end{center}
	\end{figure}

So the answer to the title of this section is ``Yes, we can accelerate without cavities, but reaching higher energies becomes very inefficient''. A solution was finally proposed by Louis Alvarez in 1946, who put the Wider\"{o}e linac into a conducting cylinder. This solved the problem of radiated energy but in addition to the synchronism condition of Eq.~(\ref{eq:sync}) one also has to make sure that each gap has the~same inherent resonant frequency, which is now determined by the diameter of the cylinder, the distance between the drift tubes, and their diameter. Figure~\ref{fig:Alvarez} shows the principle of the Alvarez linac
and we can note an important difference to the Wider\"{o}e linac. While for the latter the RF phase of the electric fields changes from gap to gap by $\pi$, in the Alvarez linac the field points in all gaps into the same direction. This comes from the fact that the field direction is given by the lowest (in frequency) resonance, which exists in the surrounding cylinder, and it turns out that for this mode the electric field points in the same direction in all gaps. One speaks of a `zero-mode' because of the zero phase difference between the gaps (more on this later). Here synchronism with the RF demands that the RF phase changes by $2\pi$ while the~particles travel from one gap to the next.

\begin{figure}[h!]
\begin{center}
\includegraphics[width=8cm]{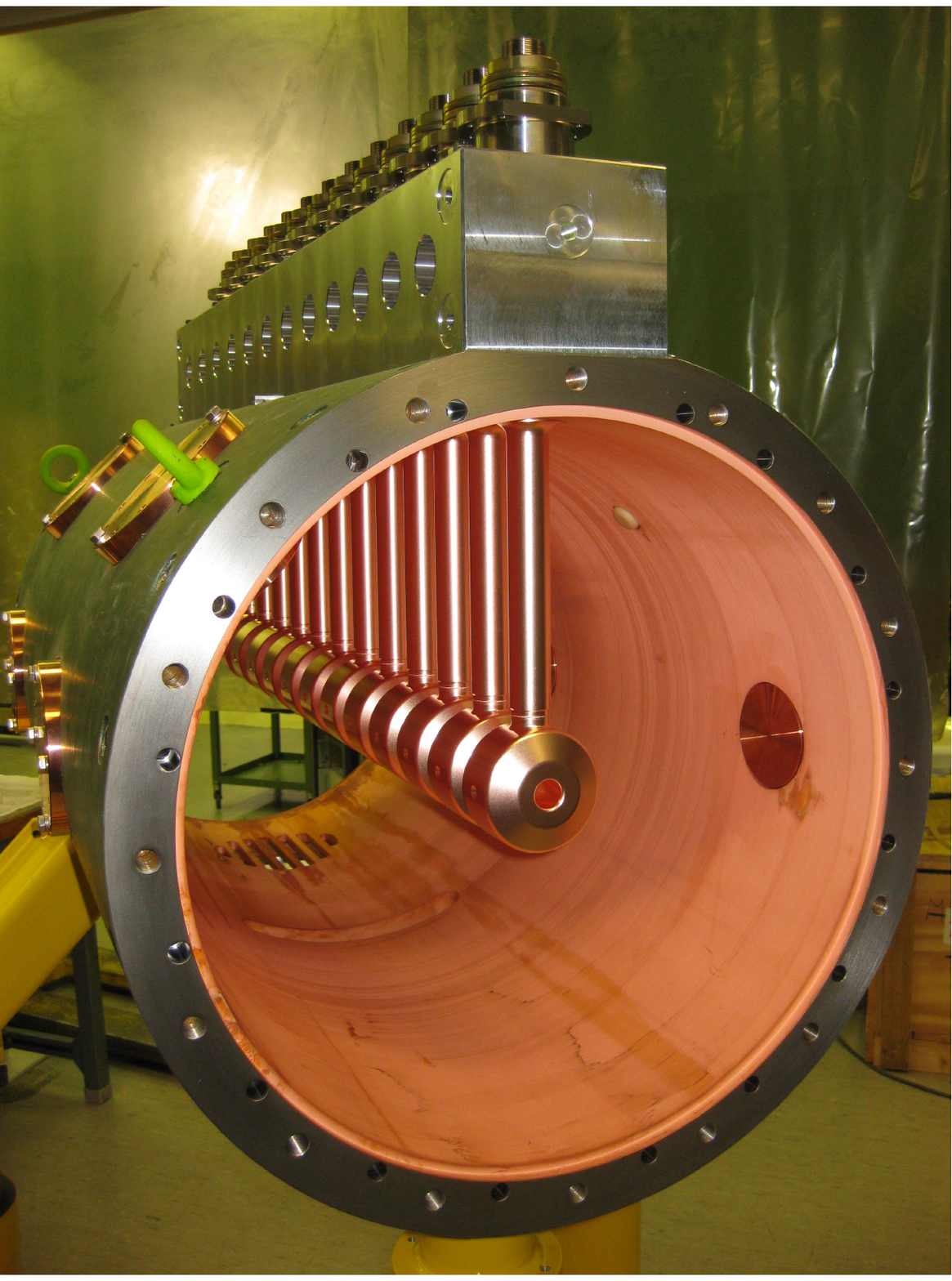}
\caption{Linac4 DTL at CERN}
\label{fig:linac4dtl}
\end{center}
\end{figure}

The Alvarez linac became possible by the development of high-power, high-frequency RF amplifiers. A technology that had just been invented during World War II to power radar systems. The typical frequency of these early radars was 200\,MHz and by coincidence this frequency results in reasonably sized cylindrical resonators with a diameter of around 1\,m. It is thus not by chance that most early linacs operate at a frequency of 200\,MHz. In Figure~\ref{fig:linac4dtl} you can see a recent implementation of a drift tube linac for Linac4 at CERN, which operates at 352\,MHz. 

\section{Introduction to Maxwell's Equations}
\subsection{Maxwell's Equations}
Maxwell's Equations were published in their earliest form in 1861-1862 in a paper called ``On Physical Lines of Force'' by the the Scottish physicist and mathematician James Clerk Maxwell. They represent a uniquely complete set of equations, which covers all areas of electrostatic and magnetostatic problems as well as electrodynamic problems of which Radio Frequency (RF) engineering is only a subset. Surprisingly they already include the effects of relativity even though they were conceived much earlier than Einstein's theories. 

In differential form Maxwell's Equations can be written as\\
\begin{align}
	\nabla \times \mathbf{H} &= \mathbf{J} + \frac{\partial\mathbf{D}}{\partial t}  \label{eq:rotH}\\
	\nabla \times \mathbf{E} &= -\frac{\partial\mathbf{B}}{\partial t} \label{eq:rotE} &&\mbox{\bf Maxwell's Equations}\\
	\nabla \cdot \mathbf{D} &= q_v \label{eq:divD} \\
	\nabla \cdot \mathbf{B} &= 0 \label{eq:divB}
\intertext{with the field components and constants being defined as:}
	\mathbf{E} &  &&\mbox{{\bf electric field} [V/m]} \nonumber \\
	\mathbf{D} &= \varepsilon_0\varepsilon_r \mathbf{E} &&\mbox{{\bf dielectric displacement} [As/m$^2$]} \\
	\mathbf{B} & &&\mbox{{\bf magnetic induction, magnetic flux density} [T]} \nonumber \\
	\mathbf{H} &= \frac{1}{\mu_0\mu_r}\mathbf{B} &&\mbox{{\bf magnetic field strength/field intensity} [A/m]} \\
	\mathbf{J} &= \kappa\mathbf{E} &&\mbox{{\bf electric current density} [A/m$^2$]} \\
	\frac{d}{dt}\mathbf{D} & &&\mbox{{\bf displacement current} [A/m$^2$]} \nonumber \\
	\varepsilon_0 &= 8.854 \cdot 10^{-12} &&\mbox{{\bf electric field constant} [F/m]} \\
	\varepsilon_r & &&\mbox{\bf relative dielectric constant} \nonumber \\
	\mu_0 &= 4\pi \cdot 10^{-7} &&\mbox{{\bf magnetic field constant} [H/m]} \\
	\mu_r & &&\mbox{\bf relative permeability constant} \nonumber\\
	\kappa & &&\mbox{{\bf electrical conductivity} [S/m]}. \nonumber
\end{align}

In the following sections we will see that most of the important RF formulas can be derived within a few lines from Maxwell's Equations. A short reminder of basic vector analysis can be found in Annex \ref{annex:basicanalysis}.

\subsection{Useful theorems by Gauss and Stokes}
The theorems of Gauss and Stokes are some of the most used transformations in this lecture and therefore we will take a moment to explain
their concept. 

\subsubsection*{Gauss' theorem}

Gauss' theorem not only saves us a lot mathematics but it also has a very useful physical interpretation when applied to Maxwell's equations. Mathematically speaking we transform the volume integral over the divergence of a vector into a surface integral over the vector itself.

\vspace*{-0.6cm}
\meqn{
	\int\limits_V \underbrace{\nabla \cdot \mathbf{a}}_{\mbox{``sources''}} dV = \oint\limits_S^{\textcolor{white}{C}} \mathbf{a} \cdot d\mathbf{S}
}{Gauss' theorem}{eq:Gauss}

\noindent The surface on the right side of the theorem is the one that surrounds the volume on the left side. Remembering that the divergence of a vector field equals its sources Gauss' theorem tells us that:

\begin{itemize}
	\item The vector flux through a closed surface equals the flux sources within the enclosed volume. 
	\item If there are no sources, the amount of flux entering and leaving a volume must be equal.
\end{itemize}

These statements can directly be applied to Maxwell's Equations. Using Eq.~\eqref{eq:divD} and applying Gauss' Theorem we get\\
\begin{figure}[h!]
\begin{minipage}{0.5\textwidth}
	\centering
	\begin{tikzpicture}
	\draw[->,very thick] (0,0) -- (2,0);
	\draw[->,very thick] (0,0) -- (-2,0);
	\draw[->,very thick] (0,0) -- (0,2);
	\draw[->,very thick] (0,0) -- (0,-2);
	\draw[->,very thick] (0,0) -- (1.41,1.41);
	\draw[->,very thick] (0,0) -- (1.41,-1.41);
	\draw[->,very thick] (0,0) -- (-1.41,1.41);
	\draw[->,very thick] (0,0) -- (-1.41,-1.41);
	\pgfdeclareradialshading{sphere}{\pgfpoint{0.5cm}{0.5cm}}%
	{rgb(0cm)=(0.9,0,0);
	rgb(0.7cm)=(0.7,0,0);
	rgb(1cm)=(0.5,0,0);
	rgb(1.05cm)=(1,1,1)}
	\pgfuseshading{sphere}
	\end{tikzpicture}
	\caption{Example of electric flux lines emanating from electric charges in the centre of a sphere}
\end{minipage}
\begin{minipage}{0.5\textwidth}
		\begin{equation}
			\int\limits_V \nabla \cdot \mathbf{E} dV = \oint\limits_S \mathbf{E} \cdot d\mathbf{S} = \frac{Q}{\varepsilon}	
		\end{equation}
		\vspace*{1cm}		
		
\end{minipage}
\end{figure}

which means that one can calculate the amount of charges within a volume simply by integrating the electric flux lines over any closed surface that surrounds the charges, or vice versa. 

The same trick can be applied to the source-free magnetic field. Here we use Eq.~\eqref{eq:divB} and get\\
\begin{figure}[h!]
\begin{minipage}{0.5\textwidth}
	\centering
	\begin{tikzpicture}
	\draw[->,very thick] (-1.5,0.8) arc (260:280:9);
	\draw[->,very thick] (-1.7,0.4) arc (263:277:15);
	\draw[->,very thick] (-1.5,-0.8) arc (100:80:9);
	\draw[->,very thick] (-1.7,-0.4) arc (97:83:15);
	\pgfdeclareradialshading{sphere}{\pgfpoint{0.5cm}{0.5cm}}
	{rgb(0cm)=(0.9,0,0);
	rgb(0.7cm)=(0.7,0,0);
	rgb(1cm)=(0.5,0,0);
	rgb(1.05cm)=(1,1,1)}
	\pgfuseshading{sphere}
	\end{tikzpicture}
	
	\vspace*{1ex}
	\caption{Example of magnetic flux lines penetrating a~sphere}
\end{minipage}
\begin{minipage}{0.5\textwidth}
		\begin{equation}
			\int\limits_V \nabla \cdot \mathbf{B} dV = \oint\limits_S \mathbf{B} \cdot d\mathbf{S} = 0
			\label{eq:GaussB}
		\end{equation}
		\vspace*{1cm}		
		
\end{minipage}
\end{figure}

Equation~\eqref{eq:GaussB} gives us the proof of what was already stated earlier: magnetic field lines have no sources ($\nabla\cdot\mathbf{B}=0$) and therefore the magnetic flux lines are always closed and have neither source nor sink. If magnetic flux lines enter a volume then they also have to leave the volume. 

\subsubsection*{Stokes' Theorem}
While Gauss' theorem is useful for equations involving the divergence of a vector, Stokes' Theorem offers a similar simplification for  equations, which contain the curl of a vector. With Stokes' Theorem we can transform surface integrals over the curl of a vector into closed line integrals over the vector itself.

\vspace*{-0.6cm}
\meqn{
	\int\limits_A \left(\nabla \times \mathbf{a} \right) \cdot d\mathbf{A} = \oint\limits_C^{\textcolor{white}{C}} \mathbf{a} \cdot d\mathbf{l}
}{Stokes' Theorem}{eq:Stokes}
One can interpret Stokes' Theorem with the help of Fig.~\ref{fig:Stokes} as follows:
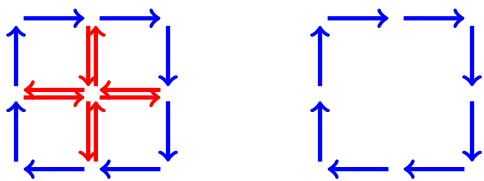
\begin{figure}[h!]
	\centering
	\begin{tikzpicture}[->,scale=1, ultra thick, blue]
	\draw (0,0.1) -- (0,0.9);
	\draw (0,1.1) -- (0,1.9);
	\draw (0.1,2) -- (0.9,2);
	\draw (1.1,2) -- (1.9,2);
	\draw (2,1.9) -- (2,1.1);
	\draw (2,0.9) -- (2,0.1);
	\draw (1.9,0) -- (1.1,0);
	\draw (0.9,0) -- (0.1,0);
	
	\draw[red] (0.1,0.95) -- (0.9,.95);
	\draw[red] (0.9,1.05) -- (0.1,1.05);
	\draw[red] (0.95,0.9) -- (0.95,0.1);
	\draw[red] (1.05,0.1) -- (1.05,0.9);
	\draw[red] (1.1,0.95) -- (1.9, 0.95);
	\draw[red] (1.9,1.05) -- (1.1,1.05);
	\draw[red] (1.05,1.1) -- (1.05,1.9);
	\draw[red] (0.95,1.9) -- (0.95,1.1);
	
	\draw (4,0.1) -- (4,0.9);
	\draw (4,1.1) -- (4,1.9);
	\draw (4.1,2) -- (4.9,2);
	\draw (5.1,2) -- (5.9,2);
	\draw (6,1.9) -- (6,1.1);
	\draw (6,0.9) -- (6,0.1);
	\draw (5.9,0) -- (5.1,0);
	\draw (4.9,0) -- (4.1,0);
	
	\end{tikzpicture}
	\caption{Illustration of Stokes' Theorem}
	\label{fig:Stokes}
\end{figure}

\begin{itemize}
	\item The area integral over the curls of a vector field can be calculated by a line integral along its closed borders, or
	\item the field lines of a vector field with non-zero curls must be closed contours. 
\end{itemize}

\begin{figure}[h!]
\parbox{6cm}{
	\begin{tikzpicture}[scale=1]
	\draw[->,ultra thick,red] (0,0) -- (2.5,0);
	\draw[->,ultra thick, blue] (2.5,0) ellipse (0.4 and 1); 
	\draw (2.5,1.1) node [above, black] {\Large H};
	\draw[->,ultra thick,blue] (2.1,0) arc (180:210:0.8 and 1);
	\draw[->,ultra thick,blue] (2.9,0) arc (0:30:0.8 and 1);
	\draw[->,ultra thick,red] (2.5,0) -- (5,0) node [right, black] {\Large I};
	\end{tikzpicture}
	\caption{\label{fig:Amp1}Illustration of Amp\`{e}res Law}}\parbox{3cm}{
	\begin{align*} \oint\limits_C^{\textcolor{white}{C}} \mathbf{H} \cdot d\mathbf{l} = I \end{align*}}\hspace{1cm}\parbox{5cm}{\raggedright\bf Amp\`{e}res Law}\hfill\parbox[b]{8mm}{\begin{equation} \label{eq:Ampere} \end{equation}}
\end{figure}
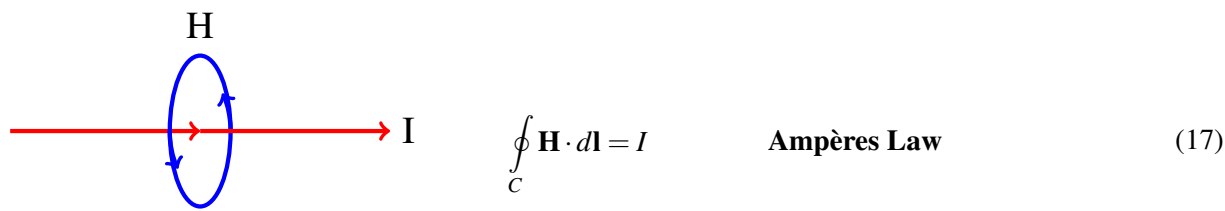

The meaning of these statements becomes immediately clear when we apply Stokes' Theorem to Maxwell's Equation~\eqref{eq:rotH}.
\begin{equation}
	\int\limits_A \left(\nabla\times\mathbf{H}\right)\cdot d\mathbf{A} = \oint\limits_C \mathbf{H} \cdot d\mathbf{l} 
	= \int\limits_A \left( \mathbf{J}+\frac{d\mathbf{D}}{d t} \right) \cdot d\mathbf{A}
\end{equation}
In the electrostatic case the time derivative disappears and the area integral over the current density is for instance the current flowing in an electric wire as shown in Fig.~\ref{fig:Amp1}. This means that with a one-line manipulation of Maxwell's equations, we have derived Amp\`{e}res Law,
which tells us that each current induces a circular magnetic field around itself, whose strength can be be calculated with a simple closed line integral along a circular path with the current at its centre. 

With similar ease we can derive Faradays' induction law, which is the basis of every electric motor or generator. All we have to do is apply Stokes' Theorem to Maxwell's Equation~\eqref{eq:rotE}.
\begin{equation}
	\int\limits_A\left(\nabla\times\mathbf{E}\right) \cdot d\mathbf{A} = 
	\underbrace{\oint\limits_c \mathbf{E} \cdot d\mathbf{l}}_{V_i} = 
	- \underbrace{\frac{d}{dt}\int\limits_A \mathbf{B}\cdot d\mathbf{A}}_{\displaystyle\frac{d\psi_m}{dt}}
\end{equation}

\noindent and again after one line we obtain one of the fundamental laws of electrical engineering. 
\begin{figure}[h!]
\parbox{6cm}{
	\centering
	\begin{tikzpicture}[scale=1, ultra thick, blue]
	\draw (0,0) circle (1);
	\foreach \x in {-1.5,-1,-0.5,0,0.5,1,1.5}
		\foreach \y in {-1.5,-1,-0.5,0,0.5,1,1.5}
			\node [red] at (\x,\y) {x};
	\draw [->] (0,1) arc (90:45:1);
	\node [fill,white] at (1.0,1.2) {\Large\textcolor{black}{$V_i$}};
	\end{tikzpicture}
	\caption{\label{fig:Faraday}Illustration of Faradays' induction law}}\parbox{3cm}{
	\begin{align*}  V_i = -\frac{d\psi_m}{dt} \end{align*}}\hspace{1cm}\parbox{5cm}{\raggedright\bf Faradays' Induction Law}\hfill\parbox[b]{8mm}{\begin{equation} \label{eq:Faraday} \end{equation}}
\end{figure}

Faradays' Law tells us that an electric voltage is induced in a loop if the magnetic flux $\psi$, penetrating the loop, changes over time as shown in Fig.~\ref{fig:Faraday}. Alternatively one can change the flux by moving the loop in and out of a static magnetic field. 
	
I hope that these examples have convinced you that Maxwell's equations are indeed very powerful and that with a bit of vector analysis we really can derive everything we need for RF engineering (maybe not always within one line...). 
	
\subsection{Displacement Current}
While usually everyone has an idea what electric and magnetic fields are, the displacement current $d \mathbf{D}/dt$ is often not so well understood. Since it is vital for the propagation of electromagnetic waves we will spend a few lines studying this quantity. We start by deriving and interpreting the continuity equation and we will then look at a simple practical example. 

We apply the divergence to Maxwell's Equation~\eqref{eq:rotH}
\begin{equation}
	\underbrace{\nabla \cdot \left( \nabla \times \mathbf{H} \right)}_{\displaystyle\equiv 0} = 
	\nabla \cdot \mathbf{J} + \underbrace{\nabla \cdot \frac{d \mathbf{D}}{dt}}_{\displaystyle\frac{d}{d t} \rho_v}.
	\label{eq:pre-continuity}
\end{equation}

\noindent Using Maxwell's Equation~\eqref{eq:divD} we have made the association between the ``sources of the displacement current ($\nabla \cdot \frac{d\mathbf{D}}{dt}$)'' and the ``rate of change of electric charges ($\frac{d}{dt}\rho_v$)''. Using the identity~\eqref{eq:vector0} we already get the continuity equation

\vspace*{-0.5cm}
\meqn{
	\displaystyle \nabla\cdot\mathbf{J} = -\frac{d}{dt} \rho_v
}{continuity equation}{eq:continuity}
to which we apply a volume integral and Gauss' theorem~\eqref{eq:Gauss}

\vspace*{-0.3cm}
\meqn{
\int\limits_V\nabla \cdot \mathbf{J} dV = \oint\limits_S \mathbf{J} \cdot d\mathbf{S} = \sum I_n = -\frac{d}{d t} \int\limits_V \rho_v dV
}{continuity equation}{eq:continuityV}
In this form the interpretation is very straight forward and we can state that:
\begin{itemize}
	\item if the amount of electric charges in a volume is changing over time, a current needs to flow; or more poignant: electric charges cannot be destroyed. 
\end{itemize}

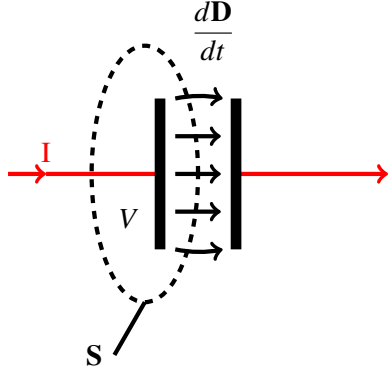
\begin{figure}[h!] 
\centering
	\begin{tikzpicture}[scale=1, ultra thick]
	\draw[->,red] (0,0) -- (0.5,0) node[red,above] {I};
	\draw[red] (0.5,0) -- (2,0);
	\draw[->,red] (3,0) -- (5,0);
	\draw[line width = 4pt] (2,-1) -- (2,1);
	\draw[line width = 4pt] (3,-1) -- (3,1);
	\draw[->] (2.2,0) -- (2.8,0);
	\draw[->] (2.2,0.5) -- (2.8,0.5);
	\draw[->] (2.2,-0.5) -- (2.8,-0.5);
	\draw[->] (2.2,1) arc (102:78:1.5);
	\draw[->] (2.2,-1) arc (258:282:1.5);
	\draw[dashed] (1.8,0) ellipse (0.7 and 1.7);
	\draw (1.6,-0.6) node {$V$};
	\draw (1.8,-1.7) -- (1.4,-2.4) node [below,left] {$\mathbf{S}$}; 
	\draw (2.7,1.9) node {$\displaystyle\frac{d\mathbf{D}}{dt}$};
	\end{tikzpicture} 
\caption{Example for displacement current: charging of a capacitor}
\label{fig:capacitor}
\end{figure}

Now it is good to know that electric charges cannot be destroyed but that does not yet help us to understand the displacement current. For this purpose we go back to Eq.~\eqref{eq:pre-continuity} and this time we do not replace the expression of the displacement current. Instead we apply a volume integral and Gauss' theorem and get
\begin{equation}
	\oint\limits_S \mathbf{J}\cdot d\mathbf{S} = \sum I_n = -\frac{d}{dt}\oint\limits_S \mathbf{D}dS = -\frac{d}{dt}\int\limits_V \rho_v dV
	\label{eq:displacement-example}
\end{equation}
which we can apply to the simple geometry of a capacitor shown in Fig.~\ref{fig:capacitor}, which is charged by a static current $I$.
 
If we assume a small volume with a surface $S$ around one of the capacitor plates, then one can directly interpret Eq.~\eqref{eq:displacement-example}: The current $I$, which enters the volume $V$ on the left, equals the flux integral of the displacement current $-\frac{d}{dt}\mathbf{D}$, which leaves the volume $V$ on the right. This means that the displacement current can be understood as a ``current without charge transport'', which in this case can only exist because of the rate of change of the electric charges ($-\frac{d}{dt}\int\limits_V \rho_v dV$) on the left capacitor plate.

\subsection{Boundary conditions}
\label{sec:boundary}
Before we try to calculate electromagnetic fields in accelerating cavities we need to understand how these fields behave close to material boundaries, e.g.\ the electrically conducting walls of a cavity. Using Stokes' and Gauss' Theorems we can quickly derive these boundary conditions. 

\subsubsection*{Field components parallel to a material boundary}
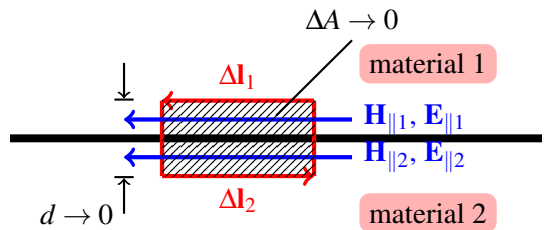
\begin{figure}[h!]
	\centering
	\begin{tikzpicture}[scale=1, ultra thick]
	\draw[line width = 3pt] (-0.5,0) -- (6.5,0);
	\draw[->,red] (3.5,0.5) -- (1.5,0.5) node [above,midway] {$\Delta \mathbf{l}_1$};
	\draw [red] (1.5,0.5) -- (1.5,-0.5);
	\draw[->,red] (1.5,-0.5) -- (3.5,-0.5) node [below,midway] {$\Delta \mathbf{l}_2$};
	\draw[red] (3.5,-0.5) -- (3.5,0.5);
	
	\draw[<-, blue] (1,0.25) -- (4,0.25) node [right] {$\mathbf{H}_{\parallel 1}$,  $\mathbf{E}_{\parallel 1}$};
	\draw[<-, blue] (1,-0.25) -- (4,-0.25) node [right] {$\mathbf{H}_{\parallel 2}$,  $\mathbf{E}_{\parallel 2}$};
	
	\draw[->|,thick] (1,1.0) -- (1,0.5);
	\draw[->|,thick] (1,-1.0) node[left] {$d \rightarrow 0$} -- (1,-0.5);
	\fill[pattern=north east lines] (3.5,0.5) -- (1.5,0.5) -- (1.5,-0.5) -- (3.5,-0.5) -- (3.5,0.5);
	\draw[thick] (3,0.3) -- (4,1.3) node[above] {$\Delta A \rightarrow 0$};
	\draw (5,0.7) node[fill=red!30,right,above,rounded corners] {material 1};
	\draw (5,-0.7) node[fill=red!30,right,below,rounded corners] {material 2};
	\end{tikzpicture}
	\caption{Boundary conditions parallel to a material boundary}
	\label{fig:boundary-parallel}
\end{figure}

We start with field components ($E_{\parallel}$, $H_{\parallel}$), which are parallel to a surface between two materials as depicted in Fig.~\ref{fig:boundary-parallel}.

We define a small surface $\Delta A$, which is perpendicular to the boundary and which encloses a small cross section of the boundary area. Then we integrate Mawell's Equations~\eqref{eq:rotH} and \eqref{eq:rotE} over this area and apply Stokes' Theorem.
\begin{align}
	\int\limits_A \nabla \times \mathbf{H} \cdot d\mathbf{A} &=  \oint\limits_C \mathbf{H}\cdot d\mathbf{l} = 
	\underbrace{\int\limits_A \mathbf{J}\cdot d\mathbf{A}}_{=i'\Delta l} + 
	\underbrace{\frac{d}{dt} \int\limits_A \mathbf{D} \cdot d\mathbf{A}}_{\rightarrow 0 \mbox{ for } \mathbf{A} \rightarrow 0}\\
	\int\limits_A \nabla \times \mathbf{E} \cdot d\mathbf{E} &= \oint\limits_C \mathbf{E} \cdot d\mathbf{l} = 
	-\underbrace{\frac{d}{dt}\int\limits_A \mathbf{B} \cdot d\mathbf{A}}_{\rightarrow 0 \mbox{ for } \mathbf{A}\rightarrow 0}
\end{align}

With Stokes Theorem the area integrals over $A$ are transformed into line integrals around the~contour $C$ of the area. If now the width $d$ of the area (see Fig.~\ref{fig:boundary-parallel}) is reduced to zero, the calculation of~the~contour integral simplifies to a multiplication of the field components $E_{\parallel}$ and $H_{\parallel}$ with the path elements $\Delta l$. The area integrals over $\mathbf{D}$ and $\mathbf{B}$ vanish and the area integral over the current density $\mathbf{J}$ is replaced by a~surface current, which may flow in the boundary plane between the two materials, times the path element $\Delta l$. This results in the following boundary condition:

\vspace*{-0.5cm}
\meqn{
	&H_{\parallel 1} - H_{\parallel 2} = i' \\
	&E_{\parallel 1} = E_{\parallel 2}
}{conditions for magnetic and electric fields parallel to a~material boundary.}{eq:generalboundaries}
In case of a wave-guide or an accelerator cavity, we generally assume one of the materials (e.g.\ material 2) to be an ideal electrical conductor and in that case the electric and magnetic field components in this material vanish, so that we get:

\vspace*{-0.5cm}
\meqn{
	H_{\parallel 1} &= i'\\
	 E_{\parallel 1} &=0
}{conditions for magnetic and electric fields parallel to ideal electric surfaces.}{eq:electricboundary}
 
\subsubsection*{Field components perpendicular to a material boundary}
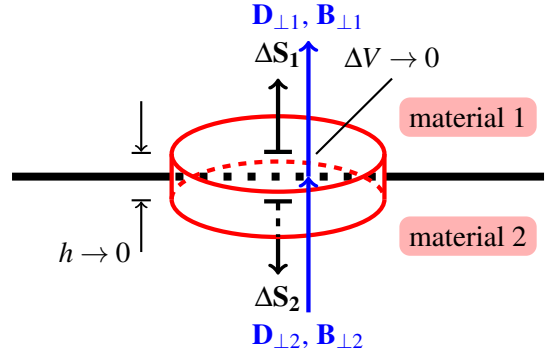
\begin{figure}[h!]
\centering
	\begin{tikzpicture}[scale=1, ultra thick]
	\draw[line width = 3pt] (-1,0) -- (1.1,0);
	\draw[line width = 3pt] (3.9,0) -- (6,0);
	\draw[loosely dashed, line width = 3pt] (1.15,0) -- (3.85,0);
	\draw[red] (2.5,0.3) ellipse (1.4 and 0.5);
	\draw[dashed,red] (1.1,-0.3) arc (180:0:1.4 and 0.5);
	\draw[->] (2.5,-0.8) -- (2.5, -1.3) node [below] {$\Delta\mathbf{S_2}$};
	\draw[red] (1.1,-0.3) arc (180:360:1.4 and 0.5);
	\draw[red] (1.1,0.3) -- (1.1,-0.3);
	\draw[red] (3.9,0.3) -- (3.9,-0.3);
	\draw[|->] (2.5,0.3) -- (2.5,1.3)node [above] {$\Delta\mathbf{S_1}$};
	\draw[|-,dashed] (2.5,-0.3) -- (2.5,-0.8);
	
	\draw[blue,->] (2.9, 0) -- (2.9, 1.8) node [above] {$\mathbf{D}_{\perp 1}$, $\mathbf{B}_{\perp 1}$};
	\draw[blue,<-] (2.9, 0) -- (2.9, -1.8) node [below] {$\mathbf{D}_{\perp 2}$, $\mathbf{B}_{\perp 2}$};
	
	\draw[->|,thick] (0.7,1.0) -- (0.7,0.3);
	\draw[->|,thick] (0.7,-1.0) node[left] {$h \rightarrow 0$} -- (0.7,-0.3);
	\draw[thick] (3,0.3) -- (4,1.3) node[above] {$\Delta V \rightarrow 0$};
	\draw (5,0.5) node[fill=red!30,right,above,rounded corners] {material 1};
	\draw (5,-0.5) node[fill=red!30,right,below,rounded corners] {material 2};
	\end{tikzpicture}
	\caption{Boundary conditions perpendicular to a material boundary}
	\label{fig:normalboundary}
\end{figure}

In a very similar way one can derive the boundary conditions for fields ($D_{\perp}$, $B_{\perp}$), which are perpendicular to a boundary surface between two materials. Only this time we don't define an area but a small cylinder with a volume $\Delta V$ around the boundary as shown in Fig.~\ref{fig:normalboundary}.

We form a volume integral of Maxwell's Equations \eqref{eq:divD} and \eqref{eq:divB} over the volume of the cylinder and we apply Gauss' Theorem to transform the volume integrals into surface integrals. 
\begin{align}
	\int\limits_V\nabla\cdot\mathbf{D} dV &= \oint\limits_S\mathbf{D} \cdot d\mathbf{S} = \int\limits_V q_v dV \\
	\int\limits_V\nabla\cdot\mathbf{B} dB &= \oint\limits_S\mathbf{B}\cdot d\mathbf{S} = 0
\end{align}

In the following step we reduce the height of the cylinder to zero, so that we end up with two surfaces, one of each side of the boundary. And now it becomes clear why one has to start with a volume integral. Since the surface element $d\mathbf{S}$ is perpendicular to the surface of the cylinder the ``dot product'' in the integrals basically reduces the vector fields $\mathbf{D}$ and $\mathbf{B}$ to the components, which are also perpendicular to the cylinder's surface. This means that above equations can now be written as\\
\meqn{
	& D_{\perp 1} - D_{\perp 2} = q_s \\
	& B_{\perp 1} = B_{\perp 2}
}{conditions for dielectric displacement and magnetic induction perpendicular to a~material boundary}{eq:perpboundary}

with $q_s$ being a surface charge (Coulomb/metre$^2$) which may exist on the boundary surface. In case material 2 is an ideal conductor, we get

\vspace*{-0.5cm}
\meqn{
	D_{\perp 1} &= q_s \\
	B_{\perp 1} &= 0
}{conditions for dielectric displacement and magnetic induction perpendicular to ideal electric surfaces.}{eq:perboundaryel}

We note that when parallel to boundary surfaces the electric and magnetic fields are used in the~boundary conditions, while when perpendicular to boundary surfaces we have a definition for the~dielectric displacement and the magnetic induction.  This means that for instance the tangential electric field $E_{\parallel}$ may be smooth across a boundary but there will be a jump in the dielectric displacement $D_{\parallel}$ if there is a different relative dielectric constant $\varepsilon_r$ in the two materials. Similarly, the component of the~magnetic induction $B_{\perp}$, which is perpendicular to a surface may be smooth, while the magnetic field $H_{\perp}$ will jump in case the two materials have different relative magnetic field constants $\mu_r$.

\section{Electromagnetic waves}
In this section we will derive the general form of the wave equation and then restrict ourselves to phenomena, which are harmonic in time. Since RF systems mostly deal with sinusoidal waves, we will be able to explain and understand most relevant phenomena with this approach. This will include the ``skin-effect'', the propagation of energy, RF losses and acceleration via travelling waves. 

\subsection{The wave equation}
We start with the simplification of only looking at homogeneous, isotropic media, meaning we assume that the electromagnetic fields ``see'' the same material constants ($\mu$, $\varepsilon$, $\kappa$) in all directions. With this assumption Maxwell's Equations can be conveniently expressed in terms of only $E$, and $H$.
\begin{align}
	\nabla \times \mathbf{H} &= \kappa\mathbf{E} + \varepsilon\frac{\partial\mathbf{E}}{\partial t}  \label{eq:rotHb}\\
	\nabla \times \mathbf{E} &= -\mu\frac{\partial\mathbf{H}}{\partial t} \label{eq:rotEb} &\hspace*{1.5cm}&\mbox{\bf Maxwell's Equations}\\
	\nabla \cdot \mathbf{E} &= \frac{q_v}{\varepsilon} \label{eq:divE} \\
	\nabla \cdot \mathbf{H} &= 0 \label{eq:divH}
\end{align}

The curl of Eq.~\eqref{eq:rotEb} together with Eq.~\eqref{eq:rotHb}, and the curl of Eq.~\eqref{eq:rotHb} together with Eqs.~\eqref{eq:rotEb} and \eqref{eq:divE} result in the general wave equations for homogeneous media.\\
\meqn{
	\nabla^2\mathbf{E} -\nabla\left(\nabla\cdot\mathbf{E}\right) &= 
	\mu\kappa\frac{d}{dt}\mathbf{E} + \mu\epsilon\frac{d^2}{dt^2}\mathbf{E}\\
	\nabla^2\mathbf{H} &= \mu\kappa\frac{d}{dt}\mathbf{H} + \mu\epsilon\frac{d^2}{dt^2}\mathbf{H}
}{wave equations in homogeneous media}{eq:waveshomogeneous}

In wave-guides and cavities we can simplify these equations even further by only considering the~inside fields, which exist in non-conducting media ($\kappa =0$) and charge-free volumes ($\nabla\cdot E = 0$).\\
\meqn{
	\nabla^2\mathbf{E} &= \mu\epsilon\frac{d^2}{dt^2}\mathbf{E}\\
	\nabla^2\mathbf{H} &= \mu\epsilon\frac{d^2}{dt^2}\mathbf{H}
}{wave equations in non-conducting and charge-free homogeneous media}{eq:wave-eq}

\subsection{Complex notation of time-harmonic fields}
The already compact wave equations in Eq.~\eqref{eq:wave-eq} can be simplified even further by taking into account that in RF engineering one usually deals with time-harmonic signals, which are sometimes modulated in phase or amplitude. We can therefore introduce the complex notation for electric and magnetic fields. We start by assuming a time harmonic electric field with amplitude $E_0$ and phase $\varphi$
\begin{align}
	E(t) &= E_0 \cos\left(\omega t+\varphi\right)\\
\intertext{which we can interpret as the real part of a complex expression.}
	E(t) &= \Re \left\{E_0 e^{i\varphi} e^{i\omega t} \right\} = \Re \left\{E_0\cos\left(\omega t + \varphi\right) + iE_0\sin\left(\omega t + \varphi\right)\right\}\\
\intertext{\indent In this form we can easily separate the harmonic time dependence $\omega t$ from the phase information $\varphi$. The phase information can be merged into the amplitude by defining a ``complex amplitude'' or ``phasor''.}
	\tilde{E} &= E_0 e^{i\varphi} \\
\intertext{We keep in mind that the physical real fields are obtained as the real part of the complex amplitude times~$e^{i\omega t}$.}
	E_0 \cos\left(\omega t + \varphi\right) &= \Re \left\{ \tilde{E} e^{i\omega t} \right\} \\
\intertext{\indent To simplify our writing we just skip the part with the harmonic time dependence and the writing of the tilde, which means that from now on all field quantities are written as complex amplitudes. In order to convince you that this really is a simplification, let us consider what happens to time derivatives when using complex notation:}
	\frac{d}{dt} \tilde{E}e^{i\omega t} &= i\omega \tilde{E} e^{i\omega t}  .
\end{align}

This means that all time derivatives in Maxwell's Equations and also in the the wave equations can simply be replaced by a multiplication with $i\omega$, and we are able to do this because the time dependence is always harmonic. Only when we are dealing with transient events, like the switching on of an RF amplifier, or the sudden arrival of a beam in a cavity, we have to go back the non-harmonic general equations. 
 
As a first application of the complex notation we re-write Maxwell's Equations.
\begin{align}
	\nabla\times\mathbf{H} &= i\omega\underline{\varepsilon}\mathbf{E} \\
	\nabla\times\mathbf{E} &= -i\omega\mu\mathbf{H}  &\hspace*{2.2cm}&\mbox{\bf Maxwell's Equations in}\\
	\nabla\cdot\mathbf{E} &= \frac{\rho_V}{\varepsilon} &&\mbox{\bf complex notation}\\
	\nabla\cdot\mathbf{H} &= 0 \\
\intertext{with the complex dielectric constant $\underline{\varepsilon}$ being defined as}
	\underline{\varepsilon} &= \varepsilon' - i\varepsilon'' = \varepsilon\left( 1-i\frac{\kappa}{\omega\varepsilon}\right).
	&&\mbox{\bf complex dielectric constant} \label{eq:complexdielectric}
\intertext{We note that $\underline{\varepsilon}$ is only complex in conducting media. One can now proceed to write the general wave equations in complex form.}
	\nabla^2\mathbf{E}-\nabla\left(\nabla\cdot\mathbf{E}\right) &= -\underline{k}^2\mathbf{E} &&\mbox{\bf general complex} \label{eq:complexwaveeqE} \\
	\nabla^2\mathbf{H} &= -\underline{k}^2\mathbf{H} && \mbox{\bf wave equations} \label{eq:complexwaveeqH}
\intertext{Also here we note that the complex wave number $\underline{k}$ becomes real in case of non-conducting media.}
	\underline{k}^2 &= \omega^2\mu\underline{\varepsilon} = \omega^2\mu\varepsilon\left( 1-i\frac{\kappa}{\omega\varepsilon}\right)
	&& \mbox{\bf complex wave number} \label{eq:complexwavenumber}
\intertext{Finally we simplify the wave equations again for the case of non-conducting charge-free media and get}
	\nabla^2\mathbf{E} &= -k^2\mathbf{E} &&\mbox{\bf complex wave equations in non-}\\
	\nabla^2\mathbf{H} &= -k^2\mathbf{H} && \mbox{\bf conducting charge-free media}
\intertext{with}
	k^2 &= \omega^2\mu\varepsilon = \frac{\omega^2}{c^2}. && \mbox{\bf free-space wave number} \label{eq:freespacek}
\end{align}

\noindent On the way we have also introduced a simple definition for the speed of light $c=1/\sqrt{\mu\varepsilon}$ in Eq.~\eqref{eq:freespacek}.

\subsection{Plane waves}
\label{sec:PlaneWaves}
As an introduction to the theory of electromagnetic waves we look at a very simple case, the so-called
plane waves. We assume again that we are in a homogenous, isotropic, linear medium and that there are no 
charges or currents, which means that Eqs.~\eqref{eq:complexwaveeqE} and \eqref{eq:complexwaveeqH} 
apply. Furthermore -- for a plane wave -- we assume that the field components only depend on one 
coordinate (e.g.\ $z$). The solution of the harmonic wave Eqs.~\eqref{eq:complexwaveeqE} and
\eqref{eq:complexwaveeqH} can then be written as the superposition of two waves

\vspace*{-0.1cm}
\meqn{
 	E_x(z) &= \underline{C}_1 e^{-\underline{\gamma}z} + \underline{C}_2 e^{+\underline{\gamma}z} \\
	H_y(z) &= \frac{1}{\underline{Z}}\left(\underline{C}_1 e^{-\underline{\gamma}z} +  \underline{C}_2
	 e^{+\underline{\gamma}z} \right)
}{}{eq:solharmwave}\\
one of which propagates in positive and one in negative $z$-direction. The complex propagation constant
 $\underline{\gamma}$ has a real component $\alpha$, which describes the damping in lossy materials, and 
 a complex component $i\beta$, which describes the propagation of the wave. The relation between the 
 propagation constant $\gamma$ and the~wavenumber $k$ is
 
\vspace*{-0.7cm} 
\meqn{
	\underline{\gamma} = \alpha + i\beta = i\underline{k} = i\omega\sqrt{\mu\underline{\varepsilon}}\mbox{.}
}{propagation constant}{eq:propconst}

We already know that time-harmonic electric and magnetic fields are linked via Maxwell's Equations, which means
that their amplitudes have a certain fixed ratio to each other. This ratio is introduced in Eq.~\eqref{eq:solharmwave}
as the wave impedance $\underline{Z}$, the ratio between the electric and magnetic field amplitude

\vspace*{-0.5cm}
\meqn{
	\underline{Z} = \frac{E_y}{H_z} = \sqrt{\frac{\mu}{\underline{\varepsilon}}}\mbox{,}
}{complex wave impedance}{}
which becomes real in the absence of lossy material. 

\vspace*{-0.5cm}
\meqn{
	Z_0 = \sqrt{\frac{\mu_0}{\varepsilon_0}} \approx 377 \Omega
}{free space wave impedance}{eq:ZfreeSpace}

\subsection{Skin depth}
\label{sec:skindepth}
\begin{figure}[h!]
	\centering
	\includegraphics[angle=270,width=0.65\textwidth]{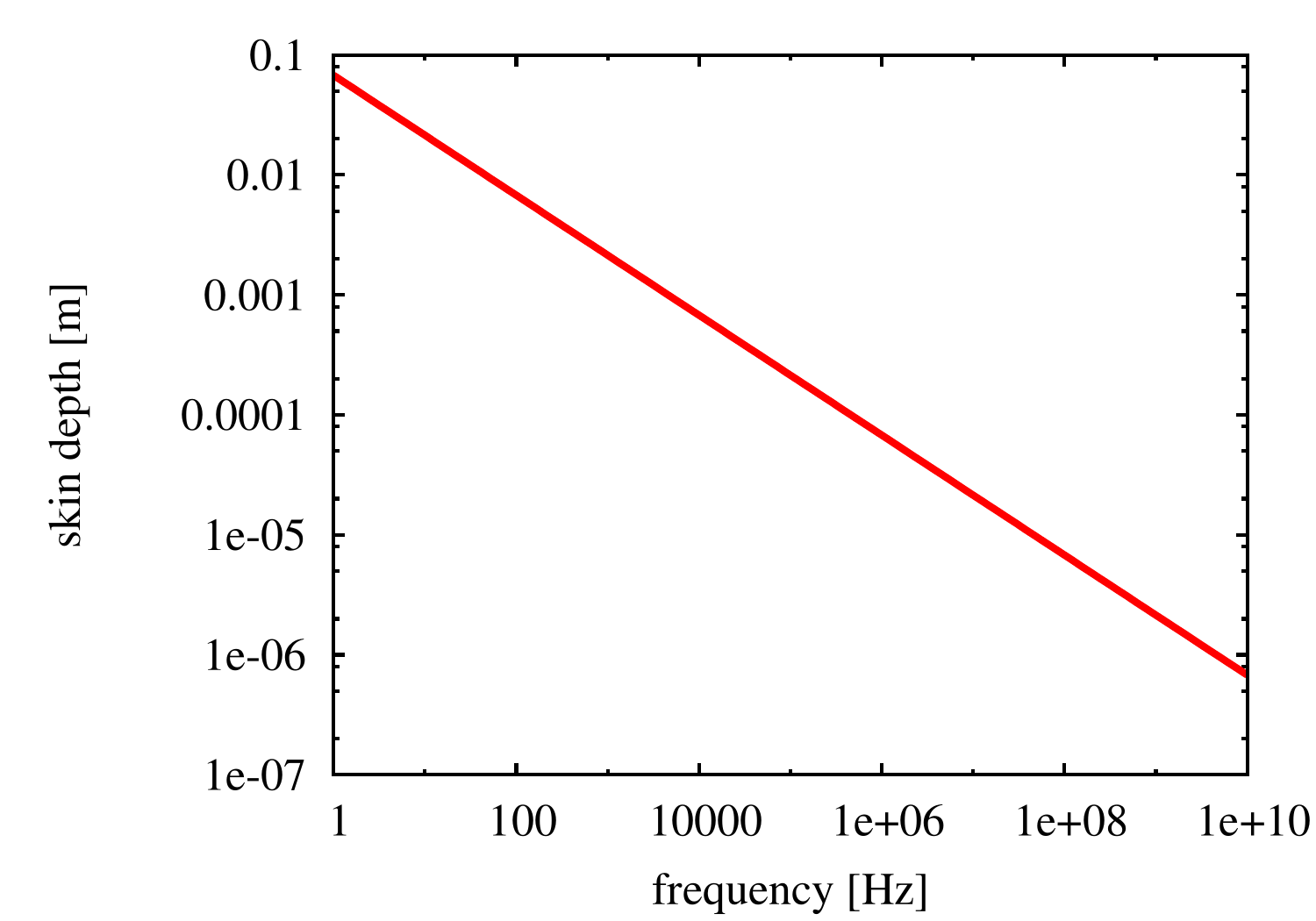}
	\caption{Skin depth versus RF frequency}
	\label{fig:skinVsFreq}
\end{figure}

When electromagnetic waves encounter a conducting (lossy) material, we have to evaluate the boundary conditions
 (Section~\ref{sec:boundary}) and we will find the wave amplitudes suddenly get attenuated with the 
 attenuation constant $\alpha$. In RF we can assume that
 \begin{align}
 	\frac{\kappa}{\omega\varepsilon} &\gg 1
 \intertext{which means that the complex wave number \eqref{eq:complexwavenumber} and obviously also the complex dielectric constant \eqref{eq:complexdielectric} are dominated by their imaginary part, so that we can write}
\underline{\varepsilon} \approx -i \varepsilon'' &= -i \frac{\kappa}{\omega} \hspace*{0.5cm}\mbox{or}\hspace*{0.5cm}
	\underline{k}^2 = -i\omega\mu\kappa
\intertext{which is actually equivalent to neglecting the displacement current. Using Eq.~\eqref{eq:propconst} we then write the~propagation constant as}
	\gamma &= \alpha + i\beta = i\underline{k} = i\omega\sqrt{\frac{-i\mu\kappa}{\omega}} = (1+i)\sqrt{\frac{\kappa\mu\omega}{2}}
\end{align}
which defines the attenuation constant $\alpha$. The so-called skin depth is then defined as the distance after which the wave-amplitudes are attenuated to $1/e \approx 36.8\%$:

\vspace*{-0.5cm}
\meqn{
	\delta_s = \frac{1}{\alpha} = \sqrt{\frac{2}{\omega\mu\kappa}}.
}{skin depth}{eq:SkinDepth}
\indent Knowing the value of the~skin depth is crucial for the design of RF equipment. Let us assume that we want to build an 
accelerating cavity, which resonates at 500\,MHz. Since high quality copper is quite expensive we consider to 
construct the cavity out of steel and then to copper plate the interior in order to have a good quality factor and 
to reduce the losses on the surface. From Eq.~\eqref{eq:SkinDepth} we calculate that the~skin depth in copper is $\approx
3\,\mu$m. Depending on how well the copper plating is mastered by the~plating company, one can now define the thickness of the copper layer, which is needed on the inside of the~cavity. Typically one uses around 10-20 times the skin depth as plating thickness. Figure~\ref{fig:skinVsFreq} shows the~dependency of the skin depth on the RF frequency.

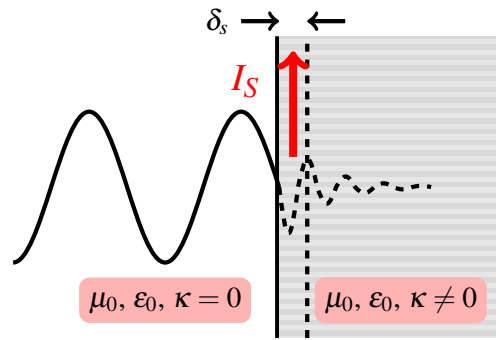
\begin{figure}[h!]
\begin{center}
\begin{tikzpicture}[scale=1, ultra thick]
\draw[line width = 3pt] (0,-2) -- (0,2);
\fill[pattern=horizontal lines light gray] (0,-2) rectangle (3,2);
\draw (-3.5,-1) cos (-3,0) sin (-2.5,1) cos (-2,0) sin (-1.5,-1) cos (-1,0) sin (-0.5,1) cos (0,0);
\draw[dashed] (0,0) sin (0.125,-0.61) cos (0.25,0) sin (0.375,0.37) cos (0.5,0) sin (0.625,-0.22) cos (0.75,0) sin (0.875,0.14) cos (1,0) sin (1.125,-0.08) cos (1.25,0) sin (1.375,0.05) cos (1.5,0) sin (1.625,-0.03) cos (1.75,0) sin (1.875,0.018) cos (2,0);
\draw[dashed] (0.37,-2) -- (0.37,2); 
\draw[->] (-0.5,2.2) -- (0,2.2);
\draw (-0.5,2.2) node[left] {$\delta_s$};
\draw[->] (0.87,2.2) -- (0.37,2.2);
\draw[->, line width = 3 pt,red] (0.185,0.4) -- (0.185,1.8);
\draw (-0.1,1.4) node[left] {\Large\textcolor{red}{$I_S$}};

\draw (1.6,-1.5) node[fill=red!30,rounded corners] {$\mu_0$, $\varepsilon_0$, $\kappa\neq 0$};
\draw (-1.5,-1.5) node[fill=red!30,rounded corners] {$\mu_0$, $\varepsilon_0$, $\kappa = 0$};

\end{tikzpicture}
\caption{\label{fig:SrfRes}Skin depth and surface resistance}
\end{center}
\end{figure}

Furthermore the skin depth allows us to easily calculate the losses in the surface. For a wave travelling along a
 conducting surface, one can define a surface resistance by assuming a constant current density within a surface
material layer, which is equivalent to the skin depth as shown in Fig.~\ref{fig:SrfRes}:

\vspace*{-0.5cm}
\meqn{
	R_{surf} = \frac{1}{\kappa\delta_s} \left[\Omega\right].
}{surface resistance.}{eq:surfResistance}
\indent This value has to be multiplied with $l/w$ to get the full RF resistance, with $l$ being the length of the conducting wall and $w$ being its width. 

\subsection{Energy and transport of energy}
We start this section by presenting Poynting's Law, and we will then explain the different components of it. 
Poynting's Law states nothing else than the conservation of electromagnetic energy:

\vspace*{-0.3cm}
\meqn{
-\frac{d}{dt}\int\limits_V w dV = \int\limits_A \mathbf{S} \cdot d\mathbf{A} + \int\limits_V \mathbf{E} \cdot \mathbf{J} dV.
}{Poynting's Law}{eq:PoyntingLaw}\\
\indent Reading the equation from left to right it states that: ``the rate of change of stored energy in a~volume equals 
the energy flow out of the volume (through a surface $\mathbf{A}$ surrounding the volume) plus the~losses within 
the volume (work performed on charges per time unit)''. In the following lines we will see that the components of 
Poynting's Law actually correspond to what is stated in the previous sentence. 

\subsubsection*{What is $\mathbf{E} \cdot \mathbf{J} $?}
In order to understand the expression $\mathbf{E} \cdot \mathbf{J}$ we follow \cite{bib:feynman} and start with the 
force acting on a charge that moves within electromagnetic fields.

\vspace*{-0.7cm}
\meqn{
	\mathbf{F} = q \left( \mathbf{E} + \mathbf{v} \times \mathbf{B} \right)
}{Lorentz Force}{eq:LorentzForce}
Multiplying the equation with $\mathbf{v}$ and knowing that $\mathbf{a} \cdot \left( \mathbf{a} \times \mathbf{b} 
\right) \equiv 0$ we get an expression for the work done on a charge per time unit
\begin{align}
	\mathbf{v}\cdot\mathbf{F} &= q \mathbf{v} \cdot \mathbf{E} .
\intertext{Assuming $N$ particles per volume unit we can write}
	N\mathbf{v}\cdot\mathbf{F} &= Nq \mathbf{v} \cdot \mathbf{E} =\mathbf{J}\cdot\mathbf{E}.
\end{align}

Therefore the expression $\mathbf{J}\cdot\mathbf{E}$ must be equal to the work done on charges per time unit and volume unit, or in other words: the loss of electromagnetic energy per volume unit.  

\subsubsection*{The Poynting vector $\mathbf{S}$ and the energy density $w$}
These quantities can be understood by manipulating Maxwell's Equations (compare e.g.~\cite{bib:Henke}).  
We multiply Eq.~\eqref{eq:rotH} with $\mathbf{E}$
\begin{align}
	\mathbf{E}\cdot\mathbf{J} &= \mathbf{E}\cdot\left( \nabla \times \mathbf{H} \right) - 
	\mathbf{E} \cdot \frac{\partial \mathbf{D}}{\partial t}  .
\intertext{Using Eq.~\eqref{eq:va1} this can be rewritten as}
	\mathbf{E}\cdot\mathbf{J} &= \mathbf{H} \cdot \left(\nabla \times\mathbf{E} \right) - 
	\nabla\cdot\left( \mathbf{E} \times \mathbf{H} \right) - \mathbf{E} \cdot \frac{\partial \mathbf{D}}{\partial t}  .
\intertext{Using the 2nd of Maxwell's Equations \eqref{eq:rotE} and assuming time invariant $\mu$ and $\varepsilon$
we can write}
		\mathbf{E}\cdot\mathbf{J} &= -\nabla\cdot \left(\mathbf{E}\times \mathbf{H}\right)
		 - \frac{\partial}{\partial t} \left(
		\frac{1}{2} \mathbf{E} \cdot \mathbf{D} + \frac{1}{2} \mathbf{H} \cdot \mathbf{B}\right).
\end{align}

Applying a volume integral together with Gauss' theorem \eqref{eq:Gauss} and rearranging the 
elements of the~equation we end up with

\vspace*{-0.1cm}
\meqn{
 	-\frac{\partial}{\partial t} \int_V \left( \frac{1}{2} \mathbf{E} \cdot \mathbf{D} + 
 	\frac{1}{2} \mathbf{H} \cdot \mathbf{B} \right) dV \hspace{3cm} \\
 	\hspace{3cm}= \int_A \left( \mathbf{E} \times \mathbf{H} \right) \cdot dA 
 	+ \int_V \mathbf{E} \cdot \mathbf{J} dV
}{Poynting's Law}{}\\
which can be directly compared with Eq.~\eqref{eq:PoyntingLaw}. On the left hand side we 
have the definition of the energy density

\vspace*{-0.7cm}
\meqn{
	w = w_{el} + w_{mag} = \frac{1}{2} \mathbf{E} \cdot \mathbf{D} + 
 	\frac{1}{2} \mathbf{B} \cdot \mathbf{H}
}{electric and magnetic energy density}{}\\
and from the right hand side we get the definition of the energy flux density or the poynting vector 
$\mathbf{S}$

\vspace*{-0.7cm}
\meqn{
	\mathbf{S} = \mathbf{E} \times \mathbf{H}. 
}{Poynting vector}{}\\
\indent The Poynting vector gives us the direction in which the electromagnetic wave transports energy, and from the cross 
product we understand that this direction is always perpendicular to the electric and magnetic field components. 
This is consistent with Paragraph~\ref{sec:PlaneWaves}, where we found that the field components ($E_x$, $H_y$) 
of a plane wave (see Eq.~\eqref{eq:solharmwave}) are perpendicular to the direction of propagation ($z$).

In the above derivation we have used Maxwell's Equations in their general form, meaning with time derivatives. In case
of the complex notation, the definitions of the energy density and Poynting vector have to be modified as follows 
(for a proof see \cite{bib:Henke} or \cite{bib:Weiland}).

\vspace*{-0.7cm}
\meqn{
	w = w_{el} + w_{mag} = \frac{1}{4} \mathbf{E} \cdot \mathbf{D}^* + 
 	\frac{1}{4} \mathbf{B} \cdot \mathbf{H}^*
}{electric and magnetic energy density in complex notation}{}

\vspace*{-0.7cm}
\meqn{
	\mathbf{S} = \frac{1}{2}\left(\mathbf{E} \times \mathbf{H}^*\right) 
}{complex Poynting vector}{eq:PoyCom}

\section{Electromagnetic waves in wave-guides}
In this section we derive the field components for electromagnetic waves, which propagate 
in waveguides. The same principle can then be used to calculate the standing wave pattern in accelerating cavities, 
which is nothing else than a superposition of two waves travelling in opposite directions. 

\subsection{Classification of modes in waveguides and cavities}
Before starting with the solution of the wave equation, we need to introduce a classification of field patterns, which 
can be found in waveguides and cavities. 

\subsubsection*{TM$_{mnp}$ modes or E$_{mnp}$ modes}
These modes have no magnetic field in the direction of propagation ($z$) and are therefore often called {\it Transverse
Magnetic} or TM-modes. On the other hand they have an electric field component, which is parallel to ($z$), and 
therefore the equivalent name {\it E-modes}.

The indices $m$, $n$, $p$ indicate the number of zeros or variations in the three directions of a coordinate system. In 
waveguides one only uses the first two indices, while in cavities -- due to the standing wave pattern along $z$ -- one 
needs all three for a complete description. In case of circular waveguides or cavities the indices indicate:

\begin{itemize}
	\item[-$m$] number of full-period variations of the field components in the azimuthal direction,
	\item[ ] for circular symmetric geometries: $\mathbf{E}$, $\mathbf{B}  \propto \cos(m\varphi)$, $\sin(m\varphi)$
	\item[-$n$] number of zeros ($x_{mn}$) of the axial field component in radial direction, 
	\item[ ] for circular symmetric geometries: $E_z$, $B_z \propto J_m(x_{mn}r/R_c)$
	\item[-$p$] number of half-period variations of the field components in longitudinal direction, 
	\item[ ] with:$\mathbf{E}$, $\mathbf{B} \propto \cos(p\pi z/l)$, $\sin(p\pi z/l).$
\end{itemize}

\begin{figure}[h!]
\centering
\includegraphics[width=0.7\textwidth]{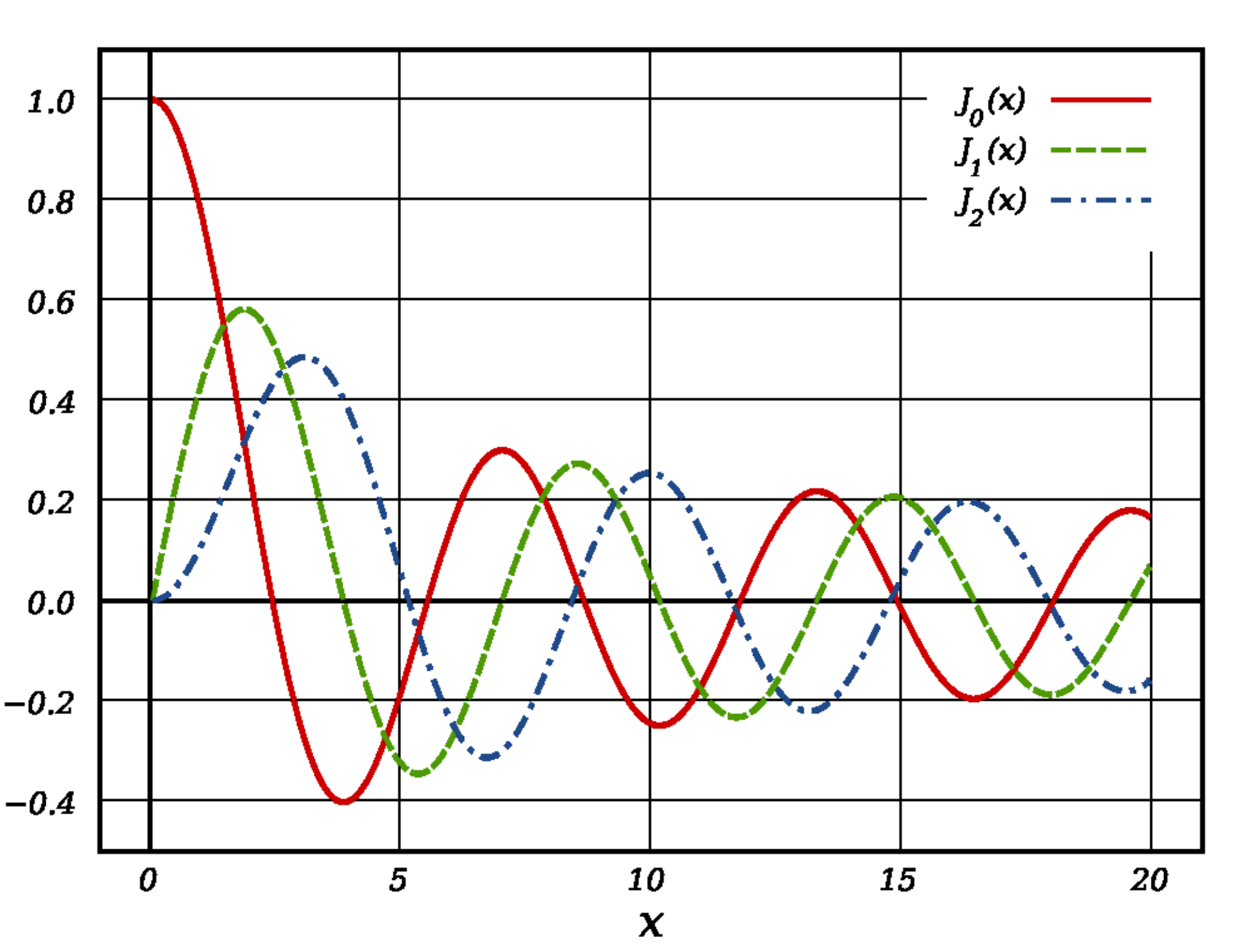}
\caption{Bessel functions of the first kind up to order 2}
\label{fig:Bessel}
\end{figure}

The functions $J_m$ which are introduced above are Bessel functions of the first kind and of m'th order and can be 
found in mathematical textbooks. The first three orders are shown in Fig.~\ref{fig:Bessel}.

\subsubsection*{TE$_{mnp}$ modes or H$_{mnp}$ modes}
Here, there is no electric field in the direction of propagation $z$, hence the name {\it Transverse Electric} or TE-modes. In analogy to the E-modes, H-modes do have a magnetic field component, which is parallel to $z$. The indices have the same meaning as above.

\subsubsection*{TEM modes}
This class of modes has neither electric nor magnetic field components in the direction of propagation. They 
can exist between two isolated conductors, e.g.\ a coaxial line. The advantage of TEM modes is that 
waves of any frequency can propagate, while TE and TM modes always have a cut-off frequency below which
they are damped exponentially (more on this later). However, the disadvantage of coaxial lines is that the losses on the two conductors
are generally higher than in rectangular or circular waveguides.  

\subsection{Solution of the wave equation in cylindric wave-guides}
Instead of trying to find solutions for all 6 vector components of the electric and magnetic fields, one can simplify 
the problem by using a vector potential $\mathbf{A}$ (without any physical meaning), which has only one component.
 Out of this vector potential one can then quickly derive all 6 field components. 
 
 It can be shown that only two types of modes can exist in waveguides: TM and TE-modes as introduced above. For 
 each mode type we introduce a vector potential $\mathbf{A}$ as follows. Since $\mathbf{H}$ and $\mathbf{E}$ 
 are divergence free and since $\nabla \cdot \left( \nabla \times \mathbf{a} \right) \equiv 0$ one can write\\
 
\vspace*{-1cm}
\begin{align}
	\mathbf{H}^{TM} &= \nabla \times \mathbf{A}^{TM}  \hspace{0.5cm} \mbox{with}& \hspace{0.5cm}  \mathbf{E}^{TM} 
	&= -\frac{i}{\omega\varepsilon}\nabla \times ( \nabla\times\mathbf{A}^{TM} )
	&& \mbox{\bf vector potential for TM waves} \label{eq:vpTM} \\
	\mathbf{E}^{TE} &= \nabla \times \mathbf{A}^{TE} \hspace{0.5cm} \mbox{with}& \hspace{0.5cm}  \mathbf{H}^{TE} 
	&= \frac{i}{\omega\mu}\nabla \times ( \nabla\times\mathbf{A}^{TE} )
	&& \mbox{\bf vector potential for TE waves} \label{eq:vpTE}
\end{align}
\noindent In both cases the vector potential fulfils the wave equation
\begin{align}
	\nabla^2 \mathbf{A} = -k^2\mathbf{A} \hspace{0.5cm} \mbox{with} \hspace{0.5cm} 
	k^2 = \omega^2 \mu\varepsilon
\end{align}
\noindent which can then be solved for the different coordinate systems and which has only one vector component in 
the direction of propagation
\begin{align}
	\mathbf{A} = A_z \mathbf{e}_z .
\end{align}
	 
\subsection*{Circular waveguides}
In circular waveguides as shown in Fig.~\ref{fig:circularwaveguide} the vector potential for TE and TM-modes is identical:
\begin{figure}[h!]
\centering
\begin{tikzpicture}[scale=0.9, ultra thick]

\draw (0,0) ellipse (0.7 and 1.4);
\draw[black,fill=gray!50] (4,1.4) arc (90:-90:0.7 and 1.4) -- (0,-1.4) arc (-90:90:0.7 and 1.4) -- (4,1.4);
\draw[->] (0,0) -- (0,1.4);
\draw (0.2,0.9) node {\Large $a$}; 
\draw[red] (-1,0) -- (0.7,0);
\draw[red, ->] (4.7,0) -- (5.5,0) node [right] {\Large $z$}; 

\end{tikzpicture}
\caption{\label{fig:circularwaveguide}Geometry of a circular waveguide}
\end{figure}
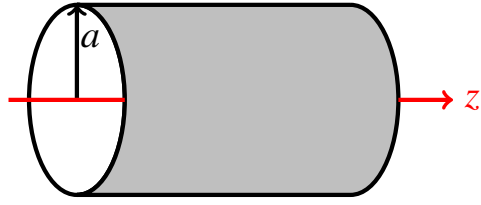

\vspace*{-1cm}
\begin{align}
	A_z^{TM/TE} &= C \mbox{J}_m (k_c r) \cos(m\varphi) \displaystyle e^{\pm i k_z z}  
	&& \mbox{\bf vector potential for circular waveguides} \label{eq:vpfcw} \\
\intertext{with}
	k_z & = \sqrt{k^2 - k_c^2}. && \mbox{\bf wave number in $z$-direction} \label{eq:kzdef}
\end{align}
\noindent Using Eq.~\eqref{eq:vpTM} we can derive the field components for TM-modes:\\

\vspace{-0.5cm}
\meqn{
\left. \begin{array}{lll}
E_r &= \displaystyle\frac{i}{\omega\varepsilon}\frac{\partial H_{\varphi}}{\partial z} &= -C \displaystyle\frac{k_z k_c}{\omega\varepsilon} \mbox{J'}_m(k_c r) \cos(m\varphi) \vspace{0.2cm} \\
E_{\varphi} &= -\displaystyle\frac{i}{\omega\varepsilon} \frac{\partial H_r}{\partial z} &= C\displaystyle\frac{m k_z}{\omega\varepsilon r} \mbox{J}_m(k_cr)\sin(m\varphi)\vspace{0.2cm}\\
E_{z} &= \displaystyle\frac{i k_c^2}{\omega\varepsilon}A_z &= C \displaystyle\frac{ik_c^2}{\omega\varepsilon} \mbox{J}_m (k_c r) \cos(m\varphi)\vspace{0.2cm}\\
H_r &= \displaystyle\frac{1}{r} \frac{\partial A_z}{\partial \varphi} &= -C\displaystyle\frac{m}{r} \mbox{J}_m (k_c r) \sin(m\varphi) \vspace{0.2cm}\\
H_{\varphi} &= -\displaystyle\frac{\partial A_z}{\partial r} &= -C k_c \mbox{J'}_m (k_c r)\cos(m\varphi)
\end{array} \right\} e^{\pm ik_z z}.
}{field components for TM modes in a circular waveguide}{eq:genfieldscirc}

Now one can use the boundary conditions to specify the cut-off wave-number $k_c$. From Section~\ref{sec:boundary} we know that electric field components parallel to the waveguide surface have to vanish at the surface, which means
\begin{equation}
	\left. \begin{array}{ll}
	E_{\varphi} (r=a) &= 0 \\
	E_z (r=a) &= 0
	\end{array} \right\} \Rightarrow J_m(k_c a) = 0 \hspace{0.5cm} \Rightarrow \hspace{0.5cm} k_c 
	= \displaystyle\frac{j_{mn}}{a}. 
\end{equation}
The n'th zeros $j_{mn}$ of the Bessel functions of m'th order are tabulated in mathematical 
textbooks \cite{bib:Abramowitz}. 

\vspace*{-0.5cm}
\noindent Using
\meqn{
	k_c = \displaystyle\frac{2\pi}{\lambda_c} = \displaystyle\frac{\omega_c}{c}
}{}{}\\
\noindent we can define the cut-off frequency of the waveguide\\

\vspace*{-1cm}
\meqn{
	\omega_c = c\displaystyle\frac{j_{mn}}{a} 
}{cut-off frequency for TM-modes in circular waveguides}{}

\noindent The mode which is most commonly used in circular waveguides is the TM$_{01}$, which has only three field components. By inserting $m=0$, and $n=1$ into \eqref{eq:genfieldscirc} and by using J'$_0(r) = -$J$_1(r)$ we get:\\
\meqn{
\left. \begin{array}{lll}
E_r &= C \displaystyle\frac{k_z k_c}{\omega\varepsilon} \mbox{J}_1(k_c r) \vspace{0.2cm} \\
E_{z} &= -C \displaystyle\frac{ik_c^2}{\omega\varepsilon} \mbox{J}_0 (k_c r) \vspace{0.2cm}\\
H_{\varphi} &= C k_c \mbox{J}_1 (k_c r)
\end{array} \right\} e^{\pm ik_z z}
}{field components of TM$_{01}$ mode in a circular waveguide}{eq:TM01circ}
with the cut-off frequency $\omega_c \approx c\frac{2.405}{a}$.

\begin{figure}[h!]
\begin{tikzpicture}
\node[inner sep=0pt,above right]{\includegraphics[height=3.5cm]{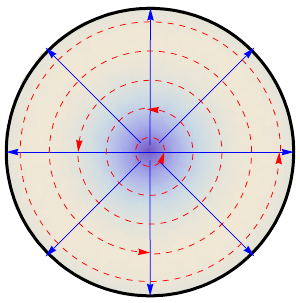}\hspace*{1cm}
\includegraphics[height=3.5cm]{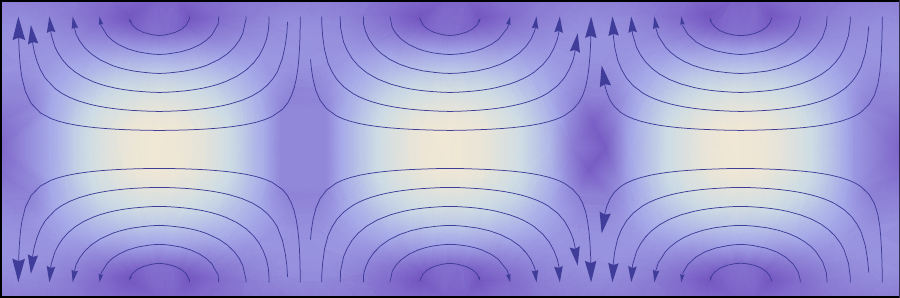}};
\draw[dashed,thick] (6.465,-0.5) -- (6.465,3.7);
\draw[dashed,thick] (9.85,-0.5) -- (9.85,3.7);
\draw[<->] (6.47,-0.2) -- node[below] {$\lambda_z/2$} (9.85,-0.2);
\node (4,3.5) {\LARGE\bf TM$_{01}$};
\end{tikzpicture}
\caption{Field lines of a TM$_{01}$ mode in a circular waveguide with $\omega=1.15*\omega_c$. Blue: electric, red (dashed): magnetic field lines. The background coloring is proportional to the norm of the field vector and light areas indicate high field regions of the magnetic field in the left plot and the electric field in the right plot.}
\label{fig:circularTM01}
\end{figure}
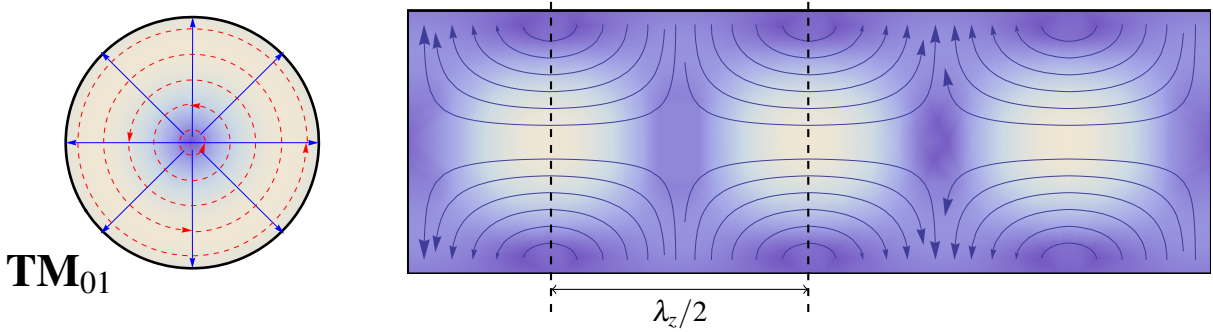

\begin{figure}[h!]
\begin{tikzpicture}
\node[inner sep=0pt,above right]{\includegraphics[height=3.5cm]{figure17a}\hspace*{1cm}
\includegraphics[height=3.5cm]{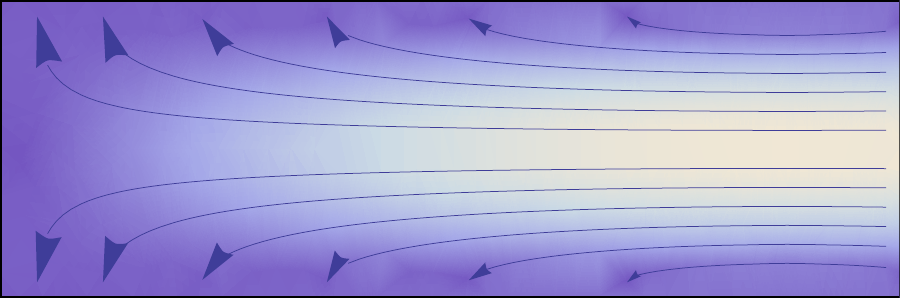}};
\node (4,3.5) {\LARGE\bf TM$_{01}$};
\end{tikzpicture}
\caption{Field lines of a TM$_{01}$ mode in a circular waveguide with $\omega=1.005*\omega_c$. Blue: electric, red (dashed): magnetic field lines. The background coloring is proportional to the norm of the field vector and light areas indicate high field regions of the magnetic field in the left plot and the electric field in the right plot.}
\label{fig:circularabovecutoff}
\end{figure}

The field pattern of the TM$_{01}$ mode is shown in Fig.~\ref{fig:circularTM01} for a mode frequency which is 
15\% above the cut-off frequency. The distance between the field minima or maxima corresponds to 0.5 times 
the~propagation wave length $\lambda_z$. With decreasing mode frequency $\lambda_z = 2\pi/k_z$ becomes 
longer and finally becomes infinite when the mode frequency equals the cut-off frequency $\omega_c$. This effect 
is shown in Fig.~\ref{fig:circularabovecutoff} where the TM$_{01}$ mode propagates with a frequency just 0.5\% 
above the cut-off frequency.  

\subsection*{Rectangular waveguides}

The derivation of fields in rectangular waveguides follows the principle we have used in the previous section for circular waveguides. In rectangular waveguides as shown in Fig.~\ref{fig:rectangularwaveguide} two different vector potentials are needed to describe TE and TM-modes: 
\begin{figure}[h!]
\centering
\begin{tikzpicture}[scale=0.9, ultra thick]

\draw (1.2,-0.8) -- (0,-1.4) -- (0,1.4) -- (1.2,2);
\draw (0,1.4) -- (1.2,1.4);
\draw (1.2,2) -- (5.2,2);
\draw[black,fill=gray!50] (1.2,2) rectangle (5.2,-0.8);
\draw[black,fill=gray!50] (0,-1.4) -- (4,-1.4) -- (5.2,-0.8) -- (1.2,-0.8) -- (0,-1.4);
\draw (-0.2,1.4) node {\Large $a$};
\draw (1.2,-1.1) node {\Large $b$};  
\draw[->,red] (0,-1.4) -- (6,-1.4) node [right] {\Large $z$}; 
\draw[->,red] (0,-1.4) -- (0,2.4) node [above] {\Large $x$};
\draw[->,red] (0,-1.4) -- (1.8, -0.5) node [right,above] {\Large $y$};

\end{tikzpicture}
\caption{\label{fig:rectangularwaveguide}Geometry of a rectangular waveguide with the transverse dimensions $a$ and $b$}
\end{figure}
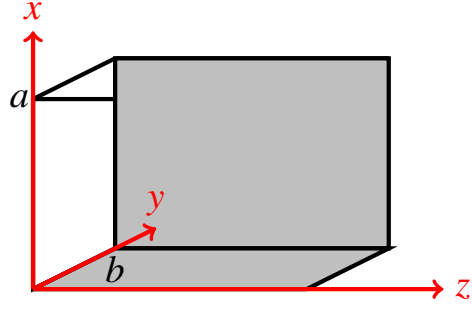

\meqn{
A_z^{TM} = C\sin(k_x x) \sin(k_y y) e^{\pm ik_z z}}
{vector potential for TM waves in rectangular waveguides}{ATMr}

\vspace{-0.7cm}
\meqn{
A_z^{TE} = C\cos(k_x x) \cos(k_y y) e^{\pm ik_z z}}
{vector potential for TE waves in rectangular waveguides}{ATEr}
\noindent with

\vspace*{-0.7cm}
\meqn{
k_z = \sqrt{k^2 - k_c^2} \hspace{0.5cm} \mbox{with} \hspace{0.5cm} k_c^2 = k_x^2 + k_y^2.}
{wavenumber in $\mathbf{z}$ direction}{}

\noindent We note that the position of the origin of the coordinate system is linked to the sine and cosine terms in Eqs.~\eqref{ATMr} and \eqref{ATEr}. The fields derived from the vector potentials have to fulfil the boundary conditions on the waveguide walls. So if we were to choose the origin for instance in the centre of the waveguide then the sine and cosine expressions would have to be exchanged to account for the changed symmetries with respect to the coordinate axes. Using again Eq.~\eqref{eq:vpTM} we derive the field components for TM-modes: 

\meqn{
\left. \begin{array}{lll}
E_x &= \displaystyle \frac{i}{\omega\varepsilon}\frac{\partial H_y}{\partial z} &= \pm C \displaystyle\frac{k_z}{\omega\varepsilon} \cos(k_x x) \sin(k_y y) \vspace{0.2cm} \\
E_y &= -\displaystyle\frac{i}{\omega\varepsilon} \frac{\partial H_x}{\partial z} &= \pm C\displaystyle\frac{k_z}{\omega\varepsilon} \sin(k_x x) \cos(k_y y) \vspace{0.2cm}\\
E_{z} &= \displaystyle\frac{i (k_z^2-k^2)}{\omega\varepsilon}A_z^{TM} &= C \displaystyle\frac{i(k_z^2-k^2)}{\omega\varepsilon} \sin(k_x x) \sin(k_y y)\vspace{0.2cm}\\
H_x &= \displaystyle \frac{\partial A_z^{TM}}{\partial y} &= C\displaystyle k_y \sin(k_x x) \cos(k_y y) \vspace{0.2cm}\\
H_y &= -\displaystyle\frac{\partial A_z^{TM}}{\partial x} &= -C k_x \cos(k_x x) \sin(k_y y)
\end{array} \right\} e^{\pm ik_z z}\hspace{-0.7cm}
}{field components for TM modes in a~rectangular waveguide\hspace{-0.8cm}}{eq:genfieldsrec}

\noindent Using the boundary conditions we can specify the wavenumbers $k_x$ and $k_y$

\begin{equation}
	\left. \begin{array}{l}
	E_y (x=a) = 0 \\
	E_z (x=a) = 0
	\end{array} \right\} \Rightarrow  k_x = \displaystyle\frac{m\pi}{a} \hspace{0.5cm} \mbox{and} \hspace{0.5cm} m=0,1,2,...
\end{equation}
\begin{equation}
	\left. \begin{array}{l}
	E_x (y=b) = 0 \\
	E_z (y=b) = 0
	\end{array} \right\} \Rightarrow  k_y = \displaystyle\frac{n\pi}{b} \hspace{0.5cm} \mbox{and} \hspace{0.5cm} n=0,1,2,...
\end{equation}

\noindent and the cut-off frequency for rectangular waveguides

\vspace*{-0.7cm}
\meqn{
\omega_c = c k_c = c \sqrt{k_x^2+k_y^2} = c\pi \sqrt{\left(\frac{m}{a} \right)^2 + \left(\frac{n}{b} \right)^2}.
}{cut-off frequency for TM-modes in rectangular waveguides}{}

The usual convention is to have $a>b$ and in that case the TE$_{10}$ mode is the mode with the lowest cut-off frequency. It is also the only mode propagating within a relatively large frequency band of $f_c^{TE,10} \rightarrow 2f_c^{TE,10}$, which is why it is the most common mode used in rectangular waveguides. The fields of TE modes can be derived using the same procedure with the TE vector potential.

\subsection{Wave propagation and dispersion relation}
In Figures~\ref{fig:circularTM01} and \ref{fig:circularabovecutoff} we have seen that the propagation wave length $\lambda_z$ of a waveguide mode is determined by its frequency and by how far the mode frequency is above the cut-off frequency of the~waveguide.  If the propagation wavelength depends on the mode frequency we can assume that also the phase velocity of a particular mode depends on the mode frequency. This relationship is called the~dispersion relation and using the definition of the wave number in Eq.~\eqref{eq:kzdef} we can write\\

\vspace*{-0.7cm}
\meqn{
k_z^2 = k^2 - k_c^2 = \frac{\omega^2 - \omega_c^2}{c^2} = \frac{\omega^2}{v_{ph}^2}}
{dispersion relation}{}
from which we can immediately see that:
\begin{itemize} 
	\item $k_z$ can only be real if the mode frequency $\omega$ is above the cut-off frequency $\omega_c$.
	\item for $\omega < \omega_c$ the mode cannot propagate and the fields are exponentially damped, 
\end{itemize}
and we also have a definition of the phase velocity, which is the speed at which the maxima and minima of the field
patterns move along the waveguide:\\

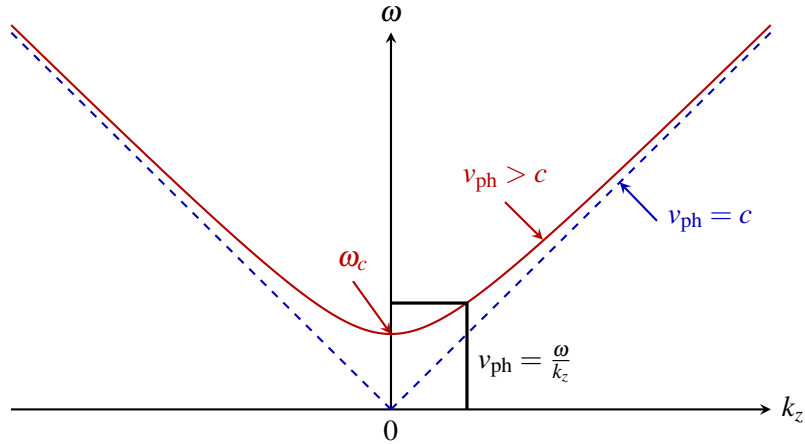
\begin{figure}[h!]
\centering
\begin{tikzpicture}[>=stealth, x=5cm/5,red!70!black,thick]
\draw[->,black](-5,0)  -- (5,0)node[right]{$k_{z}$};
\draw[->,black](0,0)node[below]{$0$}  -- (0,5)node[above]{$\omega$};

\pgfplothandlerlineto
\pgfplotfunction{\x}{0,0.01,...,5}{\pgfpointxy{\x}{sqrt(1+(\x^2))}}
\pgfusepath{stroke}

\pgfplothandlerlineto
\pgfplotfunction{\x}{0,0.01,...,5}{\pgfpointxy{-\x}{sqrt(1+(\x^2))}}
\pgfusepath{stroke}

\draw [->](1.5,2.75)node[above]{$v_{\text{ph}}>c$}--(2,2.25);
\draw [->](-0.5,1.7)node[above]{$\omega_{c}$}--(0,1);

\draw[dashed,blue!70!black,thick] (0,0) -- (5,5);
\draw[dashed,blue!70!black,thick] (0,0) -- (-5,5);
\draw [->,blue!70!black](3.5,2.5)node[right]{$v_{\text{ph}}=c$}--(3.01,3.01);

\draw [black, very thick](1,0)--(1,1.41)--(0,1.41);
\draw [black](1,0.6)node[right]{$v_{\text{ph}}=\frac{\omega}{k_{z}}$};
\end{tikzpicture}
\caption{Dispersion relation in a waveguide. The dotted line shows the case $v_{ph}=c$.}
\label{fig:DispRel}
\end{figure}

\vspace*{-0.7cm}
\meqn{
v_{ph}= \frac{\omega}{k_z} = c^2\frac{\omega^2}{\omega^2 - \omega_c^2}
}{phase velocity}{}

\noindent and which is not to be confused with the speed with which the wave actually propagates within the~waveguide. 
The dispersion relation is usually plotted in the form of the so-called ``Brillouin diagram'' as shown in Fig.~\ref{fig:DispRel}.

The slope of the dispersion relation is the group velocity\\

\vspace*{-0.7cm}
\meqn{
v_{gr} = \frac{d\omega}{d k_z}
}{group velocity}{}
\noindent which gives the velocity with which signals or energy is transported through a waveguide. 
From Fig.~\ref{fig:DispRel} we can conclude that:
\begin{itemize}
	\item Each frequency has a certain phase velocity and group velocity, which means that signals with a~certain frequency bandwidth will get deformed, while travelling along a waveguide. With the help of the dispersion relation we can easily quantify how much deformation we get.
	\item The phase velocity $v_{ph}$ is always larger than the velocity of light $c$ and at cut-off ($\omega = \omega_c$) it even becomes infinite: ($k_z = 0$, and $v_{ph} \Rightarrow \infty$).
	\item For acceleration one needs synchronism between the phase velocity (the speed of the field pattern) and the velocity of the particles $\Rightarrow$ acceleration in waveguides is impossible. 
	\item Information and therefore energy travels at the speed of the group velocity and is always slower than the speed of light. 
\end{itemize}

\subsection{Attenuation of waves (power loss method)}	
Up to now we have assumed perfect electrical conductors as the boundaries of our waveguides. Real waveguides and cavities have a certain resistance and the fields will penetrate into the conductors, which significantly complicates the solution of the wave equation. However, in Section \ref{sec:skindepth} we have seen that the~skin depth in metals is very much smaller than the RF wave length. This means that we can reasonably assume that the field patterns of waveguides with ideal boundaries or resistive metal boundaries will be practically identical (of course only for good conductors as for instance copper or aluminum). In order to calculate the attenuation of waves we will therefore use the fields of a waveguide with ideal electrical boundaries. From their magnetic fields we calculate the induced currents in the waveguide walls and then we apply the resistance of the real material to calculate the losses and then the damping of the waves. This principle is called the power loss method and is a simplified method for calculating RF losses on the surfaces of good conductors.

We start by defining the power, which is lost per length unit along the longitudinal axis of the~waveguide. 

\begin{align}
P' &= -\frac{dP}{dz} \\
\intertext{from} E,H &\propto e^{-\alpha z} \hspace{0.5cm}\Rightarrow \hspace{0.5cm} P \propto e^{-2\alpha z}\\
\intertext{we get immediately} P' &= -\frac{d P}{dz} = 2\alpha P
\end{align}
\noindent and thus the definition of the attenuation constant\\

\vspace*{-0.7cm}
\meqn{
\alpha = \frac{P'}{2P}}
{attenuation constant.}{eq:att}

In the next steps we need to derive expressions for the power $P$ transported through the waveguide, and the power loss per unit length $P'$. Using the field components of the TM$_{01}$ mode from Eq.~\eqref{eq:TM01circ} and the definition of the 
complex Poynting vector in Eq.~\eqref{eq:PoyCom} we get

\vspace*{-0.5cm}
\begin{align}
P = \frac{1}{2} \int\limits_A \left(\mathbf{E}\times\mathbf{H}^* \right) \cdot d\mathbf{A} = \frac{1}{2} \int\limits_0^a\int\limits_0^{2\pi} E_r H_{\varphi}^* rdrd\varphi &= \frac{C^2 k_z k_c^2\pi a^2 J_1^2(k_c a)}{\omega\varepsilon} ,\label{eq:Patt}\\
\intertext{where we have used}
\int\limits_0^a J_1^2(k_c r) r dr &= \frac{a^2}{2} J_1^2(k_c a).
\end{align}

\begin{figure}[h!]
\centering
\begin{tikzpicture}[scale=1, very thick]
\draw (0,0) circle (2);
\draw[dashed, thick] (0,0) circle (2.2);
\draw[thick, red] (1.32,1.32) -- (1.65,1.65); 
\draw[thick, black, dashed] (1.65,1.65) -- (2.9,2.9);
\draw[thick, red,->] (1.65,1.65) arc (45:0:2.33345);
\draw (2.33,0) node[right,red] {$H_{\varphi}=0$};
\draw[thick, red] (2.33345,0) arc (0:-45:2.33345);
\draw[thick, red] (1.65,-1.65) -- (1.32,-1.32);
\draw[thick, black, dashed] (1.65,-1.65) -- (2.9,-2.9);
\draw[thick, black, <->, dashed] (2.8,2.8) arc (45:-45:3.960);
\draw (4,0) node[right] {$a\Delta\varphi$};
\draw[thick, red, ->] (1.32,-1.32) arc (-45:0:1.8668);
\draw (1.86,0) [left,red] node {$H_{\varphi}$};
\draw[thick, red] (1.8668,0) arc (0:45:1.8668);

\draw[->|, thick] (0,2.6) -- (0,2.2);
\draw[->|, thick] (0,1.6) -- (0,2);
\draw (0,2.2) node[above, left] {$\delta_s$};

\foreach \x/\y in {0.643/0.766,0.643/-0.766,0.766/0.643,0.866/0.5,0.94/0.342,0.985/0.174,1/0,0.985/-0.174,0.94/-0.342,0.866/-0.5,0.766/-0.643}
	\fill[blue] (2.1*\x,2.1*\y) circle (0.06);

\draw (0,-2.7) node[fill=red!30,right,above,rounded corners] {$\kappa=\kappa_{Al}$};
\draw (0,-1) node[fill=red!30,right,below,rounded corners] {$\kappa=0$};
\end{tikzpicture}
\caption{\label{fig:Amp2}Amp\'{e}res Law applied to calculate the surface currents on a circular waveguide}
\end{figure}
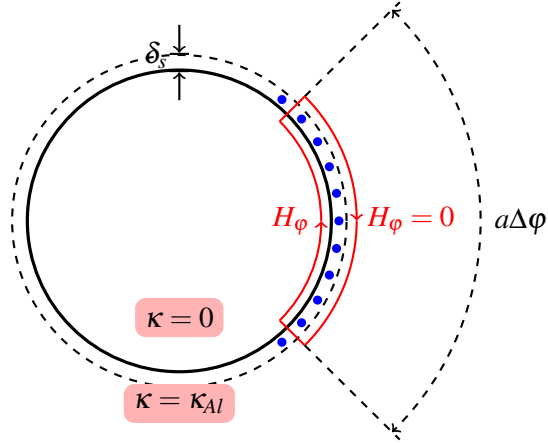

In order to calculate the losses on the waveguide surface we first need to know the surface currents which flow within the skin depth. For this purpose we make use of Amp\`{e}res Law as shown in Fig.~\ref{fig:Amp2}.

\begin{align}
\oint\limits_c \mathbf{H}\cdot d\mathbf{l} &= I = \oint\limits_c \mathbf{J} \cdot (\delta_s d\mathbf{l}) &\hspace*{0.75cm}& \mbox{\bf Amp\`{e}res Law}\\
\intertext{and since the magnetic field has only an azimuthal component we obtain}
H_{\varphi}(r=a,z) &= C k_c \mbox{J}_1 (k_c a) e^{-ik_z z} = J_z(z)\delta_s \label{eq:HpIz}.\\
\intertext{The power density in W/m$^3$ in the waveguide wall is given by}
p_v &= \frac{1}{2} \mathbf{E}\cdot \mathbf{J}^* = \frac{1}{2\kappa} J_z J_z^* = \frac{\partial^3 P}{(\partial r)(r\partial\varphi)(\partial z)} && \mbox{\bf power density} \label{eq:powden}\\
\intertext{out of which we can write an expression for the power loss per unit length. Together with Eq.~\eqref{eq:HpIz} we get}
P' &= \frac{\partial P}{\partial z} =  \int\limits_a^{a+\delta_s} \int\limits_0^{2\pi} p_v r dr d\varphi= \frac{\pi a C^2 k_c^2 J_1^2(k_c a)}{\kappa\delta_s} &&\mbox{\bf power loss per unit length} \label{eq:Ppatt}\\
\intertext{where we have used the fact that $\delta_s \ll a$ to simplify the evaluation of the integral. Now we insert Eq.~\eqref{eq:Patt} and \eqref{eq:Ppatt} into \eqref{eq:att} and get an expression for the attenuation of TM$_{01}$ modes in circular waveguides}
\alpha &= \frac{P'}{2P} = \frac{R_{surf}}{Z_0 a \sqrt{1-\left(\frac{f_c}{f}\right)^2}}. &&\parbox{5cm}{\bf attenuation of TM$_{01}$ in circular waveguide}
\end{align}

In the expression above we have used the definition of the surface resistance from Eq.~\eqref{eq:surfResistance} and the~definition of the free space wave impedance $Z_0$ of Eq.~\eqref{eq:ZfreeSpace}.

\begin{figure}[h!]
\centering
\includegraphics[angle=270,width=0.7\textwidth]{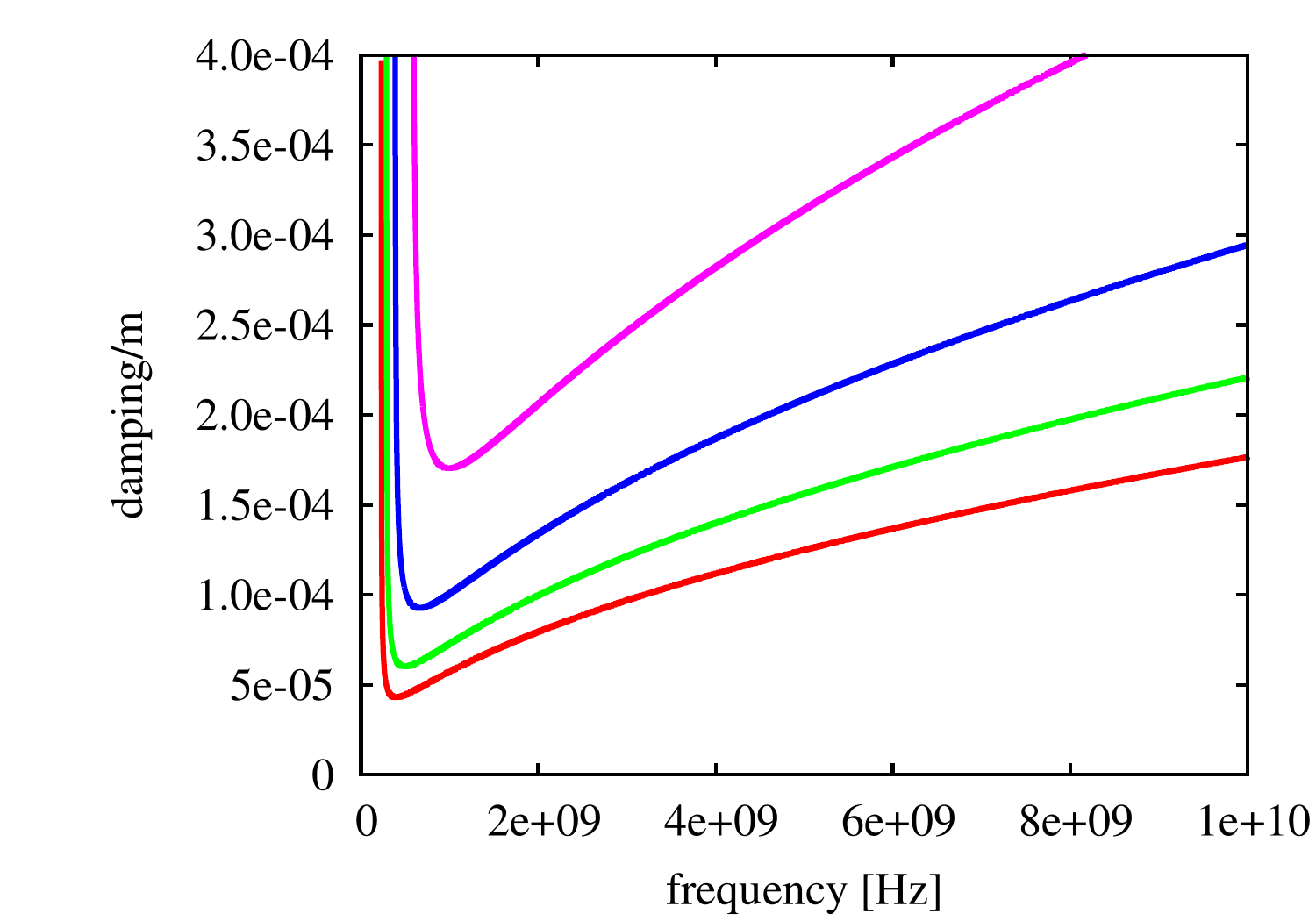}
\caption{Attenuation of a TM$_{01}$ mode in a circular aluminum waveguide with changing radius, bottom to top: 0.5\,m, 0.4\,m, 0.3\,m, 0.2\,m.}
\label{fig:AlDamping}
\end{figure}

As an example we plot the attenuation constant for an aluminum waveguide in Fig.~\ref{fig:AlDamping}, where we can see that for this type of waveguide:
\begin{itemize}
\item large diameter waveguides result in smaller losses, which means that a cost optimum has to be found between the waveguide cost, its space requirements and the losses,
\item the minimum losses occur if the operating frequency of the TM$_{01}$ mode is a factor of $\sqrt{3}$ above the cut-off frequency (try to proof this!). 
\end{itemize}

\section{Accelerating cavities}
\subsection{Travelling wave cavities}
\begin{figure}[h!]
		\begin{center}
		\begin{tikzpicture}[>=stealth, x=7cm/5,red!70!black,thick]
	\draw[->,black,dashed, line width=1pt](-2,0)  -- (2.5,0)node[right]{beam};
	\foreach \x in {-1.4,-0.7,0,0.7,1.4}
		\draw[black,fill=gray!50] (\x,0) ellipse (0.2cm and 0.8cm);
	\foreach \x in {-1.4,-0.7,0,0.7,1.4}
		\draw[black,fill=white] (\x,0) ellipse (0.12cm and 0.6cm);
	\draw[black] (-1.4,0.8) -- (1.4,0.8) (-1.4,-0.8) -- (1.4,-0.8);
	\draw[<->,black] (-1.4,-1) -- (1.4,-1) node [below,midway] {$L$};
	\draw[<->,black] (-1.8,-0.6) -- (-1.8,0.6) node [left] {$2a$};
	\draw[<->,black] (1.8,-0.8) -- (1.8,0.8) node [right] {$2b$};
	\draw[->,black] (-0.5,1) -- (-0.05,1);
	\draw[->,black] (0.5,1) -- (0.05,1) node [above,midway] {$h$};
	\end{tikzpicture}\hspace{1cm}
	\caption{Geometry of a simple travelling wave structure}
	\end{center}
	\label{fig:discloaded}
	\end{figure}
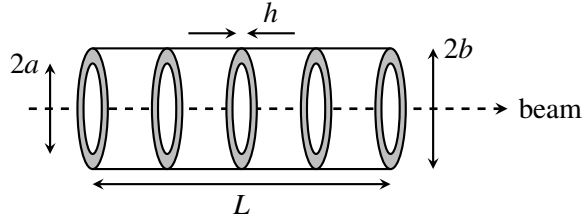

In order to accelerate particles in ``waveguide like'' structures, the phase velocity in the structures needs to be slowed down, which can be achieved by putting some ``obstacles'' into the waveguide. In Fig.~\ref{fig:discloaded} we see a simple example of a disc-loaded waveguide. 

The dispersion relation for such a structure is derived e.g.\ in \cite{bib:Wangler} as\\
\meqn{
\omega = \frac{2.405 c}{b} \sqrt{1+\kappa (1-\cos(k_z L) e^{-\alpha h})}
}{dispersion relation of disc-loaded circular waveguide}{eq:dispcirc}
\noindent
with
\begin{equation}
\kappa = \frac{4a^3}{3\pi \mbox{J}_1^2(2.405) b^2L} \ll 1\hspace*{0.5cm} \mbox{and} \hspace*{0.5cm}\alpha \approx \frac{2.405}{a}.
\end{equation}

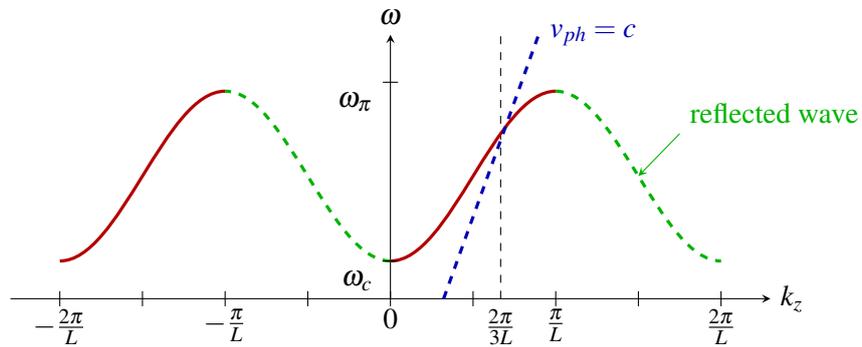
\begin{figure}[h!]
	\begin{center}
	\begin{tikzpicture}[>=stealth]
	\draw[->,black](-5,-0.5)  -- (5,-0.5)node[right]{$k_{z}$};
	\foreach \x in {-4,-3,...,4}
	\draw [x=4*1.57ex](\x,-0.6) -- (\x,-0.4);
	\draw[->,black](0,-0.5)node[below]{$0$}  -- (0,3)node[above]{$\omega$};
	\draw [x=4*1.57ex,dashed](4/3,-0.5)node[below]{$\frac{2\pi}{3L}$}--(4/3,3);
	\draw[x=-4*1.57ex,y=6.5ex,dashed,green!70!black,very thick] (0,0)cos(1,1) sin (2,2);
	\draw[x=-4*1.57ex,y=6.5ex,red!70!black,very thick](2,2)cos (3,1) sin (4,0);
	\draw[x=4*1.57ex,y=6.5ex,red!70!black,very thick] (0,0)cos(1,1) sin (2,2);
	\draw[x=4*1.57ex,y=6.5ex,dashed,green!70!black,very thick](2,2)cos (3,1) sin (4,0);
	\draw [x=4*1.57ex,y=6.5ex,->,green!70!black](3.5,1.5)node[above right]{reflected wave}--(3,1);
	\draw [dashed,blue!70!black,very thick](0.7,-0.5)--(1.95,3)node[right]{$v_{ph}=c$};
	\draw [x=4*1.57ex](-4,-0.5)node[below]{$-\frac{2\pi}{L}$};
	\draw [x=4*1.57ex](4,-0.5)node[below]{$\frac{2\pi}{L}$};
	\draw [x=4*1.57ex](-2,-0.5)node[below]{$-\frac{\pi}{L}$};
	\draw [x=4*1.57ex](2,-0.5)node[below]{$\frac{\pi}{L}$};
	\draw (0.1,0)--(-0.1,0) node[below left]{$\omega_{c}$};
	\draw (0.1,2.37)--(-0.1,2.37) node[below left]{$\omega_{\pi}$};
	\end{tikzpicture}
	\caption{Dispersion diagram for a disc-loaded travelling wave structure. Here the chosen operating point is $v_{ph} = c$ and $k_z = 2\pi/3L$}
	\label{fig:dispersion-discs}
	\end{center}
	\end{figure}

\noindent Plotting Eq.~\eqref{eq:dispcirc} gives us the Brioullin diagram in Fig.~\ref{fig:dispersion-discs} where we can see that we now get phase velocities, which are equal and even below the speed of light. We can also understand why in electron accelerators one often uses the $2\pi/3$ mode for acceleration, because for this mode (in this example) the~phase velocity is just equal to the speed of light. It should be noted that for different geometries it is possible to operate with different modes and also at velocities $v_{ph} < c$. When a structure operates in the~$2\pi/3$ mode it means that the RF phase shifts by $2\pi/3$ per cell, or in other words one RF period stretches over three cells. 

\begin{figure}[h!]
\begin{center}
\includegraphics[width=13cm]{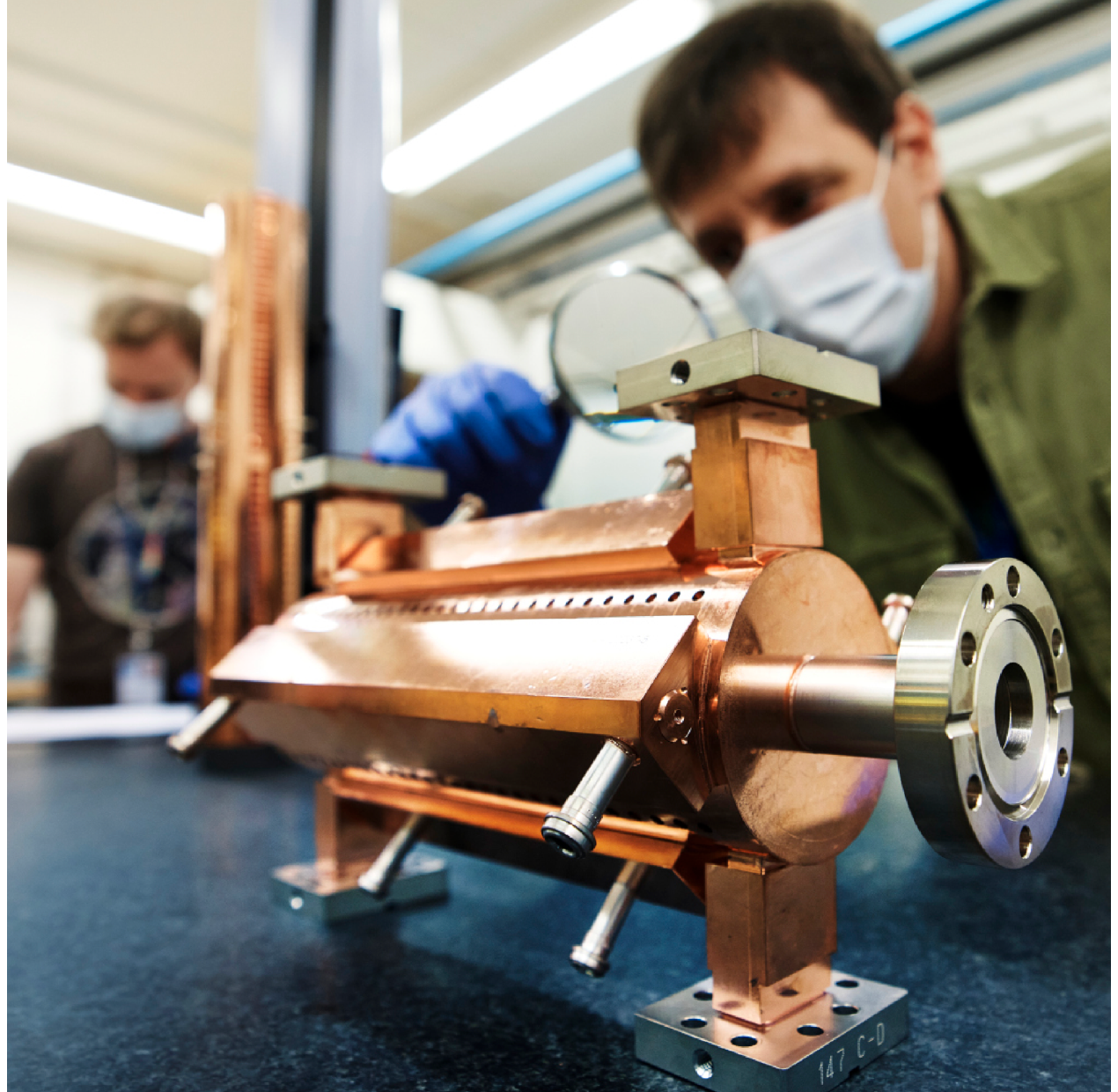}
\label{fig:travellingwave}
\caption{A CLIC prototype travelling wave structure, CERN-PHOTO-202202-013-8}
\end{center}
\end{figure}

By attaching an input and output coupler to the outermost cells of the structure we obtain a usable accelerating structure. Since the particles gain energy in every cell, the electromagnetic wave becomes more and more damped along the structure, is then extracted via the output coupler and dumped in an RF load. If one is interested to have the maximum possible accelerating gradient in each cell, then one can counteract the decreasing fields by changing the bore radius from cell to cell. The idea is to slow down the group velocity from cell to cell and to obtain a `constant gradient' structure, rather than a `constant impedance' structure where the bore radii are kept constant. Other optimizations, e.g., for maximum efficiency are also possible. Figure~\ref{fig:travellingwave} shows a travelling wave prototype structure for the CLIC study, which can operate at gradients around 100\,MV/m.

\subsection{Standing wave cavities}
One obtains a cylindrical standing wave structure by simply closing both ends of a circular wave guide with electric walls. This will yield multiple reflections on the end walls until a standing wave pattern is established. Owing to the additional boundary conditions in the longitudinal direction, we get another `restriction' for the existence of electromagnetic modes in the structure. While the longitudinally open travelling wave structure allows all frequencies and cell-to-cell phase variations on the dispersion curve, now only certain `loss-free' modes (still assuming perfectly conducting walls) with discrete frequencies and discrete phase changes can exist in the cavity. If one feeds RF power at a different frequency, then the excited fields will be damped exponentially, similar to the modes below the cut-off frequency of a~wave-guide. 

The corresponding dispersion relation for standing wave cavities can again be found in textbooks (Wangler or also Ref.~\cite{bib:CASMaurizio}). However, one should pay attention as to whether the structure under consideration has magnetic or electric cell-to-cell coupling and which kind of end-cell is used for the~analysis. The most common form of the dispersion relation is derived from a coupled circuit model with $N+1$ cells. Usually the model has half-cell terminations on both sides of the chain, representing the~behaviour of an infinite chain of electrically coupled resonators (compare with the original papers by Nagle and Knapp \cite{Knapp}):\\

\vspace{-1cm}
\meqn{
\omega_n = \frac{\omega_{0}}{\sqrt{1+k\cos\left(n\pi/N \right)}} \mbox{, }n=0,1,\ldots ,N\mbox{.}}
{dispersion relation for half-cell terminated standing wave structure}{eq:dispstandmag} 

Assuming an odd number of cells, $\omega_{\,0}$ is the frequency of the $\pi/2$ mode and of the uncoupled single cells. $k$ is the cell-to-cell coupling constant, and $n\pi/N$ the phase shift from cell to cell. For $k\ll 1$, which is usually fulfilled, the coupling constant is given by\\

\vspace*{-0.7cm}
\meqn{
 k = \frac{\omega_{\pi -mode}-\omega_{0-mode}}{\omega_{0}}\mbox{.}}
 {coupling constant}{eq:couplconstant} 

\noindent Two characteristics of the dispersion curve are worth noting:

\begin{itemize}
\item the total frequency band of the mode $\omega_{\pi - mode}-\omega_{0-mode}$ is independent of the number of cells, which means that we can determine the cell-to-cell coupling constant by measuring the complete structure (only true if all coupling constants are equal); 
\item for electric coupling the 0-mode has the lowest frequency and the $\pi$-mode has the highest frequency. In case of magnetic coupling this behaviour is reversed and one can find the corresponding dispersion curve by changing the sign before the coupling constant in Eq.~\eqref{eq:dispstandmag}.
\end{itemize}

\begin{figure}[h!]
	\begin{center}
	\begin{tikzpicture}[domain=0:7,samples=60]
	\draw[red,very thick, mark=x,mark repeat=10] plot (\x, {80/(sqrt(1-0.05*cos(180*\x /7)))});
	\draw[black] (0,77.5) -- (7,77.5);
	\draw[black] (0,77.5) -- (0,82.5);
	\draw[black] (7,77.5) -- (7,82.5);
	\draw[black] (0,77.5) node [below]{0};
	\draw[black] (7,77.5) node [below]{$\pi$};
	\draw[black] (3.5,77.7) -- (3.5, 77.5) node [below]{$\pi/2$};
	\draw[black] (-0.2,82.078) -- (0,82.078) node [left]{$\frac{\displaystyle\omega_0}{\displaystyle\sqrt{1-k}}$};
	\draw[black] (7.2,82.078) -- (7,82.078);
	\draw[black] (-0.2,78.072) -- (0,78.072) node [left]{$\frac{\displaystyle\omega_0}{\displaystyle\sqrt{1+k}}$};
	\draw[black] (7.2,78.072) -- (7,78.072);
	\end{tikzpicture}
	\caption{Dispersion diagram for a standing wave structure with seven magnetically coupled cells}
	\label{fig:dispersion-standing}
	\end{center}
	\end{figure}
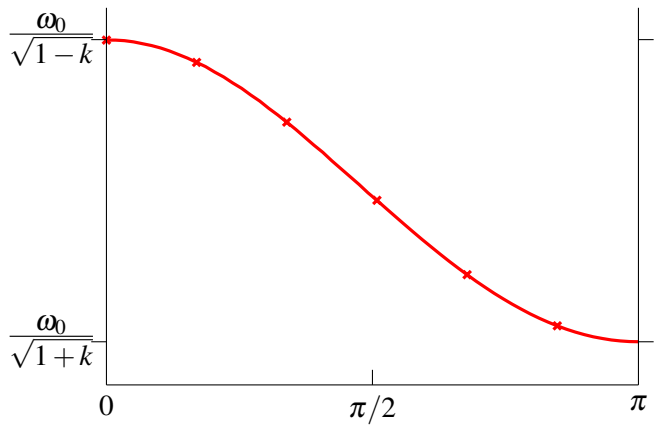

In Figure~\ref{fig:dispersion-standing} we plot the dispersion curve for a seven-cell (half-cell terminated) magnetically coupled structure according to Eq.~\eqref{eq:dispstandmag}.

In practice one usually has cavities with full-cell termination, and in this case one has to detune the frequency of the end cells to obtain a flat field distribution in the cavity \cite{Schriber01}. In this case one can have a flat field distribution for the $0$-mode or the $\pi$-mode but not for both at the same time because the end cells have to be detuned by a different amount \cite{Schriber2}. 

\subsection{Standing wave versus travelling wave}
The principal difference between the two types of cavity is how and how fast the cavities are filled with RF power. Travelling wave structures are filled `in space', which means that basically cell after cell is filled with power. For the following estimations we consider a frequency range in the 100s of MHz: the filling of a travelling wave structure typically takes place with a speed of approximately 1--3\% of the speed of light and results in total filling times in the sub-microsecond range. Standing wave structures on the other hand are filled `in time': the electromagnetic waves are reflected at the end-walls of the cavity and slowly build up a standing wave pattern at the desired amplitude. In normal-conducting cavities this process is typically in the range of 10s of microseconds. In superconducting cavities the filling process can easily go into the millisecond range (depending on the required field level, the accelerated current, and the cavity parameters). This means that for applications that require very short beam pulses ($< 1\,\mu$s) travelling wave structures are much more power efficient. For longer pulses ($> n\times 10\,\mu$s) both structures can be optimized to similar efficiencies and cost. Since one can have extremely short RF pulses in travelling wave structures, one can obtain much higher peak fields than in standing wave structures. This is demonstrated by the CLIC \cite{clic} accelerating structures, which have reached values of $\approx 100$\,MV/m (limited by breakdown of the electric field), while the design gradient for the superconducting (standing wave) ILC \cite{ilc} cavities is just slightly above 30\,MV/m (generally limited by field emission, and quenches caused by peak magnetic fields). 

Travelling wave structures can theoretically be designed for non-relativistic particles. In existing accelerators, however, they are mostly used for relativistic particles. Low-beta acceleration is typically performed with standing wave cavities.  

For lack of an obvious criterion (other than the pulse length, or particle velocity), one has to do the optimization and cost exercise for each specific application in order to decide which structure is more efficient. Two excellent papers \cite{Miller86, Moiseev00}, which perform this exercise can be used as reference. 

\subsection{The pillbox cavity}
The simplest TM-mode cavity is a simple cylinder, the so-called pill-box cavity for which we will derive the basic characteristics.

\begin{figure}[h!]
	\begin{center}
	\includegraphics[width=3.5cm]{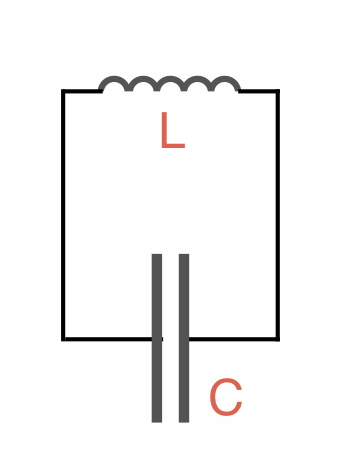}\hspace*{1cm}
	\includegraphics[width=3.5cm]{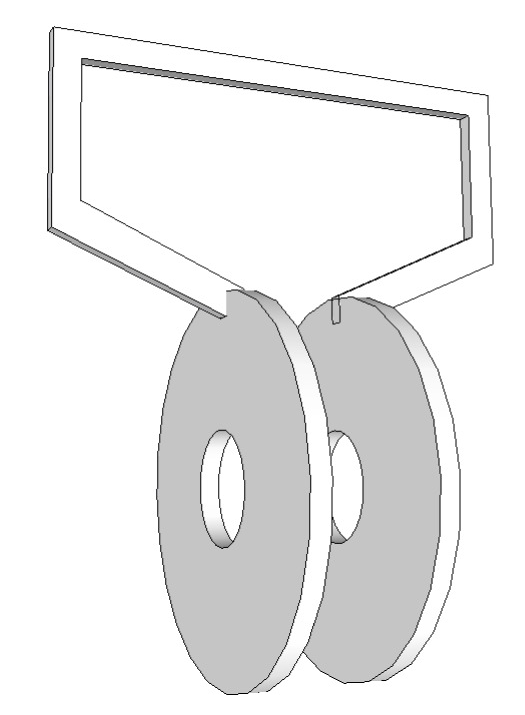}\hspace*{1cm}
	\includegraphics[width=3.5cm]{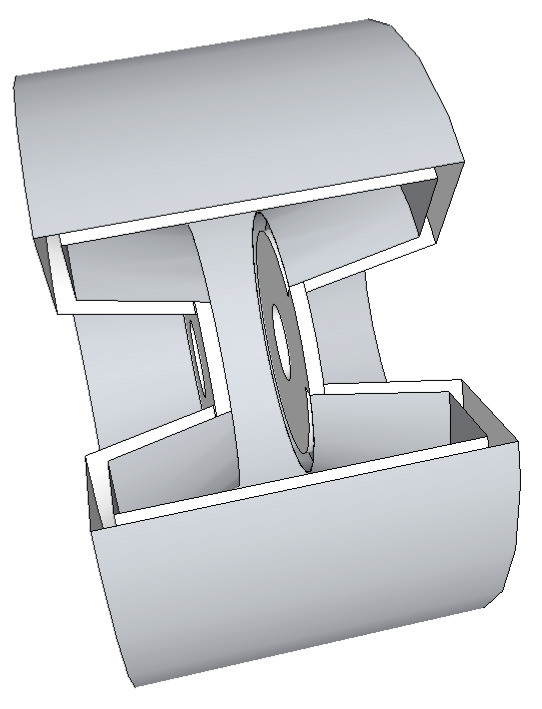}
	\caption{Transition from a lumped resonating circuit to a resonating cavity}
	\label{fig:lumped}
	\end{center}
	\end{figure}

Resonating cavities can be represented conveniently by a lumped element circuit of an inductor (storage of the magnetic energy) and a capacitor (storage of electric energy). Looking at Fig.~\ref{fig:lumped} one can easily imagine how the lumped circuit can be transformed into a cavity.

The pill-box cavity is nothing else than an empty cylinder with a conducting inner surface. The~simplest mode in this cavity is the so-called TM$_{\,010}$ mode, which has zero full-period variations in the~azimuthal direction ($\varphi$), one `zero' of the axial field component in the radial direction ($r$), and zero half-period variations in the longitudinal ($z$) direction. We can derive the general field equations by using the vector potential for circular waveguides of Eq.~\eqref{eq:vpfcw} and by simply superimposing two waves: one propagating in positive $z$-direction and one propagating in negative $z$-direction:\\
\meqn{
A_z^{TM/TE} = C \mbox{J}_m (k_r r) \cos(m\varphi) \underbrace{\left(e^{-i k_z z}+ e^{i k_z z} \right)}_{2\cos(k_z z)}.}
{vector potential for travelling waves in positive and negative $z$-direction}{}

\noindent Using Eq.~\eqref{eq:vpTM} we derive the TM field components\\ 
\meqn{
\begin{array}{lcl}
E_r &= \frac{i}{\omega\varepsilon}\frac{\partial H_{\varphi}}{\partial z} &= i2C \frac{k_z k_r}{\omega\varepsilon} \mbox{J'}_m(k_r r) \cos(m\varphi) \sin(k_z z) \vspace{0.2cm}\\
E_{\varphi} &= -\frac{i}{\omega\varepsilon} \frac{\partial H_r}{\partial z} &= -i2C\frac{m k_z}{\omega\varepsilon r} \mbox{J}_m(k_r r)\sin(m\varphi) \sin(k_z z) \vspace{0.2cm} \\
E_{z} &= \frac{i k_r^2}{\omega\varepsilon}A_z &= i2C \frac{k_r^2}{\omega\varepsilon} \mbox{J}_m (k_r r) \cos(m\varphi)\cos(k_z z) \vspace{0.2cm} \\
H_r &= \frac{1}{r} \frac{\partial A_z}{\partial \varphi} &= -2C\frac{m}{r} \mbox{J}_m (k_r r) \sin(m\varphi) \cos(k_z z)\vspace{0.2cm} \\
H_{\varphi} &= -\frac{\partial A_z}{\partial r} &= -2C k_r \mbox{J'}_m (k_r r)\cos(m\varphi) \cos(k_z z).
\end{array}}
{TM-modes in a pillbox cavity}{}

\begin{figure}[h!]
\centering
\begin{tikzpicture}[scale=0.9, ultra thick]

\draw (0,0) ellipse (0.7 and 1.4);
\draw (2,1.4) arc (90:-90:0.7 and 1.4);
\draw (0,1.4) -- (2,1.4);
\draw (0,-1.4) -- (2,-1.4);
\draw[->] (0,0) -- (0,2.2) node[above] {\Large $r$};
\draw (0.2,1.6) node {\Large $a$}; 
\draw[dashed] (0,0) -- (2.7,0);
\draw (2.9,0.35) node {\Large $L$};
\draw[->] (2.7,0) -- (3.5,0) node [right] {\Large $z$}; 

\end{tikzpicture}
\caption{\label{fig:pillbF}Pillbox cavity}
\end{figure}
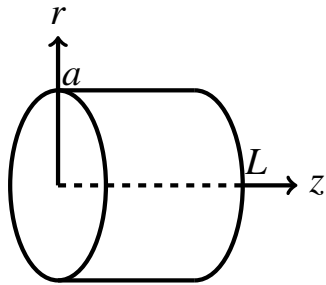

In standing wave cavities the term `cut-off' frequency does not really make sense, so we replace the~expression $k_c$ by $k_r$, indicating that we do have a radial dependency of the axial field component, which can also be interpreted as a radial wave-number. In the next step we apply the boundary conditions for a~pillbox with radius $a$ and length $L$ as shown in Fig.~\ref{fig:pillbF}.

\noindent and get

\vspace*{-0.7cm}
\begin{align}
E_r(z=0/L), E_{\varphi}(z=0/L) &= 0 \hspace{0.5cm} \Rightarrow k_z = \frac{p\pi}{L} \\
E_{\varphi}(r=a), E_z(r=a), H_r (r=a) &= 0 \hspace{0.5cm} \Rightarrow k_r = \frac{j_{mn}}{a} 
\end{align}

In the case of the circular waveguide the transverse boundary condition made a discrete quantity out of $k_c$ (which we now call $k_r$ in above equations), and thus defined the cut-off frequency. Now with the second boundary in $z$-direction we obtain a discrete solution also for $k_z$. Both of them together result in a discrete set of frequencies (the dispersion relation) for our pillbox cavity.\\

\vspace*{-0.7cm}
\meqn{
k^2 = \frac{\omega^2}{c^2} = k_z^2 + k_r^2 \hspace{0.5cm} \Rightarrow f_{mnp}^{TM}=\frac{c}{2\pi} \sqrt{\left( \frac{j_{mn}}{a} \right)^2+\left(\frac{p\pi}{L} \right)^2}}
{dispersion relation for TM-modes in a pillbox cavity}{eq:DispPillBox}

We note that the dispersion relation of a single-cell cavity as above is different from the dispersion relation that can be derived for multi-cell cavities as in the case of Eq.~\eqref{eq:dispstandmag}. The latter is derived from a~model of equivalent lumped circuits, each representing a cell resonating in the TM$_{010}$ and being coupled to its neighbours in order to model the behaviour of a multi-cell cavity, while Eq.~\eqref{eq:DispPillBox} is directly derived from Maxwell's Equations and describes a field pattern which is created by the boundary conditions of our pillbox. 

The TM-mode with the lowest frequency is the TM$_{010}$ mode with\\

\begin{figure}[h!]
\centering
\includegraphics[width=2cm]{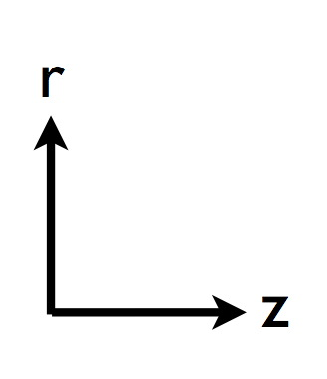}
\includegraphics[width=4cm]{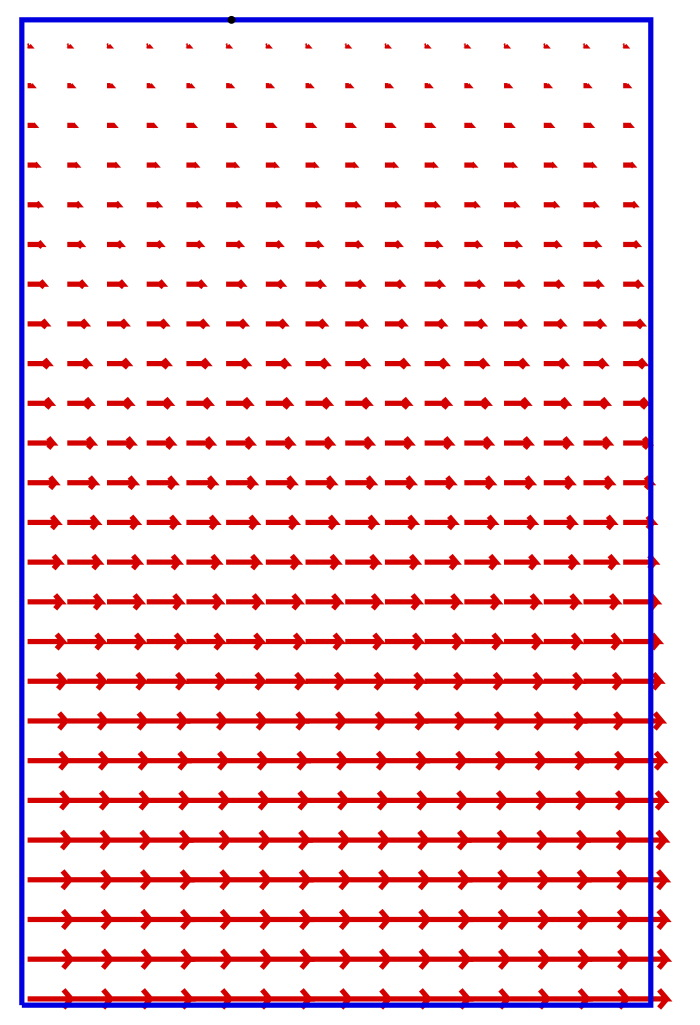}\hspace{2cm}
\includegraphics[width=4cm]{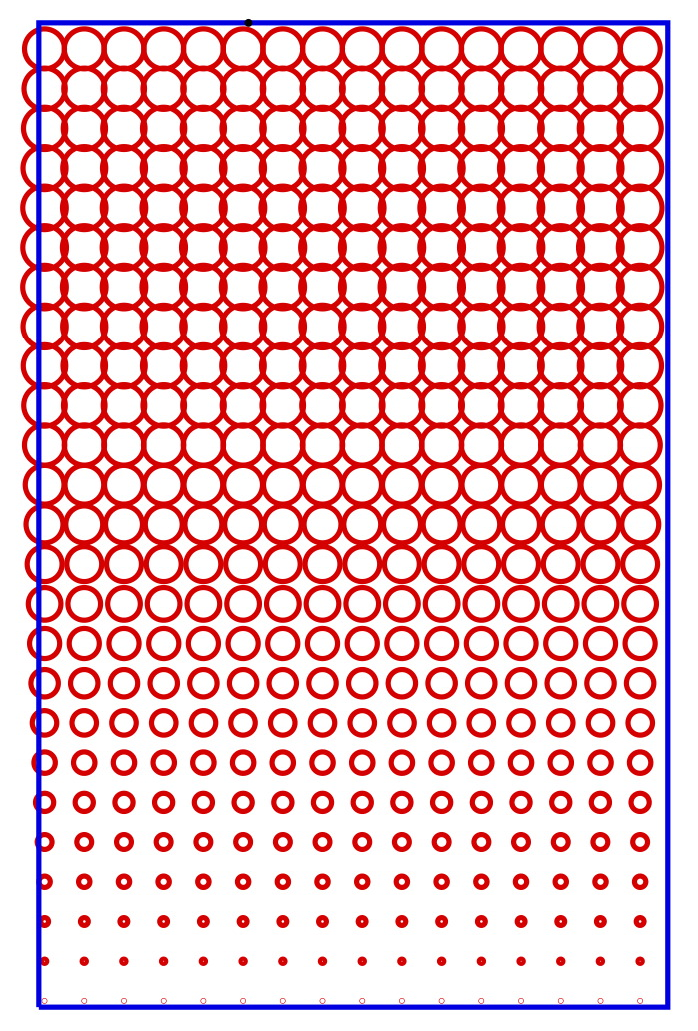}
\caption{\label{fig:PFields}Field pattern of the TM$_{010}$ mode in a pillbox cavity.}
\end{figure}

\vspace*{-0.7cm}
\meqn{
f_{010}^{TM}= \frac{2.405 c}{2\pi a}}
{frequency of the TM$_{010}$ pillbox mode}{eq:f-pillbox}
and its field components are\\
\meqn{
\begin{array}{lll}
E_z &= -i2C\frac{j_{01}^2}{a^2\omega\varepsilon}\mbox{J}_0(\frac{j_{01}}{a} r) &= E_0 \mbox{J}_0(\frac{j_{01}}{a} r)\vspace{0.2cm} \\
H_{\varphi} &= \hspace{0.5cm} 2C\frac{j_{01}}{a}\mbox{J}_1(\frac{j_{01}}{a} r) &= {\displaystyle\frac{E_0}{Z_0}} \mbox{J}_1(\frac{j_{01}}{a} r).
\end{array}}
{field components of the TM$_{010}$ pillbox mode}{}
\noindent Figure \ref{fig:PFields} shows the field pattern of the TM$_{010}$ mode simulated by Superfish$^{\copyright}$.

\begin{figure}[h!]
	\begin{center}
	\includegraphics[width=2cm]{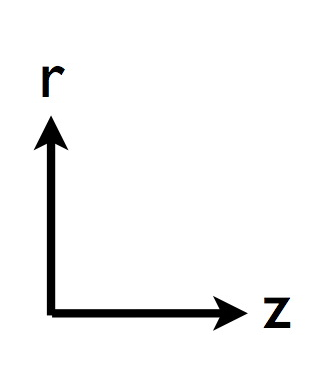}
	\includegraphics[height=4.5cm]{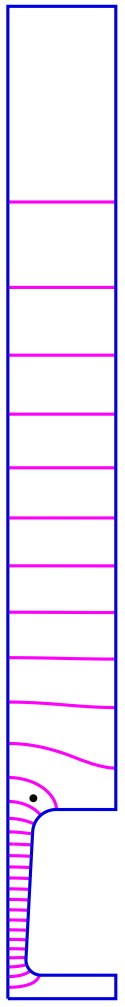}\hspace{1cm}
	\includegraphics[height=4.5cm]{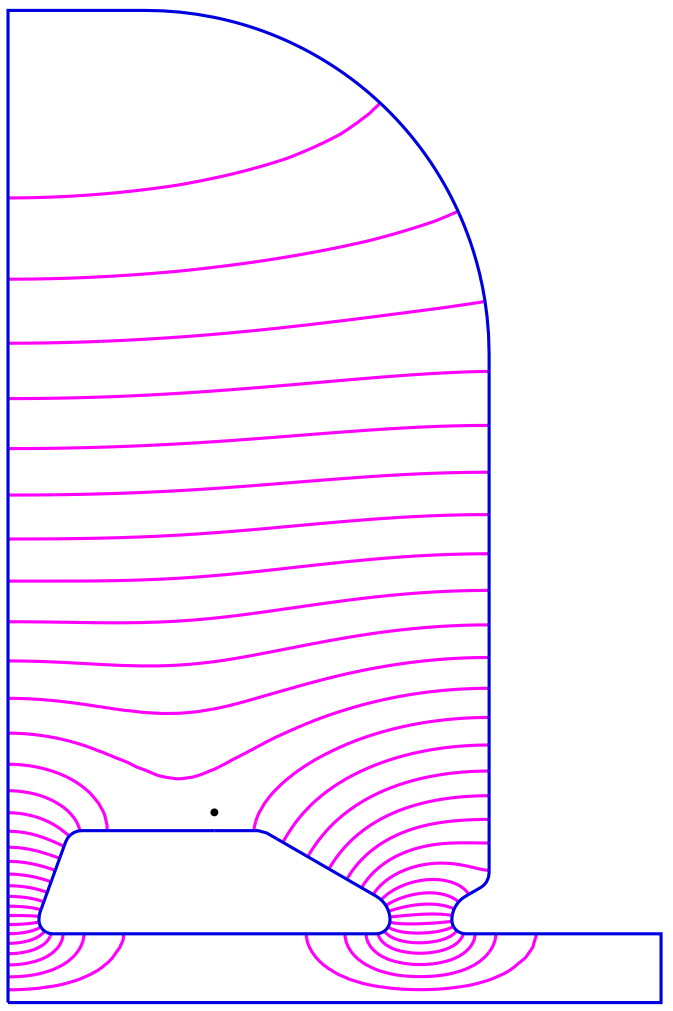}\hspace{1cm}
	\includegraphics[height=4.5cm]{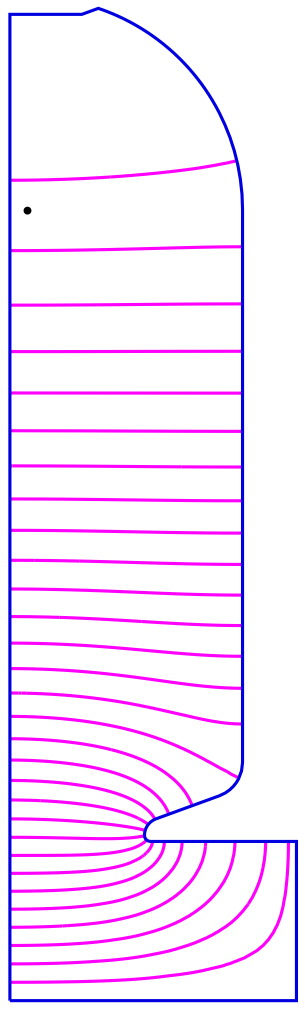}\hspace{1cm}
	\includegraphics[height=4.5cm]{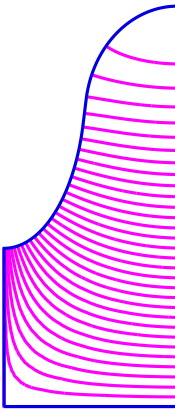}
	\caption{Typical TM mode cavities, from left to right: Drift Tube Linac (DTL), Cell-Coupled Drift Tube Linac (CCDTL), Cell-Coupled Linac (CCL), superconducting elliptical cavities.}
	\label{fig:TMtypical}
	\end{center}
	\end{figure}

\noindent Some typical examples of TM mode cavities are shown in Fig.~\ref{fig:TMtypical}.

\subsection{TE mode cavities}
\begin{figure}[h!]
	\begin{center}
	\includegraphics*[width=\textwidth]{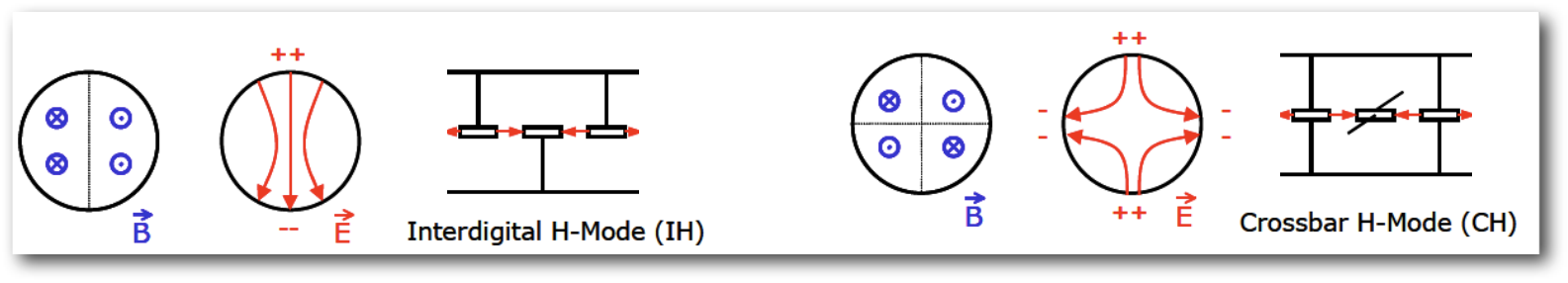}
	\caption{Commonly used TE mode cavities, left: TE$_{110}$ (interdigital TE mode), right: TE$_{210}$ (crossbar TE mode) \cite{Tiede08}}
	\label{fig:TE-modes}
	\end{center}
	\end{figure}

TE or H mode cavities are an interesting species, because by definition TE modes do not have a longitudinal electrical field component, which could be used for acceleration. On the other hand, these modes have much lower surface losses than TM modes because there is less magnetic field on the circumference of the cavity (close to the conducting walls). Less magnetic field means less surface current and hence lower losses on the inner surface of the cavity. To use the advantage of lower losses the electric field lines of the TE modes are `bent' in such a way that we get an axial electric field component, which can be used for acceleration. This is usually done by introducing drift tubes as shown in Fig.~\ref{fig:TE-modes}.

\begin{figure}[h!]
	\begin{center}
	\includegraphics[width=0.6\textwidth]{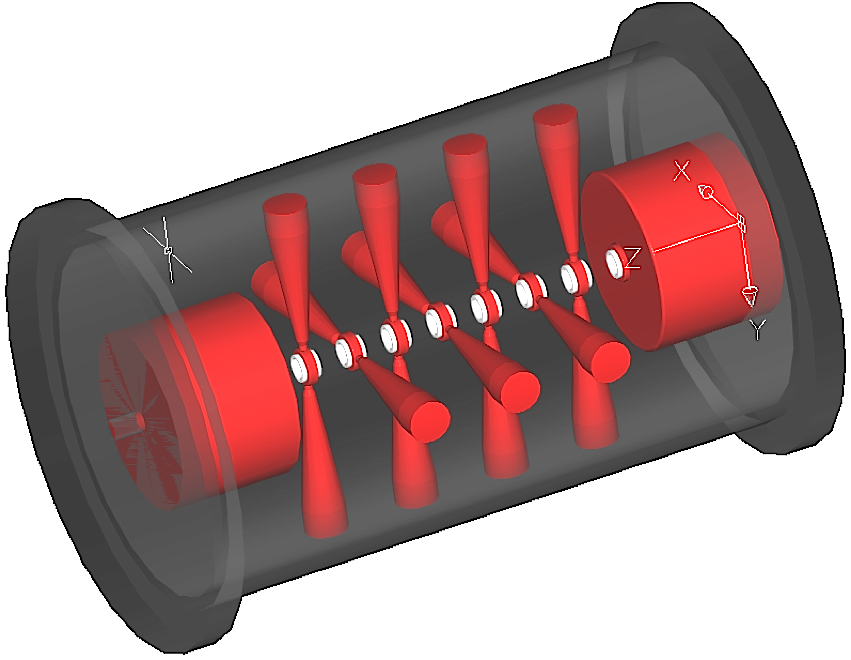}
	\caption{Crossbar H-mode cavity design example \cite{ClementeThesis}}
	\label{fig:H-mode-cavity}
	\end{center}
	\end{figure}

\begin{figure}[h!]
	\begin{center}
	\includegraphics*[width=0.85\textwidth]{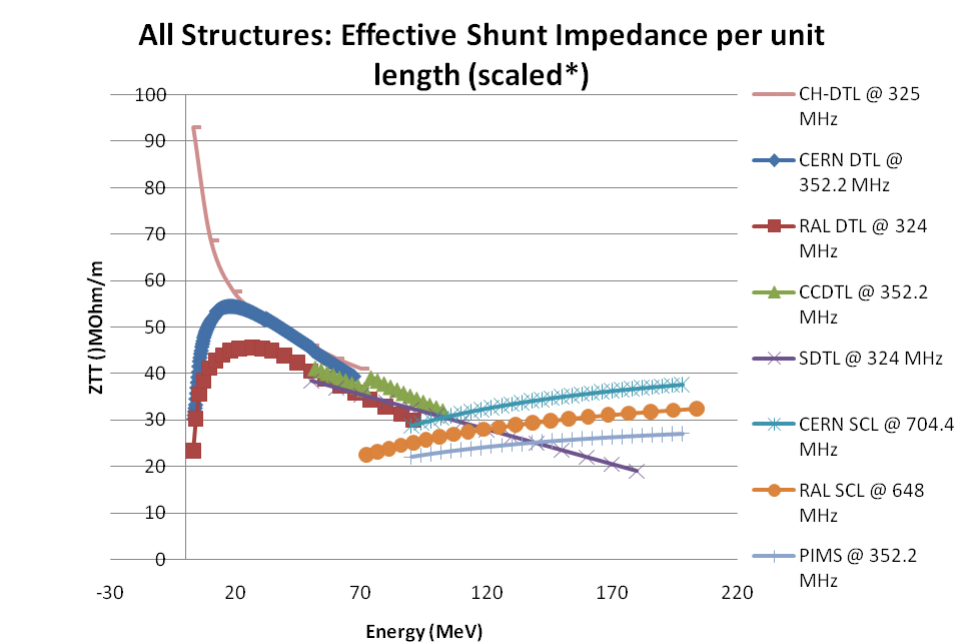}
	\caption{Shunt impedance comparison for various normal-conducting structures \cite{Ciprian08}}
	\label{fig:TEshunt}
	\end{center}
	\end{figure}

Strictly speaking these modes are no longer pure TE modes because we now have a longitudinal electric field on axis. The dominant field distribution, however, remains that of a TE mode, which means that one can take advantage of the low losses on the inner surface. The design of TE mode cavities is further complicated by the fact that a transverse electric field cannot exist parallel to the end walls of a~conducting cavity. By definition electric field lines can only have a normal orientation with respect to conducting surfaces, which means that the end-cells of TE mode cavities need a special design effort to allow for the existence of a TE mode within the cavity (see Fig.~\ref{fig:H-mode-cavity}). 
As a consequence TE mode cavities can only make use of their lower surface losses if they consist of many cells per cavity.  If we compare a typical shunt impedance curve for TE mode structures with traditional TM mode cavities (see Fig.~\ref{fig:TEshunt}), we can see that 
the advantage of lower shunt impedance only persists at low energy (for protons approximately below 20\,MeV--30\,MeV, \cite{Ciprian08}). Since at low energy space-charge forces usually enforce short transverse focusing periods, it is difficult to have a large number of accelerating cells per cavity. This problem was solved rather elegantly by the invention of the KONUS (KOmbinierte NUll grad Struktur, \cite{RatzingerHabil}) beam dynamics principle, which reduces the transverse RF defocusing and which made it possible to have longer transverse focusing periods. This principle was applied successfully to several low-duty-cycle heavy-ion accelerators, for instance Linac3 \cite{Linac3} and REX-ISOLDE \cite{Podlech99} at CERN. Up to now TE mode cavities in combination with the KONUS beam dynamics have not been used in high (average) beam power applications, where beam loss is a potentially performance limiting issue. At present several design proposals have been submitted for high-beam-power applications. 

\subsection{TEM mode cavities}
\begin{figure}[h!]
	\begin{center}
	\includegraphics[width=0.8\textwidth]{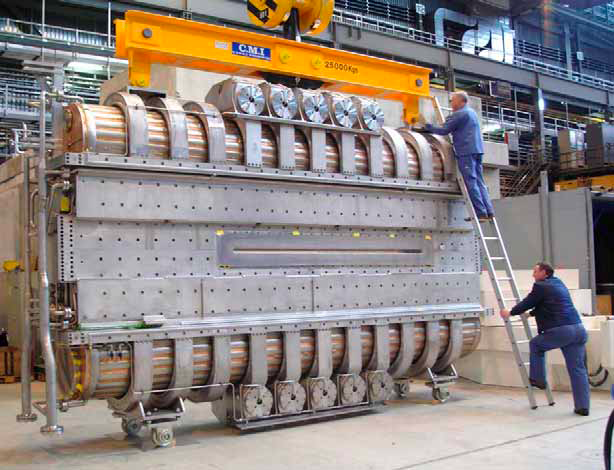}
	\caption{PSI cyclotron TM-mode cavity working at 50.6\,MHz \cite{HFitze}}
	\label{fig:PSI-cavity}
	\end{center}
	\end{figure}

The frequency of the accelerating modes used in TE and TM mode cavities is related to the transverse dimensions of these cavities and the capacitance between the drift tubes or nose cones. As a consequence, practically usable dimensions are related to a certain frequency range, which typically starts at around 100\,MHz and extends up to 10s of GHz for travelling wave structures. Lower frequencies yield prohibitively large cavities, and higher frequencies impose unrealistically tight tolerances on beam steering, alignment, and mechanical construction. An example of a very low frequency cavity is shown in~Fig.~\ref{fig:PSI-cavity}, which depicts a 50\,MHz cyclotron cavity used at PSI in Switzerland \cite{HFitze}. The cavity is 5.6\,m long, provides an accelerating voltage of 1\,MV and weighs 25\,tons.

\begin{figure}[h!]
	\begin{center}
	\includegraphics[width=0.35\textwidth]{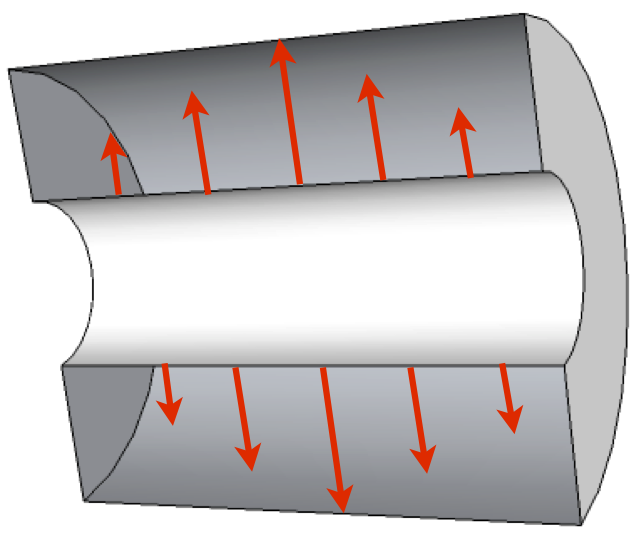}
	\includegraphics[width=0.48\textwidth]{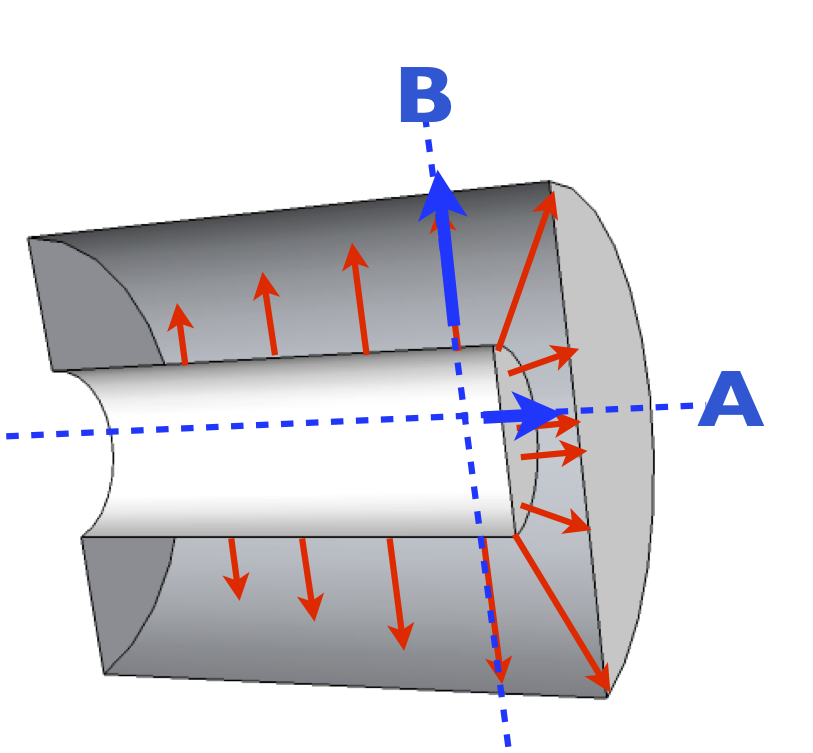}
	\caption{Left: coaxial 1/2 wave cavity, right: 1/4 wave cavity}
	\label{fig:TEM-mode}
	\end{center}
	\end{figure}

For certain applications, such as small- to medium-sized synchrotrons, one needs frequencies in the MHz range, and it is clear that TM or TE-mode cavities would become excessively large. Here one can make use of TEM cavities, were the RF frequency is no longer determined by the transverse dimensions of the cavity but rather by its length. A typical example is a coaxial cavity as shown in Fig.~\ref{fig:TEM-mode}, where the length of the cavity equals one half of a wave-length in longitudinal direction. 

Instead of using an electric wall as boundary condition on both sides of the~cavity one can also use an `open' boundary on one end of the~cavity. This has the benefits of i) shortening the length of the~cavity to a 1/4 wave-length, and of ii) bending the electric field lines into the direction of the cavity axis so that one can accelerate particles along the cavity axis (direction A in Fig.~\ref{fig:TEM-mode}). In addition to direction A one can also accelerate in direction B, which requires a tight synchronization between the particles and the~RF. Direction A is typically used in normal-conducting synchrotron cavities, while direction B is often used in normal and superconducting Quarter-Wave and Half-Wave Resonators (QWR and HWR) in low-frequency ion linacs. Quarter-wave cavities are of course no longer real TEM cavities, but since they originate from a TEM mode the classification seems somewhat justified. 

\begin{figure}[h!]
	\begin{center}
	\includegraphics[width=0.45\textwidth]{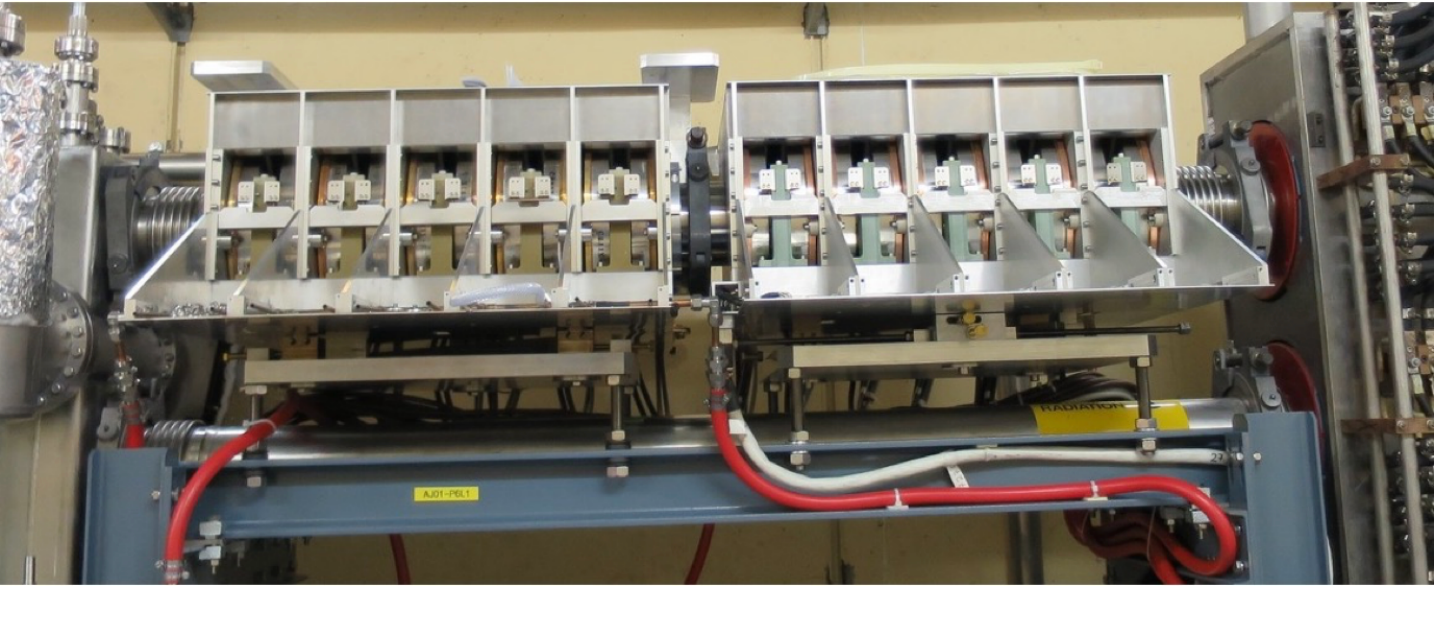}\hspace{1cm}
	\includegraphics[width=0.4\textwidth]{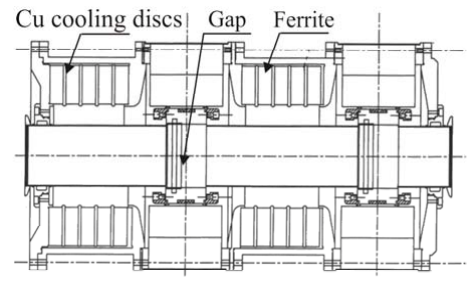}
	\caption{Left: a finemet 20\,MHz cavity in the CERN PSB, right: cross-section of a 13-20\,MHz cavity in the CERN PS \cite{Morvillo}.}
	\label{fig:filled-coax}
	\end{center}
	\end{figure}

In the case of synchrotron cavities one often fills part of the volume between inner and outer conductor with a dielectric or magnetic material as shown in Fig.~\ref{fig:filled-coax}. Both material types will shorten the cavity because of their dielectric/magnetic material constants. Magnetic materials like ferrites have the added advantage that the frequency of the cavity can be changed by the application of external fields. Because of the losses in the material the quality factor $Q$ becomes very low, which means that only a~small amount of energy is stored in the cavity enabling a fast frequency change. 

\begin{figure}[h!]
	\begin{center}
	\includegraphics[width=0.6\textwidth]{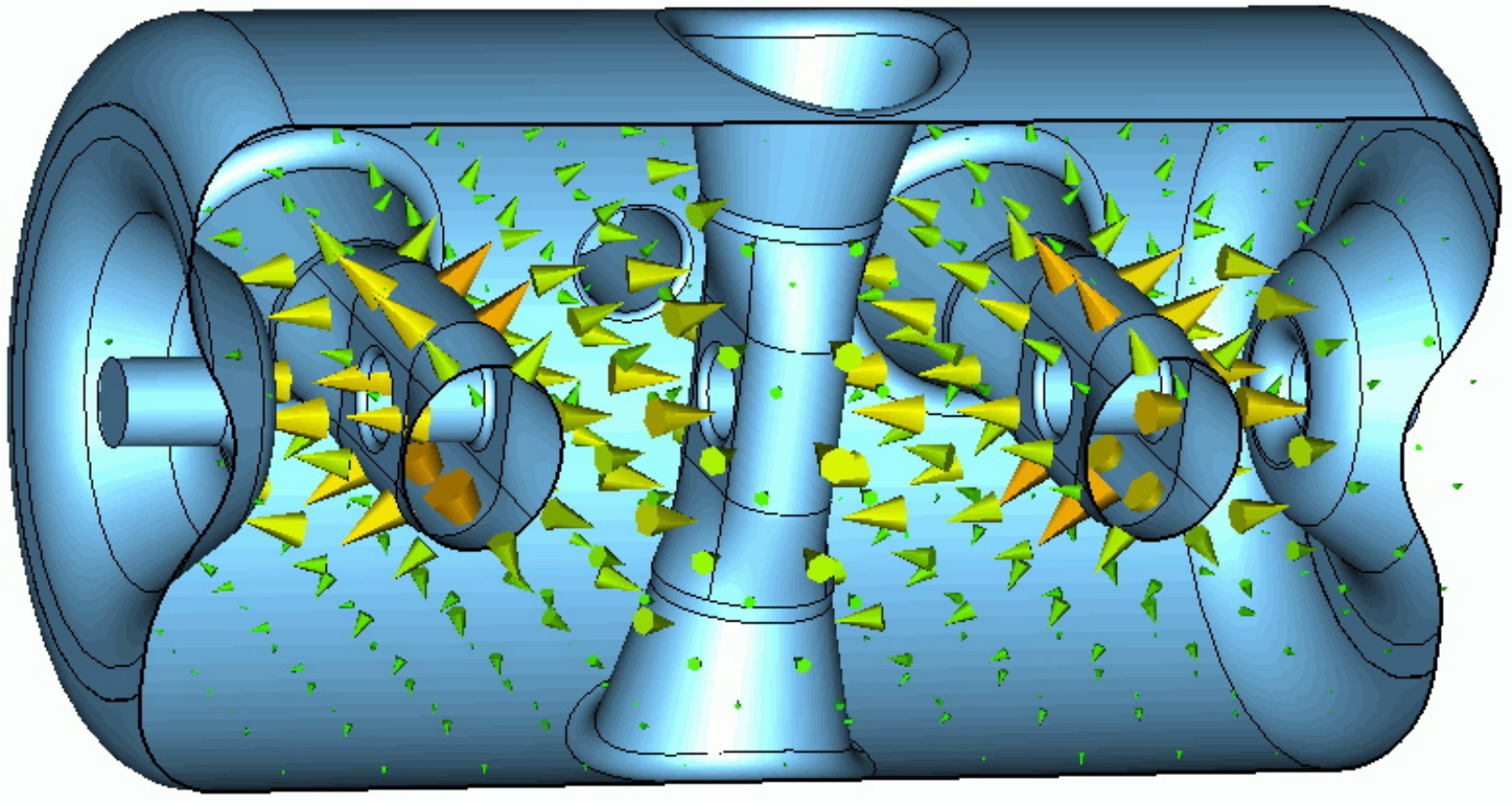}
	\caption{Triple-spoke cavity developed at FZJ \cite{Zaplatin}}
	\label{fig:spoke}
	\end{center}
	\end{figure}

An interesting variant of a TEM cavity, which is in fact difficult to identify as such, is called a~`spoke cavity' \cite{Delayen88}. It consists of $1\ldots n$ stacked half-wave cavities, with the inner conductor (the spoke) being rotated by $90^{\circ}$ from cell to cell. Figure~\ref{fig:spoke} shows an example of a triple spoke cavity developed at Forschungszentrum J\"{u}lich \cite{Zaplatin}.
At first sight spoke cavities are very similar to crossbar H-mode cavities as in Fig.~\ref{fig:H-mode-cavity}, and even the field distribution does not show striking differences. Nevertheless, the modes in both cavities are quite different: in the H-mode cavities the `bars' are quite thin in order not to disturb too much the H-mode, which is determined mainly by the diameter of the cavity. In spoke cavities the~field distribution is determined mainly by the spokes, which represent sections of short-circuited coaxial resonators, hence they are very `thick' in comparison to the H-mode cavities. Another difference is the~presence of the modified end-cells in the H-mode cavities, which are needed to make the existence of H-modes in cylindrical cavities possible. The `nose-cones', in the spoke cavity end-walls are introduced to optimize the transit time factor, but they are not needed to establish the desired field distribution. Even though many prototypes of spoke cavities have already been built, they have not yet been used in an actual accelerator. Several projects, however, have foreseen them in the low-energy parts of ion and proton accelerators, and we should soon see the first spoke cavities in operation. 

\subsection{Basic cavity parameters}
In order to characterize and optimize cavities, we need to define some commonly used figures of merit, which we will define in general terms and then apply to our simple pill-box cavity. In the following we assume that we are dealing with an axis-symmetric cavity resonating in the TM$_{010}$ mode. 

\subsubsection{Energy gain in a cavity}
For particles traversing a cavity on axis the electric field will generally have the following form\\
\begin{equation}
E_z (r=0,z,t) = E(0,z) \cos(\omega t + \varphi) \label{eq:Wgain}
\end{equation}
\noindent which we can use to calculate the energy gain of a particle when it traverses the cavity\\
\meqn{
\Delta W &= q\int\limits_{-L/2}^{L/2} E(0,z) \cos (\omega t + \varphi)\\
              &= q V_0 T \cos\varphi = qE_0 T L \cos\varphi
}{energy gain in a cavity (Panofsky equation)}{}

\noindent where the cavity voltage is given by\\
\meqn{
V_0 = \int\limits_{-L/2}^{L/2} E(0,z) dz = E_0
}{cavity voltage}{}

\noindent and where the ``difficult mathematics'' has been lumped into the so-called transit time factor\\
\meqn{
T = \frac{\int\limits_{-L/2}^{L/2} E(0,z) \cos(\omega t(z)) dz}{\int\limits_{-L/2}^{L/2} E(0,z) dz} -\underbrace{\tan\phi \frac{\int\limits_{-L/2}^{L/2} E(0,z) \sin(\omega t(z)) dz}{\int\limits_{-L/2}^{L/2} E(0,z) dz}}_{ =0 \mbox{ if } E(0,z) \mbox{ is symmetric to z=0}}
}{transit time \\factor}{}

which takes into account that the RF electric field changes during the passage of the particles. It gives the ratio between the energy gained in an RF field and a DC field and is therefore always $<1$. We note that the Panofsky equation takes into account the changing velocity of the particles when they cross the accelerating gap. This makes the integrals in the above equations difficult to solve. Assuming that the~velocity change of the beam particles during their passage is small, one can say that
\begin{equation}
\omega t \approx \omega \frac{z}{v} = \frac{2\pi z}{\beta \lambda}
\end{equation}
which changes the expression of the transit time factor to (assuming $E(0,z)$ being symmetric to $z=0$)\\
\meqn{T = \frac{\int\limits_{-L/2}^{L/2} E(0,z) \cos\left(\frac{2\pi z}{\beta\lambda}\right) dz}{\int\limits_{-L/2}^{L/2} E(0,z) dz}.
}{transit time factor for small velocity changes}{eq:Tsvc}

The accelerating voltage $V_{acc}$ is the voltage that the particle ``sees'' when crossing the cavity and should not be confused with the cavity voltage $V_0$. The thus define\\
\meqn{
V_{acc}=V_0T = E_0 L T.
}{accelerating voltage}{}

\subsubsection{Shunt impedance}
The shunt impedance tells us how much voltage a cavity will provide for a certain power, which is dissipated in the cavity walls. In cavity design this is one of the parameters to maximize, since a large shunt impedance reduces the power consumption of the RF cavities. The general definition is\\

\vspace*{-0.7cm}
\meqn{
R_s = \frac{V_0^2}{P_d}.
}{shunt impedance (linac definition)}{}

The benefit of a high shunt impedance can easily be diminished by having a small transit time factor, because the cavity voltage cannot be used efficiently to transfer energy to the beam. Therefore one usually tries to optimise the shunt impedance and the transit time factor, which explains the definition of the effective shunt impedance\\

\vspace*{-0.7cm}
\meqn{
R = \frac{(V_0 T)^2}{P_d}.
}{effective shunt impedance}{}

When comparing multi-cell structures operating at different frequencies one is less interested in the efficiency per cell (because cell sizes depend for instance on the chosen frequency) but rather in the~efficiency per unit length of the accelerating structure. For this reason one defines\\

\vspace*{-0.7cm}
\meqn{
Z = \frac{R_s}{L} = \frac{E_0^2}{P_d/L} 
}{shunt impedance per unit length}{}

\noindent and\\

\vspace*{-0.7cm}
\meqn{
ZT^2 = \frac{R}{L} = \frac{(E_0 T)^2}{P_d/L}.
}{effective shunt impedance per unit length}{}

\subsubsection{``Linac'' and ``circuit'' definition of shunt impedance}
It turns out that different communities of accelerator experts use different definitions for the shunt impedance. Linac experts usually use 
the definitions as introduced above, while people dealing with circular machines generally use a definition that is derived from the lumped 
circuit definition of a~resonator (see Section~\ref{sec:lumped}). There all shunt impedances are exactly half as large, following:\\

\vspace*{-0.7cm}
\meqn{
R_S^c = \frac{V_0^2}{2 P_d}
}{shunt impedance (circuit definition)}{}

So before you discuss shunt impedances with someone make sure that you use the same definition. In order to mark the difference clearly we use $R_s^c$ in this text to identify when the circuit is used.

\subsubsection{3db bandwidth and quality factor}
The quality factor $Q$ describes the bandwidth of a resonator and is defined as the ratio of reactive power (stored energy) to real power, which is lost in the cavity walls:\\

\vspace*{-0.7cm}
\meqn{
Q=\frac{\omega}{\Delta\omega}=\frac{\omega W}{P_d}
}{quality factor}{eq:qfactor}

\begin{figure}[h!]
\centering
\includegraphics[width=0.6\textwidth]{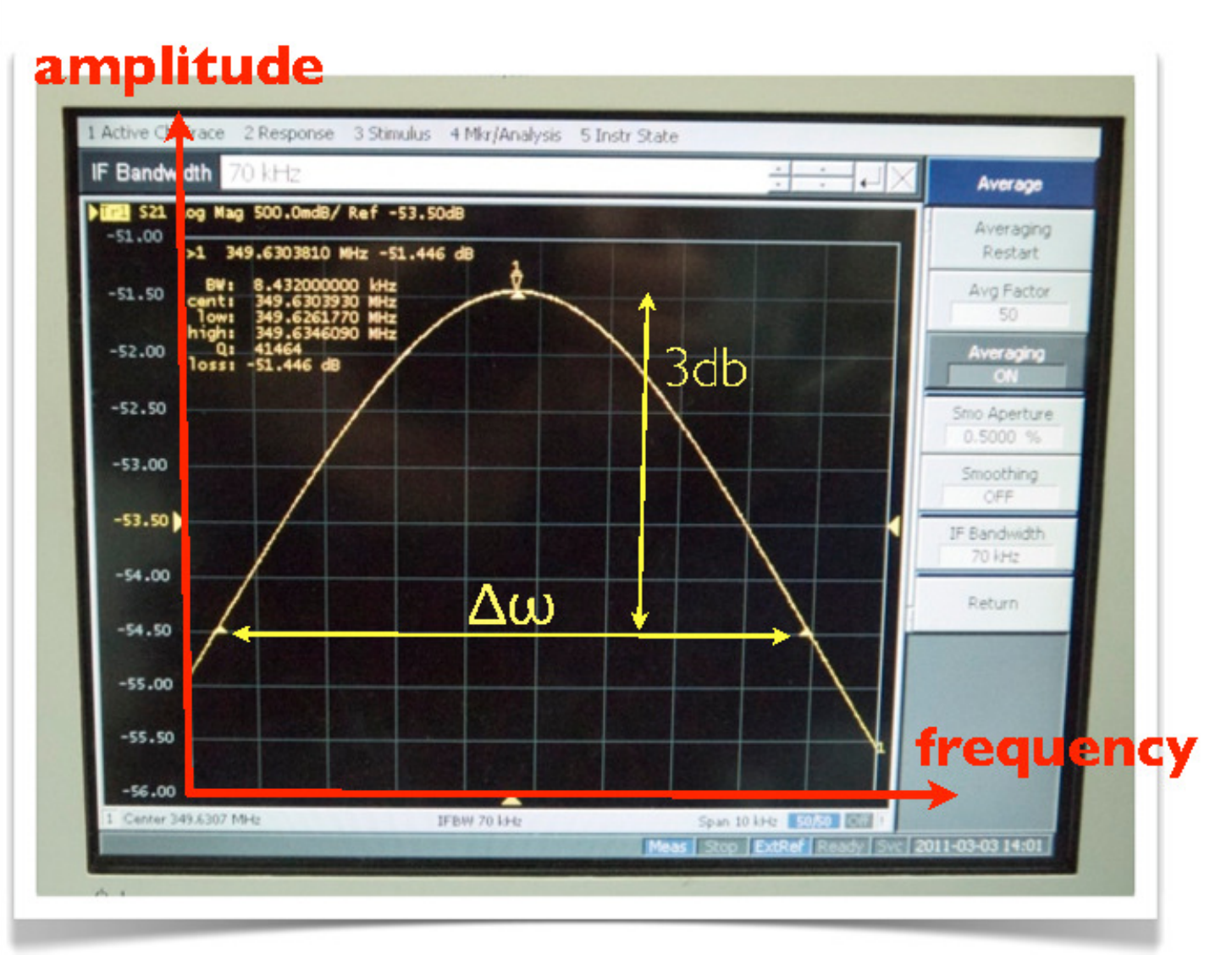}
\caption{Measurement of frequency, 3-db bandwidth, and Q-factor with a network analyser}
\label{fig:bandwidth}
\end{figure}

If a resonator were built with ideal electric walls (zero electric resistance), the resonance curve would be a delta function at the resonance frequency. So the bandwidth ($\Delta\omega$) would be zero and the~quality factor would be infinite. In reality even superconducting cavities have a certain surface resistance, which is why all our cavities have a certain bandwidth and a finite quality factor. Figure~\ref{fig:bandwidth} shows a~typical resonance curve measured with a network analyser. For this measurement two antennas penetrate the~cavity. The first one sends an RF signal with a certain frequency sweep, and the second antenna picks up the field level in the cavity. As a result we get the field level versus frequency. The bandwidth is defined as the frequency width of the resonance curve, when measuring the distance between the points where the field level has dropped by 50\% (or -3\,db) as shown in Fig.~\ref{fig:bandwidth}.

Together with the shunt impedance one defines another figure of merit $(R/Q)$, which is used to maximize the energy gain in a cavity for a certain stored energy.\\

\vspace*{-0.7cm}
 \meqn{
 \left(\frac{R}{Q} \right) = \frac{(V_0 T)^2}{\omega W}
 }{$\mathbf{(R/Q)}$}{}
 
$(R/Q)$ is independent of the surface losses of a cavity and can therefore be used to qualify the~geometry of an accelerating cavity.  

\subsubsection{Filling time of a cavity}
This section is a short extract of \cite{bib:RFCASFrank}, which can be consulted for more details. The dissipated power in a cavity must be
equal to the rate of change of the stored energy:\\

\vspace*{-0.7cm}
\meqn{
P_d = -\frac{dW}{dt} = \frac{\omega_0 W}{Q_0}
.}{}{}

\noindent A solution of above equation can be written as:\\

\vspace*{-0.7cm}
\meqn{
W(t) = W_0 e^{-\frac{2t}{\tau}}
}{}{}

\noindent which describes an exponential decay of the stored energy with the ``filling time constant'' $\tau$\\

\vspace*{-0.7cm}
\meqn{
\tau = \frac{2Q_0}{\omega_0}.
}{filling time constant}{}

In case the cavity is equipped with a power coupler, we have to consider the ``loaded $Q$'' (which will be derived later), and the filling time constant changes to\\

\vspace*{-0.7cm}
\meqn{
\tau_l = \frac{2Q_l}{\omega_0}. 
}{filling time constant for a loaded cavity}{}

In the above definition the electric fields decay exponentially with $1/\tau$, while the stored energy decays with $2/\tau$. Please be aware that you can often finds textbook definitions of the filling time constant where the stored energy decays with $1/\tau$. 

\subsection{Basic cavity parameters of a pillbox cavity}
As a small exercise we calculate in this section the cavity parameters, which were defined in the previous section, for a pillbox cavity with length $L$ and radius $a$. Since the TM$_{010}$-mode has no $z$-dependence we can simplify the expression for the transit time factor \eqref{eq:Tsvc} to\\
\meqn{
T = \frac{\int\limits_{-L/2}^{L/2} E(0,z) \cos\left(\frac{2\pi z}{\beta\lambda}\right) dz}{\int\limits_{-L/2}^{L/2} E(0,z) dz} = \frac{\sin\left(\frac{\pi L}{\beta\lambda}\right)}{\frac{\pi L}{\beta\lambda}}.
}{transit time factor in a pillbox for small velocity changes}{}

In the case of relativistic particles ($\beta\approx 1$) and a cavity length of $L=\lambda/2$, which is often chosen because then the cavity can be cascaded to a multi-cell structure, we get:\\

\vspace*{-0.7cm}
\meqn{
T = \frac{2}{\pi} = 0.64.
}{transit time factor in a pillbox for relativistic particles}{}

In real cavities one usually tries to increase the transit time factor by shortening the accelerating gap. This can be done by introducing nose cones on the cavity walls as shown for instance in Fig.~\ref{fig:lumped}.

To calculate the quality factor of our pillbox cavity we use again the power-loss method. To evaluate Eq.~\eqref{eq:qfactor} we need the stored energy and the power lost in the cavity walls. For the stored energy we get
\begin{align}
W &=W_{el}+W_{mag} = 2 W_{el} = 2 \int\limits_V \frac{1}{4}\mathbf{E}\cdot \mathbf{D}^* dV\\
\intertext{with}
E_z &= E_0 J_0 \left(\frac{j_{01}r}{a}\right)\\
\intertext{one gets}
W &= \frac{\varepsilon_0}{2} \int\limits_0^a \int\limits_0^{2\pi} \int\limits_{-L/2}^{L/2} E_0^2J_0^2\left(\frac{j_{01}r}{a}\right) rdr d\varphi dz= \frac{1}{2} E_0^2 \varepsilon_0\pi L a^2 J_1^2(j_{01}).
\intertext{To calculate the dissipated power we integrate Eq.~\eqref{eq:powden} over a volume which consists of the inner surface of our pillbox times the skin depth:}
P_d &= \frac{\delta_s}{2\kappa} \int\limits_{-L/2}^{L/2} \underbrace{J_z J_z^*}_{\frac{1}{\delta_s^2}H_{\varphi}^2(r=a,z)} 2\pi a dz + \frac{\delta_s}{\kappa} \int\limits_0^a \underbrace{J_r J_r^*}_{\frac{1}{\delta_s^2}H_{\varphi}^2(r,z=0)} 2\pi r dr \\
&=\frac{E_0^2 \pi R_{surf} a}{Z_0^2} J_1^2(j_{01}) (a+L) \label{eq:PD}\\
\intertext{where we made use of}
H_{\varphi} &= \frac{E_0}{Z_0}J_1\left(\frac{j_{01}r}{a} \right).\\
\intertext{Putting everything together we get}
Q_0 &= \frac{\omega W}{P_d} = \frac{Z_0^2 \omega}{2R_{surf}} \frac{La}{L+a} = \frac{1}{\delta_s}\frac{La}{L+a} \propto\sqrt{\omega}. \label{eq:Qpillbox}
\end{align}

As we can see the quality factor is a function of the material constants ($\kappa$ and $\mu$ within $\rho_s$), the~frequency and the geometry of the cavity. We also note that for the same cavity shape, the quality factor increases with the frequency according to $\propto \sqrt{\omega}$.

The accelerating voltage in a pillbox cavity is given by\\

\begin{figure}[h!]
\centering
\includegraphics[width=0.7\textwidth]{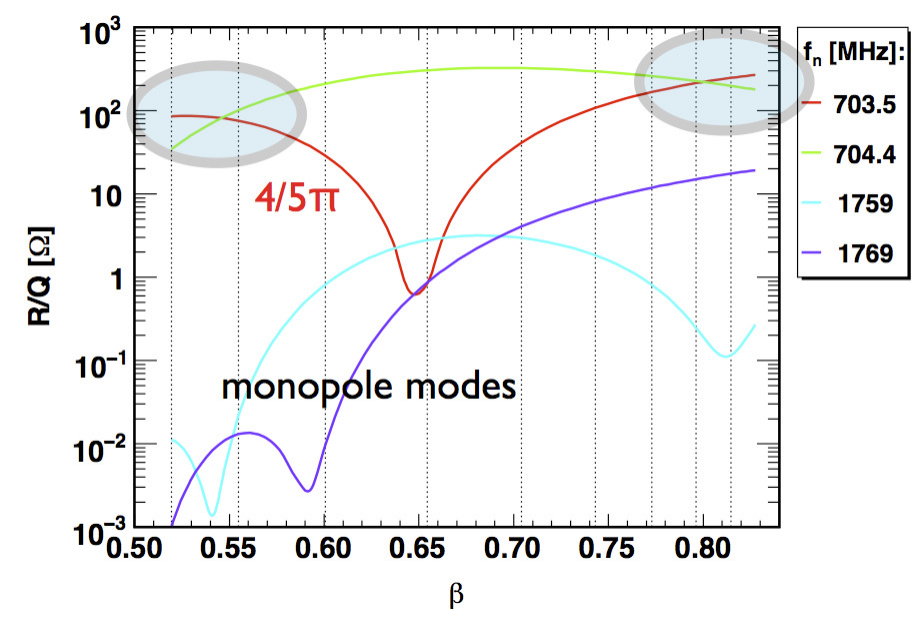}
\caption{(R/Q) dependence on particle velocity for a five-cell SC cavity with a geometric beta of $\beta=0.65$ and a~frequency of 704.4\,MHz: upper curve - $\pi$-mode (see also \cite{bib:Schuh}).}
\label{fig:rq-beta}
\end{figure}

\vspace*{-0.7cm}
\meqn{
V_{acc} = V_0 T = E_0 L T = E_0 L \frac{\sin\left(\frac{\pi L}{\beta\lambda}\right)}{\frac{\pi L}{\beta\lambda}}
}{accelerating voltage in pillbox}{eq:V0T}
and is obviously a strong function of the transit time factor. It therefore depends on the gap length ($L$) and the speed of the particle ($\beta$). Due to their high development costs superconducting cavities are often used over large velocity ranges without changing their cell length and this results in a velocity dependent acceleration efficiency. In Figure~\ref{fig:rq-beta} one can see $(R/Q) \propto (V_0T)^2$ as a function of particle velocity for a~5-cell superconducting cavity, whose geometric cell length corresponds to a particle speed of $\beta=0.65$.
If the velocity range over which the cavity is used is chose too large, one may find areas where the passband mode, which is closest to the $\pi$-mode (here the $4/5\pi$-mode), has a higher acceleration efficiency as the~accelerating mode. In Figure~\ref{fig:rq-beta} these areas are highlighted and should be avoided when designing a~linac.  
Furthermore one should be aware that also the $(R/Q)$ of the HOMs is highly dependent on the~particle velocity. 

Using the expressions for the accelerating voltage $V_0T$ (Eq.~\eqref{eq:V0T}) and the dissipated power $P_d$ (Eq.~\eqref{eq:PD}) we also get an analytical expression for the effective shunt impedance\\
\meqn{
R= \frac{(V_0 T)^2}{P_d} = \frac{Z_0}{\pi R_{surf} J_1^2(j_{01})} \frac{\sin\left( \frac{\pi L}{\beta\lambda} \right)}{\frac{\pi L}{\beta\lambda}} \frac{L^2}{a(a+L)}.
}{effective shunt impedance of a pillbox}{eq:Rpillbox}

Finally we calculate the frequency and $(R/Q)$ using Eq.~\eqref{eq:f-pillbox}, Eq~\eqref{eq:Rpillbox}, and Eq.~\eqref{eq:Qpillbox}.\\

\vspace*{-0.7cm}
\meqn{
f_{010}^{TM} = \frac{2.405 c}{2\pi a}
}{pillbox frequency}{}
\meqn{
\left( \frac{R}{Q} \right) = \frac{2c}{\omega\pi J_1^2(j_{01})}\frac{\sin\left( \frac{\pi L}{\beta\lambda} \right)}{\frac{\pi L}{\beta\lambda}} \frac{L}{a^2}
}{pillbox $\mathbf{(R/Q)}$}{}

As stated before, $(R/Q)$ is indeed independent of any material parameters. However, it does depend on the geometry of the cavity and the transit time factor. 

\subsection{A cavity as a lumped circuit}
\label{sec:lumped}
In the field of RF technology it is common practice to describe the behaviour of cavities, RF transmission lines, and couplers with equivalent lumped circuits. In this lecture we will only introduce the treatment of cavity and coupler so that one can understand how to get power into a cavity. The transmission of RF power and the associated RF transmission line theory can be found in many RF or microwave engineering text books.

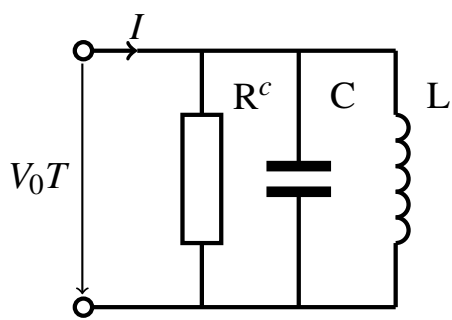
\begin{figure}[h!]
\centering
\begin{tikzpicture}[ultra thick,scale=0.85]

\draw[o-] (-1,0) -- (4,0);
\draw[o->] (-1,4) -- (0,4) node [above] {\Large $I$}; 
\draw (0,4) -- (4,4);
\draw[->, thick] (-0.85,3.8) -- (-0.85,0.2) node [left, midway] {\Large $V_0T$}; 

\draw (1,0) -- (1,1);
\draw (0.7,1) rectangle (1.3,3);
\draw (1,3) -- (1,4);

\draw (2.5,0) -- (2.5,1.8);
\draw[line width =4 pt] (2.0,1.8) -- (3,1.8);
\draw[line width =4 pt] (2.0,2.2) -- (3,2.2);
\draw (2.5,2.2) -- (2.5,4);
\draw (4,0) -- (4,1);
\draw (4,3) -- (4,4);
\foreach \x in {1,1.4,1.8,2.2,2.6}
	\draw (4,\x) arc (-90:90:0.2);

\draw (1.3,3.3) node [above, right] {\Large R$^c$};
\draw (2.8,3.3) node [above, right] {\Large C};
\draw (4.3,3.3) node [above, right] {\Large L};

\end{tikzpicture}
\caption{Lumped circuit equivalent of a resonant cavity}
\label{fig:LCR}
\end{figure}

We start with the description of a cavity with a parallel LCR-circuit as depicted in Fig.~\ref{fig:LCR}.

You may remember that one calculates the admittance of parallel circuits by adding up the~admittances of the single elements, which means that we can write the cavity impedance as\\

\vspace*{-0.7cm}
\meqn{
Z^c = \frac{1}{\displaystyle i\omega C + \frac{1}{i\omega L} + \frac{1}{R^c}}.
}{lumped circuit cavity impedance}{}

At resonance ($\omega = \omega_0$) the imaginary parts cancel each other and the cavity impedance becomes real, which means that\\

\vspace*{-0.7cm}
\meqn{
X= \omega_0 L = \frac{1}{\omega_0 C} = \sqrt{\frac{L}{C}} 
}{lumped circuit at resonance}{eq:lcX}
\noindent and that the resonance frequency is given by\\

\vspace*{-0.7cm}
\meqn{
\omega_0 = \frac{1}{\sqrt{LC}}
}{lumped circuit resonance frequency}{}
\noindent and that the power lost in the resonator is given by\\

\vspace*{-0.7cm}
\meqn{
P_d= \frac{1}{2} \frac{(V_0T)^2}{R^c}\mbox{.}
}{lumped circuit dissipated power}{}

\noindent The stored energy can be written as\\

\vspace*{-0.7cm}
\meqn{
W= \frac{1}{2} C (V_0T)^2 = \frac{1}{2} \frac{(V_0T)^2}{\omega_0^2 L}
}{lumped circuit stored energy}{eq:lcse}

\noindent and with that we get an expression for the quality factor\\

\vspace*{-0.7cm}
\meqn{
Q_0 = \omega_0 \frac{W}{P_d} = \omega_0 C R^c = \frac{R^c}{\omega_0 L}.
}{lumped circuit quality factor}{}

Our goal is to relate the lumped elements to the cavity characteristics and for this purpose we multiply Eq.~\eqref{eq:lcse} with $\omega$ and together with Eq.~\eqref{eq:lcX} we get
\begin{equation}
\frac{1}{\omega_0 C} = \sqrt{\frac{L}{C}} = \frac{(V_0 T)^2}{2\omega_0 W} = \left( \frac{R^c}{Q} \right) = \frac{1}{2} \left( \frac{R}{Q} \right)
\end{equation}
from which we can understand the difference between ``circuit Ohm'' and ``linac Ohm''  and which also provides a lumped circuit description of a cavity as summarized in Table~\ref{tab:lumpedRCL}.
\begin{table}[h!]
\caption{Lumped circuit elements of a cavity}
\centering
\begin{tabular}{lc}
\hline\hline
{\bf lumped circuit} & {\bf field description} \\
\hline
\parbox[0pt][1.2cm][c]{1cm}{$R^c$} & \parbox[0pt][1.2cm][c]{3cm}{$\displaystyle\frac{1}{2}R$} \\
\parbox[0pt][1.2cm][c]{1cm}{$C$} & \parbox[0pt][1.2cm][c]{3cm}{$\displaystyle\frac{2}{\omega_0 (R/Q)}$}  \\
\parbox[0pt][1.2cm][c]{1cm}{$L$} & \parbox[0pt][1.2cm][c]{3cm}{$\displaystyle \frac{1}{2\omega_0} \left(\frac{R}{Q} \right)$} \\
\hline\hline
\label{tab:lumpedRCL}
\end{tabular}
\end{table}

As we can see three quantities are sufficient to describe a resonator. Instead of using $R$, $L$, $C$ one can also use the parameters $\omega_0$, $Q_0$, and $(R/Q)$ to completely characterise an RF cavity as in Table~\ref{tab:3quan}.
\begin{table}[h!]
\caption{Three characteristic quantities of a cavity}
\centering
\begin{tabular}{lc}
\hline\hline
{\bf lumped circuit} & {\bf field description} \\
\hline
\parbox[0pt][1.4cm][c]{4cm}{$\displaystyle\omega_0 = \frac{1}{\sqrt{LC}}$} & \parbox[0pt][1.4cm][c]{3cm}{$\displaystyle\frac{2.405 c}{a}$ (pillbox)} \\
\parbox[0pt][1.4cm][c]{4cm}{$\displaystyle Q_0 = \omega_0 C R^c = \frac{R^c}{\omega_0 L}$} & \parbox[0pt][1.4cm][c]{3cm}{$\displaystyle Q_0 = \frac{\omega_0 W}{P_d}$}  \\
\parbox[0pt][1.4cm][c]{4cm}{$\displaystyle\left( \frac{R^c}{Q}\right) = \sqrt{\frac{L}{C}} = \frac{1}{2} \left(\frac{R}{Q}\right)$} & \parbox[0pt][1.4cm][c]{3cm}{$\displaystyle \left(\frac{R}{Q} \right) = \frac{(V_0T)^2}{\omega_0 W}$} \\
\hline\hline
\label{tab:3quan}
\end{tabular}
\end{table}

\subsection{Getting power into a cavity: couplers}
\begin{figure}[h!] 
\centering
	\begin{tikzpicture}[scale=1, ultra thick]
	\draw (-0.5,0) arc (90:100:9);
	\draw (0.5,0) arc (90:80:9);
	\draw (-0.5,-0.03) -- (-0.5,0.7);
	\draw (0.5,-0.03) -- (0.5,0.7);
	\draw[line width = 3pt] (0.47,0.7) -- (0.85,0.7);
	\draw[line width = 3pt] (-0.47,0.7) -- (-0.85,0.7);
	\draw[line width = 3pt, blue!50] (-0.47,0.78) -- (-0.85,0.78);
	\draw[line width = 3pt, blue!50] (0.47,0.78) -- (0.85,0.78);
	\draw[blue!50] (0.5, 0.78) -- (0.5, 1.5);
	\draw[blue!50] (-0.5, 0.78) -- (-0.5, 1.5);
	\draw[blue!50,fill=blue!50] (-0.1,-0.5) rectangle (0.1,1.5);
	\draw (4.5,0) arc (90:100:9);
	\draw (5.5,0) arc (90:80:9);
	\draw (4.5,-0.03) -- (4.5,0.7);
	\draw (5.5,-0.03) -- (5.5,0.7);
	\draw[line width = 3pt] (5.47,0.7) -- (5.85,0.7);
	\draw[line width = 3pt] (4.53,0.7) -- (4.15,0.7);
	\draw[line width = 3pt, blue!50] (4.53,0.78) -- (4.15,0.78);
	\draw[line width = 3pt, blue!50] (5.47,0.78) -- (5.85,0.78);
	\draw[blue!50] (5.5, 0.78) -- (5.5, 1.5);
	\draw[blue!50] (4.5, 0.78) -- (4.5, 1.5);
	\draw[line width = 7pt,blue!50] (5,-0.01) -- (5,1.5);
	\draw[line width = 7pt,blue!50] (5,0) arc (0:-178:0.5);
	\end{tikzpicture} 
\caption{Example of an antenna-type coupler (left) and loop-type coupler (right)}
\label{fig:antennaloop}
\end{figure}
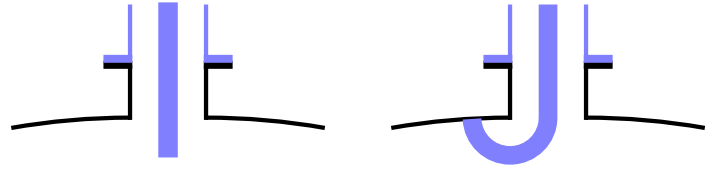

In this section we will extend the circuit model to include the power coupler and we will also extend our basic equations to describe the process of coupling power into a cavity. 
There are two basic types of couplers, which are used in standing wave cavities:
\begin{itemize}
\item antenna/loop couplers: here the coupler is usually some kind of coaxial line with the outer conductor connecting to the cavity wall and the inner conductor either penetrating into the cavity volume, or being connected in a loop to the inner surface of the cavity (see Fig.~\ref{fig:antennaloop});
\item iris couplers: here the fields in a waveguide couple to the cavity fields via an opening that connects the waveguide with the cavity.  
\end{itemize}

When designing a coupler one has to keep in mind the principle of reciprocity: the coupler has to produce a field pattern in the area of the coupling port that is very similar to the field pattern of the mode, which shall be excited in the cavity. Looking at Fig.~\ref{fig:antennaloop} one can imagine that the antenna type coupler would be very effective on the end walls of our pillbox, where it would couple electrically to the axial electric field lines. On the cylindrical surface of the pillbox the loop coupler would be a better choice with the loop oriented such that the azimuthal magnetic field penetrates the loop. 

\begin{figure}[h!]
\centering
\includegraphics[width=0.5\textwidth]{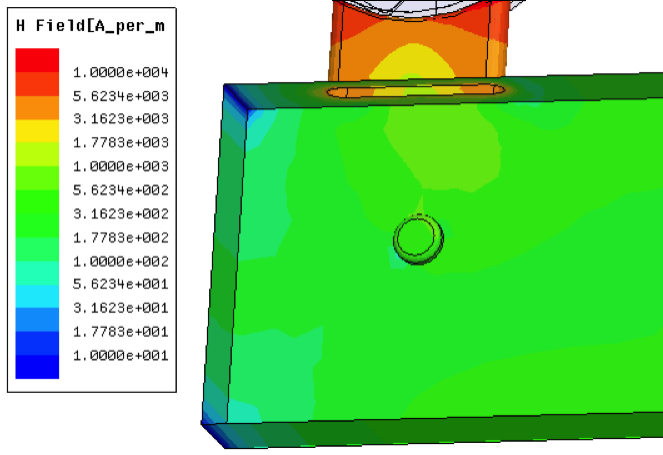}
\caption{Waveguide coupler connected to a Linac4 cavity.}
\label{fig:TaCo}
\end{figure}

Figure~\ref{fig:TaCo} shows an example of a ``Tuner Adjustable (waveguide) Coupler'' (TaCo) \cite{bib:TaCo} as it is used for the Linac4 \cite{bib:Linac4} cavities at CERN. In this case a short circuited rectangular waveguide is coupled to a standing wave cavity via a racetrack-shaped coupling iris. The coupling factor (more on this later) is here a function of: the position of the short circuit (left side), the height of the racetrack-shaped coupling channel between cavity and waveguide (on top), the size of the coupling slot, and the position of a stub tuner, which is used to fine-tune the coupling. 

\begin{figure}[h!]
\centering
\begin{tikzpicture}[ultra thick,scale=0.85]

\draw[-] (-1,0) -- (4,0);
\draw[->] (-1,4) -- (0,4) node [above] {\Large $I$}; 
\draw (0,4) -- (4,4);
\draw[->, thick] (0.5,3.8) -- (0.5,0.2) node [left, midway] {\Large $V_0T$}; 

\draw (-1,0) -- (-1,1);
\draw (-1,3) -- (-1,4);
\foreach \x in {1,1.4,1.8,2.2,2.6}
	\draw (-1,\x) arc (270:90:0.2);
\draw (-1.7,0) -- (-1.7,1);
\draw (-1.7,3) -- (-1.7,4);
\foreach \x in {1,1.4,1.8,2.2,2.6}
		\draw (-1.7,\x) arc (-90:90:0.2);
\draw[-o] (-1.7,0) -- (-2.7,0);
\draw[-o] (-1.7,4) -- (-2.7,4);
\draw (-1.35,4) node [above] {\LARGE 1:n};
\draw[->, thick] (-2.8,3.8) -- (-2.8,0.2) node [left, midway] {\Large $V_{gen}$};

\draw[dashed] (-2.7,0) -- (-5.5,0);
\draw[dashed] (-2.7,4) -- (-5.5,4);

\draw[o-] (-5.5,0) -- (-8.5,0);
\draw[o-] (-5.5,4) -- (-7.75,4) node [above] {\Large $I_{gen}$};
\draw[->] (-8.5,4) -- (-7.75,4);
\draw (-7,0) -- (-7,1);
\draw (-7,4) -- (-7,3);
\draw (-7.3,1) rectangle (-6.7,3);
\draw (-8.5,0) -- (-8.5,1.5);
\draw (-8.5,2.5) -- (-8.5,4);
\draw (-8.5,2) circle (0.5) node {RF};

\draw (1,0) -- (1,1);
\draw (0.7,1) rectangle (1.3,3);
\draw (1,3) -- (1,4);

\draw (2.5,0) -- (2.5,1.8);
\draw[line width =4 pt] (2.0,1.8) -- (3,1.8);
\draw[line width =4 pt] (2.0,2.2) -- (3,2.2);
\draw (2.5,2.2) -- (2.5,4);

\draw (4,0) -- (4,1);
\draw (4,3) -- (4,4);
\foreach \x in {1,1.4,1.8,2.2,2.6}
	\draw (4,\x) arc (-90:90:0.2);

\draw (1.3,3.3) node [above, right] {\Large $R^c$};
\draw (2.8,3.3) node [above, right] {\Large $C$};
\draw (4.3,3.3) node [above, right] {\Large $L$};
\draw (-6.7,3.3) node [above, right] {\Large $Z_0$};
\draw (-4.5,3.3) node [above, right] {\Large $Z_0$};

\draw[dashed, red] (-9.5,-1) -- (-9.5,5);
\draw[dashed, red] (-5.65,-1) -- (-5.65,5);
\draw[thick, red,<->] (-9.5,-0.8) -- (-5.65,-0.8) node [below, midway] {matched generator};
\draw[dashed, red] (-2.55,-1) -- (-2.55,5);
\draw[thick, red,<->] (-5.65,-0.8) -- (-2.55,-0.8) node [below, midway] {wave-guide};
\draw[dashed, red] (-0.7,-1) -- (-0.7,5);
\draw[thick, red,<->] (-2.55,-0.8) -- (-0.7,-0.8) node [below, midway] {coupler};
\draw[dashed, red] (5.2,-1) -- (5.2,5);
\draw[thick, red,<->] (-0.7,-0.8) -- (5.2,-0.8) node [below, midway] {cavity};
\draw[line width = 3pt, red,->] (-3.55,5.5) -- (-2.55,5.5) node [above, midway] {\Large $Z'_c$};
\draw[line width = 3pt, red,->] (-1.7,5.5) -- (-0.7,5.5) node [above, midway] {\Large $Z_c$};
\end{tikzpicture}
\caption{Lumped element circuit for RF power source, waveguide, power coupler, and cavity.}
\label{fig:pclumped}
\end{figure}
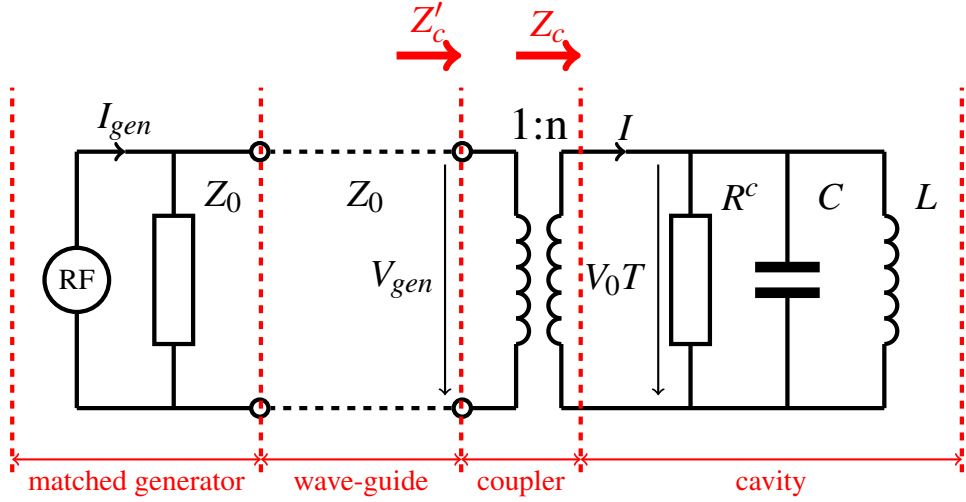

In the ideal case the power coupler is matched to the (beam-) loaded cavity, which means that there is no reflected power returning from the cavity towards the RF power source.  ``Matched'' means here that the coupler acts like an ideal transformer, which transforms the impedance $Z_c$ of the cavity into the impedance $Z_0$ of the attached waveguide. To keep things simple let us assume that the RF generator is also matched to $Z_0$ so that we can establish a lumped element circuit as shown in Fig.~\ref{fig:pclumped}.

\noindent Considering the coupler as a transformer we can write that:
\begin{equation}
\left.
\begin{array}{ll}
V_0T &= nV_{gen} \\
I &= \displaystyle\frac{I_{gen}}{n}
\end{array}
\right\} \Rightarrow Z_c = \frac{V_0T}{I} = n^2 Z'_c
\end{equation}
\noindent which means that the cavity impedance\\ 

\vspace*{-0.7cm}
\meqn{
Z_c = \frac{1}{\displaystyle i\omega C + \frac{1}{i\omega L} + \frac{1}{R^c}}
}{cavity impedance}{}
\noindent is transformed into\\

\vspace*{-0.7cm}
\meqn{
Z'_c = \frac{1}{\displaystyle i\omega n^2 C + \frac{n^2}{i\omega L} + \frac{n^2}{R^c}}
}{cavity+coupler impedance}{}
which is the impedance ``seen'' from the waveguide. The stored energy in the resonator expressed in lumped circuit values becomes\\

\vspace*{-0.7cm}
\meqn{
W= \frac{C}{2 (V_0T)^2} = n^2 \frac{C}{2 V_{gen}^2}
}{stored energy}{}
\noindent and the dissipated power can be written as\\

\vspace*{-0.7cm}
\meqn{
P_d = \frac{(V_0T)^2}{2R^c} = n^2 \frac{V_{gen}^2}{2R^c}.
}{dissipated power}{}
\noindent Now we can define the quality factor of the unloaded cavity with lumped circuit elements: \\

\vspace*{-0.7cm}
\meqn{
Q_0 = \frac{\omega_0 W}{P_d} = \omega_0 R^c C.
}{unloaded Q}{}
\noindent When the generator is switched off the stored energy of the cavity will not only be dissipated in the~cavity walls, but the power $P_{ex}$ will leak out through the power coupler.\\

\vspace*{-0.7cm}
\meqn{
P_{ex} = \frac{V_{gen}^2}{2Z_0}
}{}{eq:pex}

\noindent With $P_{ex}$ one can define the quality factor of the external load. The external $Q$ is thus defined as:\\

\vspace*{-0.7cm}
\meqn{
Q_{ex} = \frac{\omega_0 W}{P_{ex}} = n^2 \omega_0 Z_0 C.
}{external Q}{}

\subsubsection{Un-driven cavity}
In order to understand the power balance and the matching for a driven cavity with beam we start with a simple case assuming that the RF is switched off and that there is no beam in the cavity. The power balance is then\\

\vspace*{-1.1cm}
\meqn{
 P_{tot} = P_d + P_{ex}
 }{power balance of un-driven cavity}{}
 \noindent with which we can define the so-called ``loaded $Q$'' of the ensemble of cavity and coupler\\
 
 \vspace*{-0.9cm}
 \meqn{
 \frac{1}{Q_l} = \frac{1}{Q_{ex}} + \frac{1}{Q_0}.
 }{loaded $Q$}{}
\noindent The coupling between cavity and waveguide is described by the coupling factor $\beta$\\

\vspace*{-0.9cm}
\meqn{
\beta = \frac{P_{ex}}{P_d} = \frac{Q_0}{Q_{ex}} = \frac{R^c}{n^2 Z_0}.
}{coupling factor}{}
\noindent Optimum power transfer between cavity (+coupler) and waveguide takes place when the impedance at the coupler input equals the waveguide impedance at the resonance frequency of the cavity. We know that the cavity impedance becomes real at resonance, which means that
\begin{equation}
Z_c = R^c=n^2Z'_c\stackrel{!}{=}n^2Z_0 \hspace{0.5cm} \Rightarrow \hspace{0.5cm} \beta=1.
\end{equation}
 It is important to keep in mind that the ``matching condition'' $\beta=1$ is only valid for a cavity without beam. 
 
 \subsubsection{RF on, beam on}
Once we take into account the beam loading the power needed in the cavity increases and will yield a different value for the coupling factor $\beta$ at the point of optimum power transfer. A simple way to introduce the beam is to treat it as an additional loss in the cavity, which can be added to the dissipated power in the cavity walls.\\

\vspace*{-0.7cm}
\meqn{
P_{db} = P_d + P_b
}{dissipated power + beam power}{}
As in the case without beam one achieves maximum power transfer to the cavity, when the input impedance of the coupler equals the impedance of the waveguide. This condition yields zero reflections and also implies that the power needed in the cavity $P_{bd}$ 
(for losses and beam) has to be equal to $P_{ex}$ as defined in Eq.~\eqref{eq:pex}. This means that
\begin{equation}
\frac{P_{ex}}{P_{db}} = 1 = \frac{Q_{0b}}{Q_{ex}} \hspace{0.5cm} \Rightarrow \frac{P_{ex}}{P_d} = 1+\frac{P_b}{P_d}
\end{equation}
where we have introduced a quality factor $Q_{0b}$ for the cavity plus beam. For the matched condition we therefore obtain a coupling factor of\\

\vspace*{-0.9cm}
\meqn{
\Rightarrow \beta = 1+\frac{P_b}{P_d}
}{matched coupling factor with beam}{}
and the following quality factors:

\vspace*{-0.5cm}
\meqn{
Q_{ex} = Q_{0b} = \frac{\omega_0 W}{P_b+P_d}= \frac{Q_0}{1+\frac{P_b}{P_d}}= \frac{Q_0}{\beta}
}{external $Q$ with beam}{}

\vspace*{-0.7cm}
\meqn{
Q_l = \frac{Q_0}{1+\beta} =\frac{Q_0}{2+\frac{P_b}{P_d}}.
}{loaded $Q$ with beam}{}
In superconducting cavities one can generally assume that $P_b \gg P_d$, which means that the coupling factor for the matched condition can be written as

\vspace*{-0.5cm}
\meqn{
\beta = 1+ \frac{P_{b}}{P_d} \approx \frac{P_{b}}{P_d}.
}{matched coupling factor for SC cavity + beam}{}
Using

\vspace*{-0.7cm}
\meqn{
P_b = I_{beam} V_0 T\cos\phi_s
}{}{}\\
we can write a simple expression to calculate the loaded and external Q for SC cavities as

\vspace*{-0.5cm}
\meqn{
Q_l \approx Q_{ex} \approx \frac{Q_0}{P_{beam}/P_d} = \frac{V_0T}{(R/Q)I_{beam}\cos\phi_s}.
}{$\mathbf{Q_{l/ex}}$ for SC cavities}{}
The results of this paragraph can be summarized as shown in Table~\ref{tab:QPlex}.

\begin{table}[h!]
\centering
\caption{Definitions of $Q$ values and coupling factors for driven and undriven cavities}
\begin{tabular}{l|c|c}
\hline\hline
 & {\bf undriven cavity} & {\bf driven cavity} \\
 \hline  & \multicolumn{2}{c}{\parbox[0pt][1.2cm][c]{4cm}{\centerline{$\displaystyle \frac{1}{Q_l} = \frac{1}{Q_{ex}} + \frac{1}{Q_0}$}}} \\
  {\bf general} & \multicolumn{2}{c}{\parbox[0pt][1.2cm][c]{4cm}{\centerline{$\displaystyle \beta = \frac{P_{ex}}{P_d} = \frac{Q_0}{Q_{ex}}$}}} \\
  & \multicolumn{2}{c}{\parbox[0pt][1.2cm][c]{4cm}{\centerline{$\displaystyle Q_l = \frac{Q_0}{1+\beta}$}}} \\
  \hline
  & \parbox[0pt][1.2cm][c]{4.5cm}{\centerline{$\displaystyle \frac{P_{ex}}{P_d} = \frac{Q_0}{Q_{ex}} = 1 \Rightarrow \beta = 1$}} & \parbox[0pt][1.2cm][c]{4.5cm}{\centerline{$\displaystyle \frac{P_{ex}}{P_{db}} = \frac{Q_{0b}}{Q_{ex}} = 1 \Rightarrow \beta = 1+ \frac{P_b}{P_d}$}} \\
  {\bf matched case} & \parbox[0pt][1.2cm][c]{4.5cm}{\centerline{$\displaystyle Q_{ex} = Q_0 = \frac{\omega_0 W}{P_d}$}} & $\displaystyle Q_{ex} = Q_{0b} = \frac{\omega_0 W}{P_d + P_b}$ \\
  & \parbox[0pt][1.2cm][c]{4.5cm}{\centerline{$\displaystyle Q_l = \frac{Q_0}{2}$}} & $\displaystyle Q_l = \frac{Q_0}{2+\frac{P_b}{P_d}}$ \\
  \hline\hline
\end{tabular}
\label{tab:QPlex}
\end{table}

\subsection{``Matching'' a cavity}
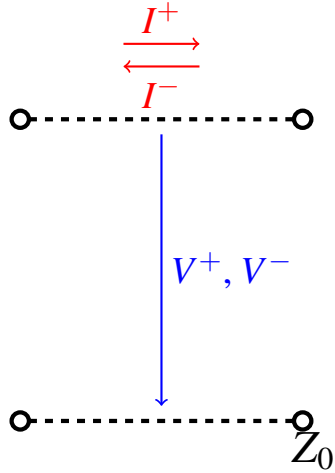
\begin{figure}[h]
\centering
\begin{tikzpicture}[ultra thick]

\draw[dashed, o-o] (0,0) -- (4,0) node [below] {\LARGE $Z_0$};
\draw[dashed, o-o] (0,4) -- (4,4);
\draw[thick,->,blue] (2,3.8) -- (2,0.2) node [right, midway] {\Large $V^{+}$, $V^{-}$};
\draw[thick,->,red] (1.5,5) -- (2.5,5) node [above, midway] {\Large $I^{+}$}; 
\draw[thick,<-,red] (1.5,4.7) -- (2.5,4.7) node [below, midway] {\Large $I^{-}$}; 

\end{tikzpicture}
\caption{\label{fig:tline}Voltages and currents along a transmission line}
\end{figure}

In the last section it was claimed that parts of an electromagnetic wave are reflected when it ``sees'' a change in impedance during its propagation. In fact the whole purpose of the power coupler was to transform the impedance of the waveguide into the impedance of the cavity. In this last section we will see why that is so. For this purpose we look at a transmission line as shown in Fig.~\ref{fig:tline}.

The transmission line is representative for a waveguide, a coaxial line or any other kind of transport geometry used to guide electromagnetic waves. Since waves can travel in positive and negative $z$-direction, one introduces a sign convention of the associated voltages and currents as shown in Fig.~\ref{fig:tline}, where the voltage vector of forward and reflected wave has the same direction while the current vectors have opposite directions. 

Using the same time- and location dependance as for electric and magnetic fields in waveguides we can write
\begin{align}
V &= V_0 e^{i(kz - \omega t)} + \Gamma V_0 e^{i(-kz - \omega t)}\\
I &= \frac{V_0}{Z_0} e^{i(kz - \omega t)} - \Gamma \frac{V_0}{Z_0} e^{i(-kz - \omega t)}
\intertext{where we have introduced a reflection coefficient $\Gamma$. If we connect a cavity with the impedance $Z'_c$ at $z=0$ the expressions above simplify to}
V &= V_0 e^{-i\omega t} (1+ \Gamma) \\
I &= \frac{V_0}{Z_0} e^{-i\omega t} (1- \Gamma)
\intertext{and the cavity impedance can be expressed in terms of the transmission line impedance $Z_0$ and the~reflection coefficient $\Gamma$}
Z'_c &= \frac{V}{I} = Z_0 \frac{1+\Gamma}{1-\Gamma}.
\intertext{We can then re-arrange the equation for the reflection coefficient and get}
\Gamma &= \frac{Z'_c-Z_0}{Z'_c+Z_0} = \frac{1-\beta}{1+\beta}.
\end{align}

From the last equation we can see that the reflections only disappear for $Z'_c = Z_0$, the ``matched condition'' where the waveguide impedance equals the cavity impedance. In the case without beam this corresponds to a coupling factor of $\beta=1$.

In the context of matching we therefore have to consider the following points:
\begin{itemize}
\item At the resonance frequency the power coupler transforms the cavity impedance into the impedance of the waveguide. 
\item If the cavity is resonating off-resonance or if the coupler is mis-matched power gets reflected and travels back to the RF source.
\item Since the cavity impedance depends on the cavity $Q$, and since in reality most cavities have different $Q$-values, each cavity needs a different matching. 
\item Beam loading increases the needed power in the cavity and changes the loaded $Q$ and the cavity impedance. Power couplers are usually matched for the case with beam loading. 
\item During the start of the RF pulse (before the arrival of the beam), when the cavity is ``filled'' with RF power, the cavity is always mismatched, which means we need to make sure that the reflected power does not damage the RF source (e.g. with a circulator between the cavity and the RF source).
\end{itemize}

\begin{figure}
\centering
\includegraphics[width=0.7\textwidth]{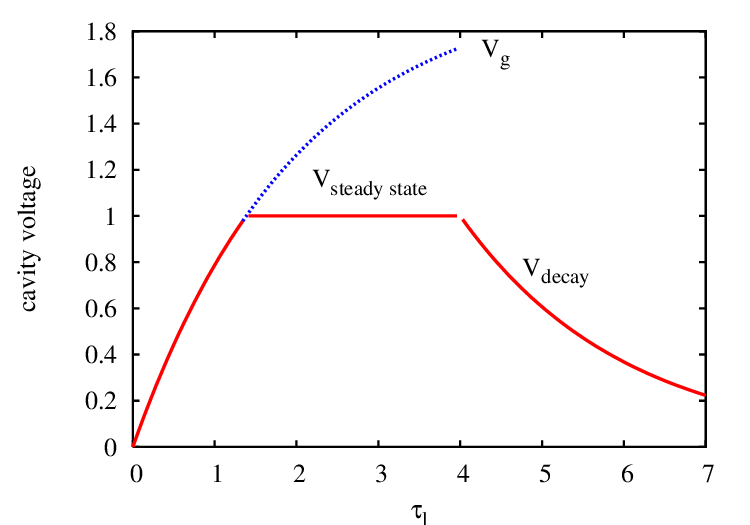}
\caption{\label{fig:SCpulsed}Voltage profile in a pulsed SC cavity.}
\end{figure}

The last point is especially pronounced in superconducting cavities, where the dissipated power is negligible with respect to the power taken by the beam. In this case one gets basically full reflection of the RF wave at the beginning of the RF pulse before the cavity field increases to its nominal level. At that point the beam should enter the cavity and from then onwards the RF generator is matched to the~power needs of the RF cavity. After switching off the RF signal the cavity voltage decays exponentially as shown in Fig.~\ref{fig:SCpulsed} (more details in \cite{bib:FGpowcon}).

\section*{Acknowledgements}

For the preparation of this lecture I have made extensive use of the material listed below in the Bibliography.

\section*{Bibliography}
\begin{itemize}
\item T. Weiland, M. Krasilnikov, R. Schuhmann, A. Skarlatos, M. Wilke: Review of Theory (I, II, III) (CAS RF Engineering 2005, Seeheim, Germany)
\item T. Wangler: Principles of RF Linear Accelerators (Wiley \& Sons)
\item A. Wolski: Theory of Electromagnetic Fields (CAS RF Engineering, 2010, Ebeltoft, Denmark)
\item H. Henke: Theoretische Elektrotechnik (German script of lectures on Electrodynamics at the Technical University of Berlin)
\item H. Henke: Basic Concepts I and II (CAS, RF Engineering, 2005, Seeheim, Germany)
\item K. Simonyi: Foundations of Electrical Engineering, 3d Volume (1963, Pergamon Press, NY), Hungarian Title: Elméleti villamossagtan Tankonyvkiado (1973, Budapest), German Title: Theoretische Elektrotechnik (VEB Deutscher Verlag der Wissenschaften, 1973)
\item Padamsee, Knobloch, Hays: RF Superconductivity for Accelerators (Wiley \& Sons). 
\end{itemize}

\appendix
\section{Cartesian coordinates ($x$, $y$, $z$)}
\subsection{Differential elements}

\eqn{$\hfill
\dd\mathbf{l} = \threevector
{\dd x} {\dd y}{\dd z}\hfill $}{path element}{}\\
\eqn{$\hfill
\dd V = \dd x \dd y \dd z
\hfill $}{volume element}{}

\subsection{Differential operators}
\eqn{$\hfill\nabla\phi=\threevector{\frac{\partial\phi}{\partial x}}{\frac{\partial\phi}{\partial y}}{\frac{\partial\phi}{\partial z}}\hfill $}{gradient}{}\\
\eqn{$\hfill\nabla\cdot\mathbf{a}= \displaystyle\frac{\partial a_x}{\partial x} +\frac{\partial a_y}{\partial y} + \frac{\partial a_z}{\partial z}\hfill $}{divergence}{}

$\left. \right.$\\
\eqn{$\hfill\nabla\times\mathbf{a} = \threevector{\frac{\partial a_z}{\partial y}-\frac{\partial a_y}{\partial z}}{\frac{\partial a_x}{\partial z}-\frac{\partial a_z}{\partial x}}{\frac{\partial a_y}{\partial x}-\frac{\partial a_x}{\partial y}}\hfill $}{curl}{}\\
\eqn{$\hfill\Delta\phi = \displaystyle\frac{\partial^2\phi}{\partial x^2} + \frac{\partial^2\phi}{\partial y^2} + \frac{\partial^2\phi}{\partial z^2}\hfill $}{Laplace}{}

\section{Cylindrical coordinates ($r$, $\phi$, $z$)}
\label{sec:cylindrical}
\subsection{Transformations}

\meqn{
x &= r \cos\varphi \\
y &= r \sin\varphi \\
z &= z
}{}{}

with $ 0 \le r \le \infty$, $0 \le \varphi \le 2\pi $

\subsection{Differential elements}
\eqn{$\hfill
\dd\mathbf{l} = \threevector{\dd r}{r \dd\varphi}{\dd z}\hfill $}
{path element}{}\\
\eqn{$\hfill
\dd V = r \dd r \dd\varphi \dd z
\hfill $}{volume element}{}

\subsection{Differential operators}

$\left. \right.$\\
\eqn{$\hfill
\nabla\phi = \threevector{
	\frac{\partial \phi}{\partial r}}
{
	\frac{1}{r}\frac{\partial \phi}{\partial \varphi}}
{
	\frac{\partial \phi}{\partial z}}
\hfill $}{gradient}{}\\
\eqn{$\hfill
\nabla \cdot \mathbf{a} =
\displaystyle
\frac{1}{r}\frac{\partial (r a_r)}{\partial r} +
\frac{1}{r}\frac{\partial a_{\varphi}}{\partial\varphi} +
\frac{\partial a_z}{\partial z}
\hfill $}{divergence}{}

$\left. \right.$\\
\eqn{$\hfill
\nabla\times \mathbf{a} = \threevector{
	\frac{1}{r}\frac{\partial a_z}{\partial\varphi} - \frac{\partial a_{\varphi}}{\partial z}}
{
	\frac{\partial a_r}{\partial z} - \frac{\partial a_z}{\partial r}}
{
	\frac{1}{r} \left(
		\frac{\partial (ra_{\varphi})}{\partial r} -
		\frac{\partial a_r}{\partial\varphi}
\right)}
\hfill $}{curl}{}\\
\eqn{$\displaystyle\hfill
\Delta\phi = \frac{\partial^2 \phi}{\partial r^2} + \frac{1}{r}\frac{\partial\phi}{\partial r} + \frac{1}{r^2}\frac{\partial^2\phi}{\partial \varphi^2} + \frac{\partial^2\phi}{\partial z^2}\hfill $}{Laplace}{}

\section{Spherical coordinates ($r$, $\vartheta$, $\varphi$)}
\label{sec:spherical}
\subsection{Transformations}
\meqn{
x &= r\sin\vartheta\cos\varphi  \\
y &= r\sin\vartheta\sin\varphi \\
z &= r\cos\vartheta
}{}{}

with $0 \le r \le \infty$, $0\le \vartheta \le \pi$, $0\le\varphi \le 2\pi$

\subsection{Differential elements}
\eqn{$\hfill\dd\mathbf{l} = \threevector{\dd r}{r \dd\vartheta}{r\sin\vartheta\dd\varphi}\hfill $}{path element}{}\\
\eqn{$\hfill\dd V = r^2\sin\vartheta \dd r \dd\vartheta \dd\varphi\hfill $}{volume element}{}

\subsection{Differential operators}
\eqn{$\hfill
\nabla\phi = \threevector{\frac{\partial\phi}{\partial r}}
{\frac{1}{r}\frac{\partial\phi}{\partial\vartheta}}
{\frac{1}{r\sin\vartheta}\frac{\partial\phi}{\partial\varphi}}
\hfill $}{gradient}{}\\
\eqn{$\hfill
\nabla\cdot\mathbf{a} = \displaystyle\frac{1}{r^2}\frac{\partial(r^2a_r)}{\partial r} + \frac{1}{r\sin\vartheta}\frac{\partial(a_{\vartheta}\sin\vartheta)}{\partial\vartheta} + \frac{1}{r\sin\vartheta}\frac{\partial a_{\varphi}}{\partial\varphi}
\hfill $}{divergence}{}\\
\eqn{$\hfill
\nabla\times\mathbf{a} = \threevector{\frac{1}{r\sin\vartheta}\left( \frac{\partial(a_{\varphi}\sin\vartheta)}{\partial\vartheta} - \frac{\partial a_{\vartheta}}{\partial\varphi}\right)}{\frac{1}{r}\left( \frac{1}{\sin\vartheta}\frac{\partial a_r}{\partial\varphi} - \frac{\partial(r a_{\varphi})}{\partial r}  \right)}{\frac{1}{r}\left( \frac{\partial (r a_{\vartheta})}{\partial r} - \frac{\partial a_r}{\partial\vartheta}   \right)}
\hfill $}{curl}{}\\
\meqn{
\Delta\phi &= \displaystyle\frac{\partial^2\phi}{\partial r^2} + \frac{2}{r}\frac{\partial\phi}{\partial r} + \frac{1}{r^2\sin\vartheta}\frac{\partial}{\partial\vartheta}\left( \sin\vartheta\frac{\partial\phi}{\partial\vartheta} \right)\\ &+ \frac{1}{r^2\sin^2\vartheta}\frac{\partial^2\phi}{\partial\varphi^2}
}{Laplace}{}

\section{Useful relationships}

\begin{align}
\nabla\cdot\left(\mathbf{a}\times\mathbf{b}\right) &= \mathbf{b} \cdot\left(\nabla\times\mathbf{a} \right) - \mathbf{a}
\cdot\left(\nabla\times\mathbf{b}\right) \label{eq:va1} \\
\nabla \cdot \left( \nabla \times \mathbf{a} \right) &= 0 
\end{align}

\section{Basic Vectoranalysis and its application to Maxwell's Equations}
\label{annex:basicanalysis}
In order to make efficient use Maxwell's Equations some basic Vectoranalysis is needed, which is introduced in this section. More detailed introductions can be found in a number of textbooks as for instance in the excellent Feynman lectures on Physics \cite{bib:feynman}.
\subsubsection*{Gradient of a potential}
The gradient of a potential $\phi$ is the derivation of the potential function $\phi(x,y,z)$ in all directions of a~particular coordinate system (e.g.\ $x$, $y$, $z$). The result is a vector which tells us by how much a potential changes in the different directions. Applied to the geographical profile of a mountain landscape the~gradient describes the slopes of the landscape in all directions. The mathematical sign for the gradient of a~potential one uses the ``nabla operator'', and applied to a cartesian coordinate system one can write\\
\meqn{
	\nabla \Phi = \begin{pmatrix} \frac{\partial}{\partial x} \\[1ex] \frac{\partial}{\partial y} \\[1ex] \frac{\partial}{\partial z} \end{pmatrix} 
	\Phi = \begin{pmatrix} \frac{\partial\Phi}{\partial x} \\[1ex] \frac{\partial\Phi}{\partial y} \\[1ex] \frac{\partial\Phi}{\partial z} \end{pmatrix}.
}{gradient of a potential}{gradient}   

\noindent The gradient expressions for cylindrical and spherical coordinate systems are given in Annex \ref{sec:cylindrical} and \ref{sec:spherical}.

\subsubsection*{Divergence of a vector field}

The divergence of a vector field $\mathbf{a}$ tells us if the vector field has a source. If the resulting scalar expression is zero we have a ``source-free'' vector field like for instance the magnetic field. From basic physics we know that there are no magnetic monopoles, which is why magnetic field lines are always closed.  In Maxwell's Equations \eqref{eq:divB} this is property included by the fact that the divergence of the magnetic induction $\mathbf{B}$ equals zero. 

In cartesian coordiates the divergence of a vector field is defined as\\
\meqn{\displaystyle
	\nabla \cdot \mathbf{a} = \begin{pmatrix} \frac{\partial}{\partial x} \\[1ex] \frac{\partial}{\partial y} \\[1ex] \frac{\partial}{\partial z}
	 \end{pmatrix} \cdot \begin{pmatrix} a_x \\[1ex] a_y \\[1ex] a_z \end{pmatrix} = \frac{\partial a_x}{\partial x} + 
	 \frac{\partial a_y}{\partial y} + \frac{\partial a_z}{\partial z}.
}{divergence of a vector}{eq:divergence}

\noindent The expressions for cylindrical and spherical coordinate systems are given in Annex \ref{sec:cylindrical} and \ref{sec:spherical}.

\subsubsection*{Curl of a vector field}
When forming the curl of a vector we are interested to know if there are any curls or eddies in the field. Let us imagine that we look at the flow of water in a cooling pipe. To check for curls we can use a stick around which a ball can rotate freely. We position a cartesian coordinate system at an arbitrary origin and align the stick first with the $x$-axis, and then with the $y$ and $z$ axes. If the ball starts rotating in any of these positions, then we know that the curl of the vector field describing the water flow is non-zero in the respective axis directions. The curl of a vector $\mathbf{a}$ is therefore also a vector because its information is direction specific. Its mathematical form in cartesian coordinates is defined as\\
\meqn{
	\nabla \times \mathbf{a} &= \begin{pmatrix} \frac{\partial}{\partial x} \\[1ex] \frac{\partial}{\partial y} \\[1ex] \frac{\partial}{\partial z}
	 \end{pmatrix} \times \begin{pmatrix} a_x \\[1ex] a_y \\[1ex] a_z \end{pmatrix} \\
	 &= det \begin{pmatrix} \mathbf{u_x} & \mathbf{u_y} & \mathbf{u_z} \\[1ex] 
	 \frac{\partial}{\partial x} & \frac{\partial}{\partial y} & \frac{\partial}{\partial z} \\[1ex] a_x & a_y & a_z \end{pmatrix} 
	 =  \begin{pmatrix} \frac{\partial a_z}{\partial y} - 
	 \frac{\partial a_y}{\partial z} \\[1ex] \frac{\partial a_x}{\partial z} - 
	 \frac{\partial a_z}{\partial x} \\[1ex] \frac{\partial a_y}{\partial x} - \frac{\partial a_x}{\partial y} \end{pmatrix}.
}{curl of a vector}{eq:curl}

\noindent The \textit{unity} vectors $\mathbf{u_n}$ have no physical meaning and simply point in the $x$, $y$, and $z$ directions. They have a constant length of 1. The expressions for cylindrical and spherical coordinate systems can be found in Annex \ref{sec:cylindrical} and \ref{sec:spherical}.  

\subsubsection*{2nd derivations}
In some instances we have to make use of 2nd derivations. One of the expressions that is used regularly in electrodynamics is the Laplace operator $\Delta = \nabla^2$, which --- since the operator itself is scalar--- can be applied to scalar fields and vector fields

\vspace*{-0.6cm}
\meqn{
	\displaystyle \Delta \phi = \nabla \cdot \left( \nabla \phi \right) = \nabla^2 \phi
	 = \frac{\partial^2 \phi }{\partial x^2} + \frac{\partial^2 \phi}{\partial y^2} + \frac{\partial^2 \phi}{\partial z^2} .
}{Laplace operator}{eq:laplace}
The expressions for cylindrical and spherical coordinate systems can be found in Annex \ref{sec:cylindrical} and \ref{sec:spherical}.

Furthermore we introduce two interesting identities: 
\begin{align}
	\nabla \times (\nabla \phi) &= 0 \label{eq:potential0}\\
	\nabla\cdot (\nabla \times \mathbf{a}) &= 0. \label{eq:vector0}
\end{align}
Equation~\eqref{eq:potential0} tells us that if the curl of a vector equals zero, then this vector can be written as the~gradient of a potential. This feature can save us a lot of writing, when we deal with complicated three dimensional expressions for the electric and magnetic fields and we will use this principle later on to define non-physical potential functions that can describe (via derivations) complete three dimensional vector functions.

In the same way Eq.~\eqref{eq:vector0} can (and will...) be used to describe divergence-free fields with simple ``vector potentials''.



\end{document}